%% file: thesis.tex
\documentclass[12pt]{report}   
\usepackage{graphicx}  
\usepackage[letterpaper, left=1in, right=1in, top=1in, bottom=1in]{geometry}
\usepackage{setspace}  
\usepackage{times}  
\usepackage[explicit]{titlesec}  
\usepackage[titles]{tocloft}  
\usepackage[utf8]{inputenc} 
\usepackage[backend=bibtex, sorting=none, bibstyle=ieee]{biblatex}  
\usepackage{appendix}  
\usepackage{rotating}  
\usepackage[normalem]{ulem}  
\usepackage{textcomp} 
\usepackage{indentfirst} 
\usepackage{booktabs,array,arydshln} 
\usepackage{amsmath} 
\usepackage[T1]{fontenc} 

\usepackage[hidelinks]{hyperref}
\usepackage{amssymb}   
\usepackage{physics}   
\usepackage{verbatim} 
\usepackage{multirow} 
\usepackage[compat=1.1.0]{tikz-feynman} 
\usepackage{pdfpages} 
\usepackage{caption}   
\usepackage{subcaption}  
\usepackage{slashed} 

\usepackage[colorinlistoftodos,prependcaption,textsize=large]{todonotes}    

\DeclareUnicodeCharacter{2212}{FIX ME!!}  
\DeclareUnicodeCharacter{202F}{\,}

\presetkeys%
    {todonotes}%
    {inline,bordercolor=white,textcolor=red,backgroundcolor=pink}{}

\bibliography{references}

\AtEveryBibitem{\clearfield{issn}}
\AtEveryBibitem{\clearlist{issn}}

\AtEveryBibitem{\clearfield{language}}
\AtEveryBibitem{\clearlist{language}}

\AtEveryBibitem{\clearfield{doi}}
\AtEveryBibitem{\clearlist{doi}}

\AtEveryBibitem{\clearfield{url}}
\AtEveryBibitem{\clearlist{url}}

\AtEveryBibitem{%
  \ifentrytype{online}
    {}
    {\clearfield{urlyear}\clearfield{urlmonth}\clearfield{urlday}}}

\begin{document}
\doublespacing  

\input{titlePage.tex}


\input{CopyrightPage.tex}

\pagenumbering{gobble}
\input{Abstract.tex}

\pagenumbering{roman}
\setcounter{page}{1} 
\renewcommand{\cftchapdotsep}{\cftdotsep}  
\renewcommand{\cftchapfont}{\normalfont}  
\renewcommand{\cftchappagefont}{}  
\renewcommand{\cftchappresnum}{Chapter }
\renewcommand{\cftchapaftersnum}{:}
\renewcommand{\cftchapnumwidth}{5em}
\renewcommand{\cftchapafterpnum}{\vskip\baselineskip} 
\renewcommand{\cftsecafterpnum}{\vskip\baselineskip}  
\renewcommand{\cftsubsecafterpnum}{\vskip\baselineskip} 
\renewcommand{\cftsubsubsecafterpnum}{\vskip\baselineskip} 

\titleformat{\chapter}[display]
{\normalfont\bfseries\filcenter}{\chaptertitlename\ \thechapter}{0pt}{\large{#1}}

\renewcommand\contentsname{Table of Contents}

\begin{singlespace}
\tableofcontents
\setlength{\cftparskip}{\baselineskip}
\listoffigures
\listoftables
\end{singlespace}

\clearpage

\phantomsection
\addcontentsline{toc}{chapter}{Acknowledgments}
\input{acknowledgements.tex}

\phantomsection
\addcontentsline{toc}{chapter}{Dedication}
\input{Dedication.tex}



\clearpage
\pagenumbering{arabic}
\setcounter{page}{1} 

\phantomsection
\addcontentsline{toc}{chapter}{Introduction}
\input{Preface.tex}

\titleformat{\chapter}[display]
{\normalfont\bfseries\filcenter}{}{0pt}{\large\chaptertitlename\ \large\thechapter : \large\bfseries\filcenter{#1}}  
\titlespacing*{\chapter}
  {0pt}{0pt}{30pt}	
  
\titleformat{\section}{\normalfont\bfseries}{\thesection}{1em}{#1}

\titleformat{\subsection}{\normalfont}{\thesubsection}{0em}{\hspace{1em}#1}



\input{chapter1.tex}


\input{chapter2.tex}

\input{chapter2.1.tex}


\input{chapter3.tex}


\input{chapter4.tex}


\input{chapter5.tex}


\clearpage
\phantomsection
\addcontentsline{toc}{chapter}{Conclusion}
\input{Conclusion.tex}

\clearpage
\phantomsection 
\titleformat{\chapter}[display]
{\normalfont\bfseries\filcenter}{}{0pt}{\large\bfseries\filcenter{#1}}  
\titlespacing*{\chapter}
  {0pt}{0pt}{30pt}

\begin{singlespace}  
\setlength\bibitemsep{\baselineskip}  
\addcontentsline{toc}{chapter}{References}  

\printbibliography[title={References}]

\end{singlespace}


\titleformat{\chapter}[display]
{\normalfont\bfseries\filcenter}{}{0pt}{\large\chaptertitlename\ \large\thechapter : \large\bfseries\filcenter{#1}}  
\titlespacing*{\chapter}
  {0pt}{0pt}{30pt}	
  
\titleformat{\section}{\normalfont\bfseries}{\thesection}{1em}{#1}

\titleformat{\subsection}{\normalfont}{\thesubsection}{0em}{\hspace{1em}#1}

\input{appendix.tex}

\end{document}

%% file: titlePage.tex

\begin{titlepage}
\begin{center}

\begin{singlespacing}
\vspace*{6\baselineskip}
MicroBooNE investigations on the photon interpretation of the MiniBooNE\\ low energy excess \\
\vspace{3\baselineskip}
Guanqun Ge\\
\vspace{18\baselineskip}
Submitted in partial fulfillment of the\\
requirements for the degree of\\
Doctor of Philosophy\\
under the Executive Committee\\
of the Graduate School of Arts and Sciences\\
\vspace{3\baselineskip}
COLUMBIA UNIVERSITY\\
\vspace{3\baselineskip}
\the\year
\vfill

\end{singlespacing}

\end{center}
\end{titlepage}

%% file: CopyrightPage.tex
\begin{titlepage}
\begin{singlespacing}
\begin{center}

\vspace*{35\baselineskip}

\textcopyright  \,  \the\year\\
\vspace{\baselineskip}	
Guanqun Ge\\
\vspace{\baselineskip}	
All Rights Reserved
\end{center}
\vfill

\end{singlespacing}
\end{titlepage}

%% file: Abstract.tex

\begin{titlepage}
\begin{center}

\vspace*{5\baselineskip}
\textbf{\large Abstract}

MicroBooNE investigations on the photon interpretation of the MiniBooNE\\ low energy excess 

Guanqun Ge
\end{center}
The MicroBooNE experiment is a liquid argon time projection chamber with 85-ton active volume at Fermilab, operated from 2015 to 2020 to collect neutrino data from Fermilab's Booster Neutrino Beam. One of MicroBooNE's physics goals is to investigate possible explanations of the low-energy excess observed by the MiniBooNE experiment in $\nu_{\mu}\rightarrow \nu_{e}$ neutrino oscillation measurements. MicroBooNE has performed searches to test hypothetical interpretations of the MiniBooNE low-energy excess, including the underestimation of the photon background or instrinic $\nu_{e}$ background. This thesis presents MicroBooNE's searches for two neutral current (NC) single-photon production processes that contribute to the photon background of the MiniBooNE measurement: NC $\Delta$ resonance production followed by $\Delta$ radiative decay: $\Delta \rightarrow N\gamma$, and NC coherent single-photon production. Both searches take advantage of boosted decision trees to yield efficient background rejection, and a high-statistic NC $\pi^0$ measurement to constrain dominant background, and make use of MicroBooNE's first three years of data. The NC $\Delta \rightarrow N\gamma$ measurement yielded a bound on the $\Delta$ radiative decay process at 2.3 times the predicted nominal rate at 90\% confidence level
(C.L.), disfavoring a candidate photon interpretation of the MiniBooNE low-energy excess as a factor of 3.18 times the nominal NC $\Delta$ radiative decay rate at the 94.8\% C.L. The NC coherent single photon measurement leads to the world's first experimental limit on the cross-section of this process below 1 GeV, of $1.49 \times 10^{-41} \text{cm}^2$ at 90\% C.L., corresponding to 24.0 times the nominal prediction.  
\vspace*{\fill}
\end{titlepage}

%% file: acknowledgements.tex

\clearpage
\begin{center}

\vspace*{5\baselineskip}
\textbf{\large Acknowledgements}
\end{center}

First, I would like to thank my advisor, Georgia Karagiorgi, for her guidance throughout my Ph.D. Georgia has been an incredible mentor, always providing insightful and creative research suggestions and ideas, patiently reminding me of the big picture when I get lost, and encouraging me to be proactive. I couldn't ask for a better mentor than her. Georgia is not only a role model for me in research but also in life. I always remember the snow walk at the river bank in 2021 when I was inspired by her to start self-development, which led me to discover the world of violin.

I also want to thank Mark Ross-Lonergan for his guidance on many projects we have collaborated on, for regular meetings, and for hopping on calls whenever I have questions. I would like to thank Jos\'{e} Crespo-Anad\'{o}n for sharing knowledge on readout electronics and introducing me to the word ``power-cycle''.

I would like to thank Mike Shaevitz and Leslie Camilleri for their constructive suggestions on my analysis especially when I am stuck. I want to thank my current and past labmates, Yeon-jae, Kathryn, Iris, Keng, Ibrahim, Akshay, Nupur, Seokju and Karan, with whom I share discussions and laughter. I thank Bill Seligman for his support on computing issues at Nevis, and thank Amy Garwood and Grace Ho for their efforts in making Nevis such a lovely place to work.

I want to thank my boyfriend, Zihuai, for the laughter we shared, for his support, patience, and resilience,  which inspire me to be a better person, and for being my emotional safety net. 

I appreciate the support from my family when I needed it the most. Thank you for supporting my decisions despite our differing opinions, and for reminding me that I am always unconditionally loved. I love you so much and I feel so lucky to be a part of this family. 

I would also like to acknowledge the National Science Foundation (NSF) for their support. This work was supported by the NSF under Grant Nos. PHY-2310080, PHY-2013070, and PHY-1707971.
\clearpage


%% file: Dedication.tex

\begin{center}

\vspace*{5\baselineskip}
\end{center}

\centerline{\textit{To my parents, sis, and grandma.}}
\centerline{\textit{To my dear grandpa, in loving memory.}}


%% file: Preface.tex

\begin{center}
\vspace*{5\baselineskip}
\textbf{\large Introduction}
\end{center}

The field of neutrino physics has evolved significantly in the recent decades in both experimental and theoretical fronts, and has witnessed several milestones over the past century: postulation of neutrino, integration of neutrino in the Standard Model (SM), experimental discovery of neutrinos, and the confirmation of neutrino oscillation. While the neutrinos are massless in the SM theory, experimental observation of neutrino oscillation implies non-zero neutrino mass therefore physics Beyond the Standard Model (BSM). Besides the origin of neutrino mass, there are other remaining questions such as, what is the mass hierarchy of neutrinos, do neutrinos and anti-neutrinos behave the same or differently (CP conservation or CP violation), why is our universe dominated with matter, and do neutrinos have something to do with this? etc. Answering these questions requires precise measurements of neutrino interaction cross-sections and neutrino oscillation parameters.

For muon neutrino ($\nu_{\mu}$) to electron neutrino ($\nu_{e}$) oscillation measurements, neutral current single photon (NC $1\gamma$) events have been an important background, especially for experiments with difficulty in distinguishing electrons from photons. One such example is the MiniBooNE experiment using a Cherenkov detector, which was designed to measure $\nu_{e}$ appearance at $\mathcal{O}(1)\;\text{eV}^2$ neutrino mass-squared difference with a neutrino beam at $\mathcal{O}(1)$ GeV energy regime. MiniBooNE observed an excess above the predicted background in the low energy region, and suffered from large contributions of NC $1\gamma$ backgrounds due to the similarity in Cherenkov rings from electrons and photons. 

The NC $\Delta$ radiative decay ($\Delta \rightarrow N\gamma$) process dominates NC $1\gamma$ production in the sub-GeV energy regime. This process is a very important background to the MiniBooNE $\nu_{e}$ measurement and has not been directly measured in MiniBooNE. With liquid argon time projection chamber (LArTPC) technology, which offers excellent calorimetric and spatial resolution and enables electron versus photon separation, MicroBooNE is well suited to investigate whether the MiniBooNE excess is truly a photon excess or electron excess. MicroBooNE performed a search for NC $\Delta \rightarrow N\gamma$ events, and I was involved in the validation of $\pi^0$ background modeling and signal fits for the final result. The result of this search with MicroBooNE's first three years of data has been published in Physical Review Letters and selected as editors' suggestion.

The NC coherent $1\gamma$ process is a sub-dominant contribution for NC $1\gamma$ production, with predicted rate at one tenth of that of NC $\Delta$ radiative decay, but understanding the rate of NC coherent $1\gamma$ is important for future precise neutrino measurements. This process has never been measured experimentally, and I have been leading the search for it in MicroBooNE. 
As of this writing, this analysis has been unblinded with MicroBooNE's first three years of data and is in the stage of preparing a paper for submission to Physical Review D. 

The operation of the MicroBooNE detector has facilitated exploring the LArTPC technology and R\&D studies that are critical for future experiments, such as the upcoming Short-Baseline Near Detector (SBND) as part of the Short-Baseline Neutrino (SBN) program at Fermilab, and the future Deep Underground Neutrino Experiment (DUNE). As part of the Columbia neutrino group at Nevis Labs, I participated in the production test of the TPC readout front-end modules (FEMs) for SBND. I have also studied the sensitivity of the SBN program to neutrino oscillation due to eV-scale sterile neutrinos, and currently been involved with the development of new-generation fitting software for this purpose~\cite{Schulz:2020gnu}. For the DUNE experiment, I was involved in TPC self-trigggering studies for detecting neutrinos from supernova bursts, based on convolutional neural networks.

This thesis focuses on MicroBooNE's searches for the NC incoherent $1\gamma$ production, more specifically NC $\Delta$ resonance production followed by $\Delta$ radiative decay, and NC coherent $1\gamma$. Chapter~\ref{chap:1} overviews the history of neutrinos, neutrino oscillation and the MiniBooNE excess. Chapter~\ref{chap:2} introduces the MicroBooNE experiment, including neutrino beam and the MicroBooNE detector, followed by a description of the modelling and event reconstruction in MicroBooNE in Ch.~\ref{chap:2.1}. Chapter~\ref{ch3:ncdelta_LEE} describes the search for NC $\Delta$ radiative decay. Chapter~\ref{ch:signal_model} discusses the NC coherent $1\gamma$ process and its simulation in MicroBooNE, and Ch.~\ref{ch:5} describes the NC coherent $1\gamma$ search in MicroBooNE.


%% file: chapter1.tex
\chapter{Neutrinos and Neutrino Oscillation}\label{chap:1}

Neutrinos are one of the elementary matter particles and are known as the most ``elusive'' ones because they are extremely hard to detect\footnote{For 1GeV neutrinos, interaction cross section is $\sim 10^{-38} \text{cm}^{2}$ per nucleon.}. They are believed to have been generated in abundance in the early universe, and studying the properties of neutrinos would help us probe one of the most fundamental questions: ``How did our universe evolve?'' This chapter first gives a brief overview of the history of neutrino theory and experimental discovery in Sec.~\ref{ch1:nu_theory}, then introduces neutrino oscillation and its empirical formalism in Sec.~\ref{ch1:solar_nu_problem},~\ref{ch1:nu_mass} and~\ref{ch1:nu_osc_formula}, as well as recent experimental neutrino oscillation anomalies in Sec.~\ref{ch1:exp_anomaly}. 

\section{Neutrinos in the Standard Model}\label{ch1:nu_theory}
\subsection{Evolution of Neutrino Theory}
Neutrino was first postulated in 1930 by W. Pauli as an electrically neutral particle with spin $\frac{1}{2}$ to explain the continuous energy spectrum of electrons emitted from nuclear $\beta$-decay, 
\begin{equation}
    n \rightarrow p + e^{-} + \overline{\nu_{e}} 
\end{equation}
in an ``desperate'' attempt to rescue energy conservation law \cite{Pauli_letter}. Later, in 1934, Fermi published his theory of beta decay in conformity with Pauli's proposal and named this new, neutral charge and lightweight particle ``neutrino''\footnote{Though Fermi named this neutral particle as ``neutrino'',  now we know it is actually an antineutrino because of conservation of lepton number.}.  

Fermi proposed that the electron and neutrino are emitted during beta decay similarly to photon emission from radioactive nucleus.  He assumed that electrons and neutrinos can be created or annihilated and all four particles of the decay interact at the same vertex, and developed the theory for beta decay~\cite{Fermi_beta_decay_original, Fermi_beta_decay_translation} in complete analogy with radiation theory \cite{radiation_theory}.   The Lagrangian for the interaction part of four particles is \cite{lesov2009weak}:
\begin{equation}
\mathcal{L}_{int} = G_{F}(\overline{u_{p}}\gamma_{\mu}u_{n})(\overline{u_{e}}\gamma^{\mu}u_{\nu})
\end{equation} 
Where $G_{F}$ is the Fermi coupling constant, $u_{x}$ is the Dirac spinor of particle $x$, $\overline{u_{x}} = u_{x}^{\dag}\gamma^{0}$ is the conjugate Dirac spinor and $\gamma_{\mu}/\gamma^{\mu}$ are Dirac gamma matrices. Fermi concluded that neutrino has a mass of either zero or very small in comparison to the mass of the electron. Fermi theory is not only useful for the description of the beta decay but also applies to other processes, for example, $\mu^- \rightarrow e^- + \nu_{\mu} + \overline{\nu_{e}}$ and $\pi^+ \rightarrow \mu^+ + \nu_{\mu}$\footnote{For these interactions, the same strength constant $G_{F}$ applies, hence this class of interactions are also called as ``universal Fermi interactions'' \cite{gauge_theory}.}.

While many processes can be understood within Fermi's framework,  this theory faced several challenges\footnote{In Fermi's theory, only charge-current neutrino interaction was accommodated, not neutral current neutrino scattering. There are also other difficulties such as violation of unitarity for cross section of first-order scattering process, and divergence of higher-order processes, please see Ref. \cite{gauge_theory} for more detailed overview. 
}, one of which is the violation of parity conservation observed experimentally. Parity transformation (P)  is a mirror reflection: 
\begin{equation}
    P \varphi(t, \vec{x}) P^{-1} = \eta_{a}\gamma^{0} \varphi(t, -\vec{x})
\end{equation}
where $\varphi$ is Dirac spinor and $\eta_{a}$ is a phase term. Parity is a multiplicative quantum number, and particles have intrinsic parity. $\overline{\varphi} \gamma^{\mu} \varphi$ transforms like a vector under Parity transformation \cite{gauge_theory}, so Fermi's theory conserves parity. However, experimental findings on K mesons in the 1950s presented a very puzzling situation: two charged mesons $\theta^+ $ and $\tau^{+}$ with different decay modes ($\theta^+  \rightarrow \pi^{+} + \pi^0$ and $\tau^+  \rightarrow \pi^{+} + \pi^{+} + \pi^{-}$ respectively) were found to have nearly identical mass \cite{k_mass1, k_mass2} and lifetime \cite{k_lifetime1, k_lifetime2}. This indicated that $\theta^+ $ and $\tau^{+}$ are the same particle\footnote{$\theta^+$ and $\tau^+$ are the same particles, now referred to as $K^+$.}, while parity conservation would suggest otherwise: with the intrinsic parity of pion being -1, the parity of $\theta^+ $ is thought to be +1, and -1 for $\tau^+$. 

In 1956, T. D. Lee and C. N. Yang pointed out that while parity conservation is tested to be experimentally true for strong and electromagnetic interactions, it remains a hypothesis for the weak interaction and is not yet backed up by experimental evidence, and proposed several experiments to test parity conservation \cite{lee_yang}. A few months after Lee and Yang's paper, C. S. Wu and her collaborators showed that the angular distribution of emitted electron from the beta decay of polarized $\text{Co}^{60}$ nuclei is asymmetric \cite{cswu_parity_violation}, evidence that parity conservation is violated in beta decays. Soon after Wu's finding, parity violation was confirmed in muon decay too \cite{parity_violation_muon_decay}.

In light of the parity non-conservation, in 1957, Lee and Yang proposed a possible two-component theory of neutrinos, where the state of neutrino can be described by a wave function with only two components instead of the usual four \cite{lee_yang_2component}. This theory requires the neutrino mass to be zero, and shows that the neutrino always enters the beta-decay interaction in a polarized state (the spin of neutrino will \textit{always} be parallel to its momentum while vice-versa for the antineutrino)\footnote{At the same year, A. Salam \cite{Salam1957} and L. Landau \cite{Landau_1957} both also proposed theory that violates parity conservation and requires zero neutrino mass.}. 
One limitation of the proposed theory is that it can not explain weak decays that do not involve neutrinos (for example the $\theta-\tau$ puzzle). This is acknowledged at the end of their paper: ``Perhaps this means that a more fundamental theoretical question should be investigated: the origin of all weak interactions. ''

The 1950's witnessed a significant breakthrough in the theoretical development of weak interactions. In 1958, R. P. Feynman and M. Gell-mann went further and suggested a universal (V, A) interaction \cite{feynman_universal_VA}. Universal (V, A) theory was also independently postulated by E. C. G. Sudarshan and R. E. Marshak~\cite{marshak1958proceedings, Sudarshan_Marshak} in the same year; here we will briefly discuss Feynman and Gell-mann's model following the same notation. 

In general, the Lagrangian for the four-particle interaction can be described by: 
\begin{equation}
\mathcal{L}_{int} = \sum_{i}C_{i}(\overline{\varphi_{n}}O_{i}\varphi_{p})(\overline{\varphi_{\nu}}O_{i}\varphi_{e})
\end{equation}
where $\varphi_{x}$ is the four-component Dirac spinor of particle $x$, $\overline{\varphi_{x}}$ is the conjugate Dirac spinor, and $O_{i}$ represents appropriate operators characterizing the decay and weighted by constant $C_{i}$. Table~\ref{tab_operator} shows all the possible forms of $O_{i}$ such that $(\overline{\varphi}O_{i}\varphi)$ transforms correctly under Lorentz transformation and $(\overline{\varphi_{n}}O_{i}\varphi_{p})(\overline{\varphi_{\nu}}O_{i}\varphi_{e})$ is Lorentz scalar.

\begin{table}[h!]
    \centering
    \begin{tabular}{|c|c|}
    \hline 
      $O_{i}$   &  Transformation property of $(\overline{\varphi}O_{i}\varphi)$\\
      \hline
      $O_{S} = 1$   & Scalar (S) \\
      $O_{V} = \gamma^{\mu}$ & Vector (V) \\
      $O_{T} = \sigma^{\mu\nu} = \frac{i}{2}[\gamma^{\mu}, \gamma^{\nu}]$ & Tensor (T) \\
      $O_{A} = \gamma^{\mu}\gamma_{5}$ & Axial Vector (A) \\ 
      $O_{P} = \gamma_{5}$ & Pseudoscalar (P)\\
      \hline 
    \end{tabular}
    \caption[Elementary fermion transition operators.]{Five possible transition operators $O_{i}$ where $\gamma^{\mu}$ are Dirac matrices, and $\gamma_{5} \equiv -i\gamma_{0}\gamma_{1}\gamma_{2}\gamma_{3}$. Taken from Tab. 1.2 from Ref.~\cite{gauge_theory} with modification.}
    \label{tab_operator}
\end{table}
Instead of replacing only $\varphi_{e}$ by $\frac{1}{2}(1 \pm \gamma_{5})\varphi_{e}$, Feynman and Gell-mann supposed that all particles entering the interaction should be treated in the same way.  Projection operator $a$ is defined\footnotemark:
\begin{equation}
    a= \frac{1}{2}(1 \pm \gamma_{5}) 
\end{equation}\label{eq:def_a} 
\footnotetext{The sign of `+' or `-' for the projection operator was not determined yet when the theory was proposed. Note also that in the original paper, operator a was defined as $a= \frac{1}{2}(1 \pm i\gamma_{5})$; this was due to a slight change in the definition of $\gamma_{5}$ matrix. Here we are following modern notation of Gamma matrices.}$\gamma_{5}$, the chirality operator has eigenvalues $\pm1$ corresponding to particles' chirality: left-handed (-1) or right-handed (+1). Operator $\frac{1}{2}(1 - \gamma_{5})$ would project Dirac field $\varphi$ to its left-hand component, and $\frac{1}{2}(1 + \gamma_{5})$ leads to particles' right-hand component. 

In this case, the Lagrangian takes the form of
\begin{equation}
\mathcal{L}_{int} = \sum_{i}C_{i}(\overline{a\varphi_{n}}O_{i}a\varphi_{p})(\overline{a\varphi_{\nu}}O_{i}a\varphi_{e})
\end{equation}
Because operators $O_{S}$, $O_{T}$, and $O_{P}$ all commute with $\gamma_{5}$, only contributions with $O_{A}$ and $O_{V}$ coupling remain. Furthermore, since $\gamma_{5}^2 = \text{I}$ and $\gamma_{5}a = \pm a$ (depending on the $\pm$ sign in $a$), this leads to the same coupling for $O_{A}$ and $O_{V}$ and the Lagrangian becomes:
\begin{align}
\mathcal{L}_{int} & \sim (\overline{\varphi_{n}}\gamma_{\mu}a\varphi_{p})(\overline{\varphi_{\nu}}\gamma^{\mu}a\varphi_{e}) \\ 
& \sim (\overline{\varphi_{n}}\gamma_{\mu}(1 \pm \gamma_{5})\varphi_{p})(\overline{\varphi_{\nu}}\gamma^{\mu}(1 \pm \gamma_{5})\varphi_{e})
\end{align}
Further applying this rule universally to other weak decays, such as $\mu$ decay, yields the conclusion that the neutrino only enters the interaction in state $a\varphi_{\nu}$, i.e. only neutrinos in polarized spin can exist. 

This theory yields a two-component neutrino with polarized spin and, due to the mixture of vector and axial vector contribution, gives a violation of the parity conservation. The helicity of neutrinos\footnote{For massless particles, helicity and chirality are the same. } was reported to be negative in 1958~\cite{neutrino_helicity}, i.e. the neutrino is a left-handed particle, and its spin is antiparallel to its momentum, consistent with V-A (vector ``minus'' axial) theory. Neutrinos have been assumed to be massless, spin $\frac{1}{2}$, ``left-handed'' particles in the theory development since. 

The (V-A) theory is a starting point for the formulation of the unified electroweak gauge theory. J. Schwinger first suggested that there could be a synthesis between electromagnetism and weak interactions by introducing an "isotopic" triplet of vector fields, which comprise two electrically-charged massive particle fields mediating the weak interaction -- now known as the $W^{+}$ and $W^{-}$ boson -- and a neutral massless photon field \cite{Schwinger}.  Extending Schwinger's work, later, S. L. Glashow showed that, with the requirement of ``partially-symmetric''\footnote{Partial symmetry refers to the invariance of only part of the Lagrangian under a group of infinitesimal transformations in Ref.~\cite{GLASHOW1959107}.} Lagrangian and experimental consideration, at least one neutral vector boson $Z^0$ is needed besides the three fields (including photon) \cite{Glashow_1961}, though the masses of $W^{\pm}$and $Z^0$ remain arbitrary. 

A. Salam and J. C. Ward shared the same idea as Glashow and pointed out that the mass of the neutral vector boson $Z^0$ should not be massless, and might be at least as massive as the charged vector bosons $W^{\pm}$ \cite{Salam_Ward}. In 1967 S. Weinberg proposed a theory that unites the electromagnetism and weak interaction but spontaneously breaks the symmetry between the two and gives masses to intermediate boson fields (W, Z)~\cite{Weinberg}, by incorporating Brout-Englert-Higgs mechanism \cite{Higgs_Mechanism, Brout_higgsMechanism, Kibble_higgsMechanism}. This is now known as electroweak theory, or Glashow-Salam-Weinberg theory, which is an important part of the Standard Model (SM) of particle physics.

\subsection{Neutrinos in the Standard Model}
The SM is a gauge theory developed in the 1950 - 1970s describing electromagnetic, weak and strong interactions. Here we will limit our discussion to electroweak theory related to neutrinos following Ref. \cite{gauge_theory}. 

In the SM, neutrinos are assumed to be massless ``left-handed'' particles, and there are three generations of neutrinos ($\nu_{e}$, $\nu_{\mu}$ and $\nu_{\tau}$), which together with charged leptons ($e$, $\mu$, $\tau$) form three generations of lepton families. The left-handed neutrino field forms an ``isospin'' doublet with the left-handed component of the charged lepton field, and the right-handed component of the charged lepton field by itself forms a singlet: 
\begin{equation}
    L_{l} = \frac{1 -\gamma_{5}}{2}\begin{pmatrix}
 \varphi_{\nu_{l}} \\
  \varphi_{l}
\end{pmatrix}, 
\; 
R_{l}  = \frac{1+\gamma_{5}}{2}\varphi_{l}, \;\;
l = (e, \mu, \tau)
\end{equation}

\begin{figure}[ht!]
\centering
\includegraphics[bb = 0 0 100 100, trim = 2.5cm 13cm 0 13cm, clip=true]{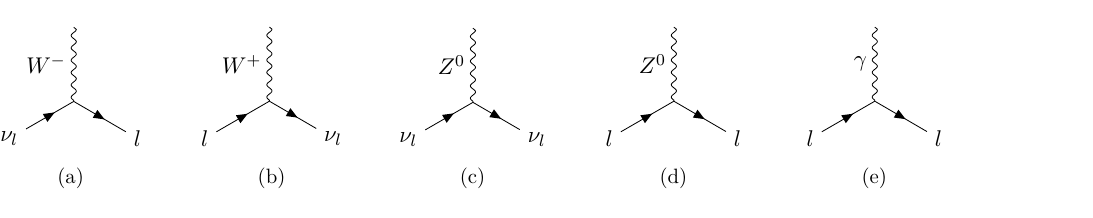}
\caption[Feynman diagrams for leptonic interactions through $W^{\pm}$, $Z^0$ and photon exchange.]{Feynman diagrams correponding to Eq.~\ref{eq:lep_boson_int}. Direction of time in the Feynman histogram is from left to right.}
\label{fig:feynman}
\end{figure}

The $W$ and $Z$ boson fields are mixtures of vector fields $\vec{A_{\mu}} = (A_{\mu}^1, A_{\mu}^2, A_{\mu}^3)$ and singlet $B_{\mu}$ with respective coupling constants of $g$ and $g^{'}$:
\begin{align}
    \text{photon field:}\; A_{\mu} &= \cos{\theta}B_{\mu} + \sin{\theta} A_{\mu}^3  \\
    Z^0 \text{ field:}\; Z_{\mu} &= -\sin{\theta}B_{\mu} + \cos{\theta} A_{\mu}^3  \\
    W \text{ fields:} \; W_{\mu}^{\pm} &= \frac{1}{\sqrt{2}}(A_{\mu}^1 \mp iA_{\mu}^2)
\end{align}
where mixing angle $\theta$ is the Weinberg angle, and defined as: 
\begin{equation}
    \sin{\theta} = \frac{g^{'}}{\sqrt{g^2 + g^{'2}}}, \; \cos{\theta} = \frac{g}{\sqrt{g^2 + g^{'2}}}
\end{equation}
The Lagrangian for lepton-boson interaction writes as:
\begin{equation}\label{eq:lep_boson_int}
\begin{split}
    \mathcal{L}^{int} &= \frac{g}{2\sqrt{2}}[\overline{\varphi_{l}}\gamma^{\mu}(1-\gamma_{5})\varphi_{\nu_{l}}W_{\mu}^{(-)} + \overline{\varphi_{\nu_{l}}}\gamma^{\mu}(1-\gamma_{5})\varphi_{l}W_{\mu}^{(+)}] \\
    &+ \frac{g}{4\cos{\theta}}[\overline{\varphi_{\nu_{l}}}\gamma^{\mu}(1-\gamma_{5})\varphi_{\nu_{l}} - \overline{\varphi_{l}}\gamma^{\mu}(1- 4\sin^2{\theta} - \gamma_{5})\varphi_{l}]Z_{\mu} \\
    &- e\overline{\varphi_{l}}\gamma^{\mu}\varphi_{l}A_{\mu}
\end{split}
\end{equation}
The Lagrangian prescribes two types of weak interactions neutrinos can participate in: 
\begin{itemize}
    \item Charged current (CC) interaction,  mediated by either charged $W^{+}$ or charged $W^{-}$ boson as indicated by the first and second terms in the Lagrangian. 
    \item Neutral current (NC) interaction, mediated by neutral $Z^{0}$ boson, presented by the third term.
\end{itemize}
The fourth term in the Lagrangian represents the NC interaction of leptons through the Z boson, and the last term corresponds to charged lepton scattering through photon exchange. Figure \ref{fig:feynman} shows the Feynman diagrams of these interactions. 

Physical $W$ and $Z$ fields gain masses through symmetry-breaking by the Higgs mechanism: 
\begin{equation}
    M_{W} = \frac{g\lambda}{2}, \; 
    M_{Z} = \frac{g\lambda}{2\cos{\theta}} = \frac{M_{W}}{\cos{\theta}}
\end{equation}
where $\frac{\lambda}{2}$ is vacuum expectation of the Higgs field. The Lagrangian Eq.~\ref{eq:lep_boson_int} also indicates some decay modes of $W$ and $Z$ bosons\footnote{$W$ and $Z$ bosons also couple to quarks, and can decay to hadrons as well, which in fact has overall larger branching ratio than decay into leptons.}: 
\begin{equation}
    W^{+} \rightarrow l^{+} + \overline{\nu_{l}}, \; W^{+} \rightarrow l^{+} + \overline{\nu_{l}}, \; Z^{0} \rightarrow l^+l^-, \; Z^{0} \rightarrow \nu_{l} \overline{\nu_{l}}
\end{equation}
Especially for $Z^{0} \rightarrow \nu_{l} \overline{\nu_{l}}$ decay, since neutrinos are hard to detect, this is an ``invisible'' decay mode. By theoretically calculating the decay width of $Z^{0}$ into neutrinos, and experimentally measuring the overall decay width of $Z^{0}$ boson and the decay width of $Z^{0}$ in other visible decay channels, the number of neutrino flavors that interact with $Z$ (with mass $< \frac{M_{Z}}{2}$) can be determined. This is measured to be $N_{\nu} = 2.9963 \pm 0.0074$ by the Large Electron Position experiments \cite{JANOT2020135319}, and agrees with three lepton families.

\subsection{Experimental Discovery of Neutrinos}
After the postulation of the neutrino, it took more than 20 years before it was experimentally observed. The first electron antineutrino was observed by F. Reines and C. Cowan in 1956 through inverse beta-decay \cite{reines_cowan}. They made use of two 200-liter water targets located near a fission reactor with a large flux of electron antineutrinos and observed the signal from positron annihilation and neutron capture, both of which are products of the inverse beta decay: 
\begin{equation}\label{eq:inverse_beta}
    \overline{\nu_{e}} + p \rightarrow e^{+} + n
\end{equation}

The second neutrino species was discovered by L. Lederman, M. Schwartz and J. Steinberger \cite{mu_nu_discovery} in 1962 in an experiment at Brookhaven Alternating Gradient Synchrotron accelerator. The neutrino beam was created by the decay-in-flight of pions produced by bombarding 15 GeV protons on the Beryllium target. Muon tracks produced by the neutrino interactions were observed instead of electrons expected from electron neutrino interactions, thus showing that there are at least two types of neutrinos distinct from each other: electron neutrino and muon neutrino.

Tau neutrino, the third species, was not experimentally found until 2000 by the DONUT experiment \cite{tau_nu_discovery} at Fermilab, though its existence was implied by the discovery of tau lepton in 1975 \cite{tau_lepton_discovery}, both of which form the third-generation lepton family. Tau neutrino interaction was identified by the tau lepton from the neutrino interaction vertex. Because the tau lepton is very heavy with a rest mass of 1.8GeV, the DONUT experiment made use of a proton beam with energy as high as 800 GeV colliding with a tungsten target, producing charged mesons, which subsequently decay and generate high-energy neutrinos. Four tau neutrino interactions were observed by DONUT, with an estimated background of 0.34 events, establishing the evidence of the existence of tau neutrino.

\section{Evidence of Non-zero Neutrino Mass}\label{ch1:solar_nu_problem}
While it was widely believed that neutrinos were massless particles through the 1970s, the first hint of non-zero neutrino mass came from a series of solar neutrino measurements starting in the late 1960s, which led to the so-called ``solar neutrino problem''~\cite{particle_phys_review}. 

The first solar neutrino measurement was made by the Homestake experiment of R. Davis and his collaborators~\cite{homestake}. They utilized a cylindrical tank of 520 tones of chlorine to detect neutrinos from the Sun through CC interaction: $^{37}Cl + \nu_{e} \rightarrow$  $^{37}Ar + e^{-}$. 
A total of 108 solar neutrinos were observed during 25 years (1970 - 1994), and the combined result yielded a reaction rate of neutrinos from $B^8$ decay in the Sun at $2.56 \pm 0.16 \text{(statistical)} \pm 0.16 \text{(systematic)}$ solar neutrino units (SNU, defined as one interaction per $10^{36}$ atoms per second) \cite{Cleveland_1998}. This was much lower than the standard solar model prediction of $8.5 \pm 1.8$ SNU \cite{SSM_Bahcall}.

Other experiments also observed a reduced neutrino flux from the Sun: radiochemical gallium-based experiments (GALLEX \cite{GALLEX} and SAGE~\cite{SAGE}) that measure neutrinos through gallium interaction $^{71}Ga + \nu_{e} \rightarrow$  $^{71}Ge + e^{-}$, and water-based experiments (Kamiokande~\cite{Kamiokande_solar_neutrino} and Super-Kamiokande~\cite{SK_solar_neutrino}) that measure neutrino-electron scattering. While these experiments differ in detector material and technologies as well as the neutrino energy they are sensitive to, all reported that the measured neutrino flux to be one-half or a third of the standard solar model prediction. 

While many explanations were raised to explain the observed deficit (such as errors in the solar model assumption or unknown experimental error), a lot was ruled out. One plausible explanation was neutrino oscillation, an idea first proposed by B. Pontecorvo \cite{Pontecorvo:1957qd} in 1957, where he pointed out that like the $K^{0}$ meson, a neutrino could be a mixed particle with the possibility of neutrino oscillation. In 1985 S. P. Mikheyev and A. Yu. Smirnov discovered that in the matter of the Sun, the mixing of neutrinos could be resonantly enhanced and thus could explain the measured deficit even with a very small mixing angle \cite{Mikheyev:1985zog}. 

The solar neutrino problem was finally resolved in 2002 when Sudbury Neutrino Observatory (SNO) published result of measured solar neutrino flux using the NC interaction~\cite{SNO_solar_nu}: $$\nu_{x} + d \rightarrow p + n + \nu_{x}$$ where $x$ could be any type of neutrinos. The total flux measured with NC interaction was found to be consistent with the solar model prediction. This, together with different levels of deficits observed in the CC and elastic-scattering channels, provided strong evidence that solar electron neutrinos experience flavor transitions between their generation in the core of the Sun and their arrival on Earth\footnote{The oscillation of electron neutrinos is enhanced by the Mikheyev–Smirnov–Wolfenstein (MSW) effect in the medium of the Sun~\cite{MSW_effect}.}.

The first firm evidence of neutrino oscillation came from the Super-Kamiokande (Super-K) experiment in 1998 \cite{SK_atmo_nu_oscillation}. Super-K measured atmospheric muon neutrinos and observed a deficit in the number of muon neutrinos with zenith angle dependence -- the number of upward-going neutrinos (which arrive at the detector after traveling through the earth) was half of the number of downward-going neutrinos, while assuming no neutrino oscillation the prediction is flat over the zenith angle. 

The experimental observations of the SNO and Super-K experiments established neutrino flavor transformation -- electron neutrino disappearance by the SNO experiment and muon neutrino disappearance by the Super-K experiment, respectively -- and definitively confirmed that neutrinos have non-zero mass; these pioneering observations were recognized with the 2015 Nobel Prize in Physics~\cite{nobel_prize}.

\section{Introducing Non-zero Neutrino Mass in Minimal Extension to the Standard Model}\label{ch1:nu_mass}
This section briefly describes how the SM can be extended to accommodate neutrinos with masses. Two possible mass terms could be constructed for neutrinos: Dirac mass term or Majorana mass term, where neutrinos are treated either as Dirac particles or Majorana particles\footnote{Please refer to \cite{ParticleDataGroup:2020ssz, mass_model_overview, radiative_mass_model, Bilenky_numass} for more detailed overview. }.
\subsection{Dirac Neutrino Mass}
In the SM electroweak theory, all quarks and charged fermions get their masses through the Yukawa couplings with the Higgs field, and the mass term connects the left-handed field with the right-handed field. The most straightforward way to add neutrino mass in the theory is to introduce a right-handed neutrino, in the same way as other fermions gain mass from Yukawa coupling. The right-handed neutrino would be an SU(2) singlet, $\nu_{i R}$, and couple to the Higgs boson through Yukawa coupling. We can add a Dirac mass term to the Lagrangian \cite{Bilenky_numass}: 
\begin{equation}
    \mathcal{L}^{\nu}_{Yukawa} = -\sqrt{2} \overline{\varphi}_{\alpha L} Y_{\nu}^{\alpha i}\nu_{i R}\tilde{\phi} + h.c.
\end{equation}
where $\alpha$ runs through $e, \nu, \tau$ and $i = 1, 2, .. n$ where $n$ is the number of singlet fermion fields; $\varphi_{\alpha L}$ is the left-handed lepton doublet, $ Y_{\nu}^{\alpha i}$ are Yukawa couplings and $\tilde{\phi}$ is the conjugated Higgs doublet. This, after spontaneous symmetry breaking and diagonalization of the matrix $Y$, would give a neutrino mass term:
\begin{equation}
    \mathcal{L}^{\nu} = - \sum_{\alpha} m_{\alpha} \overline{\nu_{\alpha L}}\nu_{\alpha R} + h.c. = - \sum_{\alpha} m_{\alpha} \overline{\nu_{\alpha}}\nu_{\alpha}
\end{equation}
This is called the Dirac mass term. Here, the left-handed and right-handed neutrino fields combine to form the four-component neutrino field $\nu_{\alpha} = \nu_{\alpha L} + \nu_{\alpha R}$ with mass of $m^{\alpha} = y^{\alpha}\lambda$, where $y^{\alpha}$ is Yukawa coupling and $\frac{\lambda}{\sqrt{2}}$ is the vacuum expectation of the Higgs field \cite{gauge_theory}. 

While this mechanism allows incorporating neutrino mass, it does not explain why neutrinos are much lighter than the charged leptons, since $y^{\alpha}$ are input parameters. Moreover, to accommodate small neutrino mass consistent with experimental limits, coupling $y^{\alpha}_{\nu}$ has to be at least ten orders of magnitude smaller than Yukawa couplings of other fermions, which seems unnatural.

\subsection{Majorana Neutrino Mass}
Instead of introducing a right-handed neutrino, which is experimentally not observed, it's possible to introduce neutrino mass with only left-handed neutrino fields, provided that lepton number conservation is violated \cite{Pontecorvo:1957qd, GRIBOV1969493}. This could be done by realizing that the particle-antiparticle conjugation operation $\hat{C}$ defined as:
\begin{equation}
    \hat{C}: \varphi \rightarrow \varphi^{c} = C\overline{\varphi}^{T}, \; C = i\gamma_{2}\gamma_{0}
\end{equation}
flips the chirality of particles \cite{Akhmedov:1999uz}:
\begin{equation}
    \hat{C}: \varphi_{L} \rightarrow (\varphi_{L})^{c} = (\varphi^{c})_{R}, \; \varphi_{R} \rightarrow (\varphi_{R})^{c} = (\varphi^{c})_{L}
\end{equation}
Thus, a Majorana mass term can be added to the Lagrangian \cite{Bilenky_numass}: 
\begin{equation}
        \mathcal{L}^{M} = - \frac{1}{2} \overline{\nu_{\alpha L}}M^{\alpha \beta}(\nu_{\beta L})^c + h.c.
\end{equation}
where $\alpha, \beta = e, \nu, \tau$ and $\nu_{L}$ represents the left-handed neutrino field, and $M$ is complex, symmetric $3\times3$ matrix\footnote{If there are n different flavors of neutrinos, M will be $n\times n$ matrix.}. Since $M$ is symmetric, it can be diagonalized by a unitary matrix $$M = UM^{diag}U^{T}, \; M^{diag} = \text{diag}(m_1, m_2, m_3)$$ and the Lagrangian takes the form of: 
\begin{equation}
        \mathcal{L}^{M} = - \frac{1}{2}\sum_{i = 1} ^{3}m_{i}\overline{\nu_{i}}\nu_{i}
\end{equation}
\begin{equation}
    \nu_{i} = \sum_{\alpha}U^{\dag}_{i\alpha}\nu_{\alpha L} + \sum_{\alpha}(U^{\dag}_{i\alpha}\nu_{\alpha L})^{c}, \; \alpha = e, \nu, \tau
\end{equation} 
Here $\nu_{i}$ ($i = 1,2,3$) is the mass eigenstate and is a Majorana field $\nu_{i} = \nu_{i}^{c}$.

The flavor neutrino fields can thus be derived as a mixture of different mass fields:
\begin{equation}
    \nu_{\alpha L}(x) = \sum_{i} U_{\alpha i}\nu_{iL}(x)
\end{equation}
$U$ is the mixing matrix, also called Pontecorvo–Maki–Nakagawa–Sakata (PMNS) matrix \cite{Maki:1962mu}. In this approach, the lepton number is broken by $\Delta L= 2$ units, and the neutrino masses are also free parameters, meaning this also does not explain neutrinos' lightness.

\subsection{See-saw Mechanism}
The see-saw mechanism offers a very simple and satisfactory explanation of the lightness of neutrinos by introducing right-handed neutrinos with large mass. In type-I seesaw mechanism \cite{Yanagida:1980xy, Yanagida_seesaw_1979, Gell-Mann:1979vob, PhysRevLett.44.912, PhysRevD.23.165, Fukugita:2003en}, a right-handed SU(2) singlet $\nu_{R}$ is introduced and the Lagrangian has a Yukawa coupling term of the right-handed neutrino with the Higgs boson and lepton doublet $\varphi_{L}$ (first term), and a Majorana mass term for the right-handed neutrino (second term): 
\begin{equation}
\mathcal{L}^{M} = -\sqrt{2} \overline{\varphi}_{\alpha L} D_{\nu}^{\alpha i}\nu_{i R}\tilde{\phi} - \frac{1}{2} \overline{\nu_{i R}}M^{i j}_{R}(\nu_{j R})^c + h.c. 
\end{equation}
Here $D_{\nu}^{\alpha i}$ and $M^{i j}_{R}$ are complex Dirac and Majorana mass matrices. The Lagrangian can be reorganized into:
\begin{equation}
\mathcal{L}^{M} = -\frac{1}{2}n^{T}_{L}C\mathcal{M}n_{L} + h.c. 
\end{equation}
where $n_{L} = (\nu_{L}, (\nu_{R})^c)$ and matrix $\mathcal{M}$ has the form of 
\begin{equation}
    \mathcal{M} = \begin{pmatrix}
0 & D\\
D^{T} & M
\end{pmatrix}
\end{equation}
and can be diagonalized by a unitary matrix $U$ to give mass eigenvalues $m_{i}$ and corresponding mass-eigenstates $\nu_{i}$. Again the mass-eigenstates $\nu_{i}$ are Majorana fields, and the flavor neutrino fields (as well as right-handed neutrino fields) can be written as: 
\begin{equation}\label{eq:nu_mass_mix}
    \nu_{\alpha}(x) = \sum_{i} U_{\alpha i}\nu_{i}(x)
\end{equation} In the simple one-flavor case, $D, M$ are numbers and the mass eigenvalues are: 
\begin{equation}
    m_{1, 2} = \frac{M}{2} \mp \sqrt{(\frac{M}{2})^2 + D^2}
\end{equation}
In the case of $M \gg D$, $m_{1} \approx \frac{D^2}{M}$ and $m_{2} \approx M$. The small neutrino mass is thus achieved.

Note that Eq.~\ref{eq:nu_mass_mix} implies that for three light Majorana neutrinos, the PMNS matrix is not a $3\times 3$ matrix, and its top-left $3\times 3$ submatrix will not be unitary. However the violation of unitarity is very small, on the order of $\mathcal{O }(D/M)$~\cite{ParticleDataGroup:2020ssz}, so for the three neutrino case, the PMNS matrix is treated unitary. 

The PMNS matrix can be parametrized as the product of three rotations through mixing angles $\theta_{23}$, $\theta_{13}$ and $\theta_{12}$ with phase $\delta_{CP}$ and a diagonal phase matrix:
\begin{align}
    U &= \begin{pmatrix}
U_{e1} & U_{e2} & U_{e3} \\
U_{\mu 1} & U_{\mu 2} & U_{\mu 3} \\
U_{\tau 1} & U_{\tau 2} & U_{\tau 3} \\ 
\end{pmatrix}  \\
   &= \begin{pmatrix}
1 & 0 & 0 \\
0 & c_{23} & s_{23} \\
0 & -s_{23} & c_{23} \\
\end{pmatrix} 
\begin{pmatrix}
c_{13} & 0 & s_{13}e^{-i\delta_{CP}} \\
0 & 1 & 0  \\
-s_{13}e^{i\delta_{CP}} & 0 & c_{13} \\
\end{pmatrix}
\begin{pmatrix}
c_{12} & s_{12} & 0 \\
-s_{12} & c_{12} & 0 \\
0 & 0 & 1
\end{pmatrix}
\begin{pmatrix}
e^{i\eta_{1}} & 0 & 0 \\
0 & e^{i\eta_{2}} & 0\\
0 & 0 & 1
\end{pmatrix}
\end{align}
where $c_{ij} = \cos{\theta_{ij}}$ and $s_{ij} = \sin{\theta_{ij}}$. The two phases $\eta_{1}$ and $\eta_{2}$ are associated with the Majorana neutrinos and are zero for the case of Dirac neutrino.

\section{Neutrino Oscillation in Vacuum}\label{ch1:nu_osc_formula}
With the mixing PMNS matrix relating neutrino flavor eigenstates with mass eigenstates, neutrino oscillation can be formalized. Neutrinos are produced in pure flavor eigenstates from a weak interaction, which can be written as a superposition of the mass eigenstates \footnotemark:
\begin{equation}
   |\nu(t = 0)\rangle =  |\nu_{\alpha}\rangle  = U^{*}_{\alpha i}|\nu_{i}\rangle  
\end{equation}
\footnotetext{Note complex conjugate of PMNS matrix is used here different from Eq. \ref{eq:nu_mass_mix}, this is because the neutrino field annihilates neutrino state, while creation operator is needed to act on the vacuum to create neutrino state $|\nu_{\alpha}\rangle$. \cite{Giganti:2017fhf}}
When a neutrino travels, the mass eigenstates evolve with time: 
\begin{equation}
    |\nu(t)\rangle =  U^{*}_{\alpha i}|\nu_{i}(t)\rangle =  U^{*}_{\alpha i}e^{-iE_{i}t}|\nu_{i}\rangle 
\end{equation}
After time $t$, the probability of detecting a neutrino in $\beta$ flavor $|\nu_{\beta}\rangle$ is:
\begin{align}
    P_{\alpha \beta} &= |\langle \nu_{\beta} | \nu(t) \rangle|^2 \\
    &= |\sum_{i, j} U^{*}_{\alpha i}U_{\beta j} e^{-iE_{i}t}\langle \nu_{j} |\nu_{i}\rangle |^2 \\
    &= |\sum_{i}U^{*}_{\alpha i}U_{\beta i} e^{-iE_{i}t}|^2 \\
    &= \sum_{i,j} U^{*}_{\alpha i}U_{\beta i}U_{\alpha j}U^{*}_{\beta j}e^{i(E_j - E_i)t}
\end{align}
We treat mass eigenstates as plane waves with well-defined momenta $\vec{p}$ \footnote{Please see Ref. \cite{Akhmedov:2019iyt} for a detailed review of quantum mechanical aspects in the theory of neutrino oscillations.}, and because neutrinos are ultra-relativistic, the energy of the neutrino can be expanded: $E_{i} = \sqrt{p^2 + m_{i}^2} \simeq p + \frac{m_{i}^2}{2p} \simeq p + \frac{m_{i}^2}{2E}$ (using $E \simeq p$) and time $t \simeq L/c$, with $L$ being the distance the neutrino travels. The oscillation probability takes the form of: 
\begin{multline}\label{eq:prob_formula}
    P_{\alpha \beta} = \delta_{\alpha \beta} - 4\sum_{i<j}\Re(U^{*}_{\alpha i}U_{\beta i}U_{\alpha j}U^{*}_{\beta j})\sin^2{(\frac{\Delta m^2_{ji} L}{4E})} 
    + 2\sum_{i<j}\Im(U^{*}_{\alpha i}U_{\beta i}U_{\alpha j}U^{*}_{\beta j})\sin{(\frac{\Delta m^2_{ji} L}{2E})}
\end{multline}
where $E$ is the energy of the neutrino and $\Delta m^2_{ji} = m^2_{j} - m^2_{i}$ is the mass-squared difference between the $i^{th}$ and $j^{th}$ mass eigenstates. Note that this probability formula also applies to $n\times n$ neutrino scenarios. 

The term $$\frac{\Delta m^2_{ji} L}{4E} = 1.267\frac{\Delta m^2_{ji}}{eV^2}\frac{L/E}{m/MeV}$$ in Eq.~\ref{eq:prob_formula} controls the oscillation frequency of probability. Moreover, to enable flavor oscillation, elements of the mixing matrix shall not be zero, and there should be at least two non-zero, different masses (for the $3\times 3$ neutrino case) such that $\Delta m^2_{ji} \neq 0$ for any $i < j, i, j \in [1,2,3]$. By experimentally controlling neutrinos' energy and travel distance, and observing the oscillation effect, oscillation amplitude and mass square difference $\Delta m^2$ can be measured. Table \ref{tab:LE_to_dm} shows the range of $L$ and $E$ of different experiments with different neutrino sources and the corresponding $\Delta m^2$ they are sensitive to.

\begin{table}[h!]
    \centering
    \begin{tabular}{|p{5em}|cc|c|c|}
    \hline 
    Experiment & & L (m) & E (MeV) & $|\Delta m^2|$ (eV$^{2}$) \\
    \hline 
    \hline 
    Solar &  &$10^{10}$ & 1 & $10^{-10}$ \\
    \hline 
    Atmospheric & &  $10^4$ - $10^7$ & $10^2$ - $10^5$ & $10^{-1}$ - $10^{-4}$ \\
    \hline 
    \multirow{2}{*}{Reactor} & SBL & $10^2$ - $10^3$ & 1 & $10^{-2}$ - $10^{-3}$ \\
    & LBL & $10^4$ - $10^5$ & 1 & $10^{-4}$ - $10^{-5}$ \\
    \hline 
    \multirow{2}{*}{Accelerator} & SBL & $10^2$ & $10^3$ - $10^4$ & > 0.1 \\
    & LBL & $10^5$ - $10^6$ & $10^3$ - $10^4$ & $10^{-2}$ - $10^{-3}$ \\
    \hline 
    \end{tabular}
    \caption[Characteristic values of $L$ and $E$ for different types of experiments performing neutrino oscillation measurements.]{Characteristic values of $L$ and $E$ for experiments performed using various neutrino sources and the corresponding ranges of $|\Delta m^2|$ to which they can be most sensitive for flavor oscillations in vacuum. SBL stands for Short Baseline, which is used to indicate that experiments are located close to the site of neutrino production, and LBL for Long Baseline. It should be noted that the solar neutrino measurements are probing the MSW matter effect inside the Sun rather than neutrino oscillation occurring during travel through the vaccum. Taken from Tab.~14.1 from Ref.~\cite{ParticleDataGroup:2020ssz}.}
    \label{tab:LE_to_dm}
\end{table}

\section{Short-Baseline Neutrino Anomalies}\label{ch1:exp_anomaly}

``If there's one thing you can count on physicists for, it's to keep an eye out for an anomaly''.\footnote{quoted from E. Siegel's article at \url{https://www.forbes.com/}}

As highlighted in Tab.~\ref{tab:LE_to_dm}, in general, experiments at shorter baselines are sensitive to larger mass-squared differences. Anomalies have been observed in both solar neutrino experiments employing calibration sources of neutrinos, for example the GALLEX experiment~\cite{ANSELMANN1995440,1998114,KAETHER201047} and the SAGE experiment~\cite{PhysRevC.80.015807,PhysRevLett.77.4708}, and experiments using neutrino beams. In this section, we review two experimental anomalies observed in accelerator-based short-baseline experiments: the LSND anomaly and the MiniBooNE anomaly. The MiniBooNE anomaly motivated the proposal for the MicroBooNE experiment~\cite{MicroBooNE_proposal}, on which the searches presented in this thesis are performed.

\subsection{Liquid Scintillator Neutrino Detector Experiment}
The Liquid Scintillator Neutrino Detector (LSND) was a short-baseline experiment located at Los Alamos National Laboratory, constructed in the 1990s to search for $\overline{\nu}_{\mu}$ to $\overline{\nu}_{e}$ oscillation in the $\Delta m_{21}^2 \sim 1 \text{eV}^2$ region. The LSND detector was located about 30m from the neutrino source, and saw a neutrino beam dominated by $\nu_{\mu}$, $\overline{\nu}_{\mu}$ and $\nu_{e}$ produced with a proton beam of 798 MeV kinetic energy (KE) producing pions and muons decaying at rest. 
The $\overline{\nu}_{e}$ flux was only $\sim 8 \times 10^{-4}$ as large as the  $\overline{\nu}_{\mu}$ flux in the neutrino energy range of [20., 52.8] MeV \cite{LSND_1995}. 

\begin{figure}[ht!]
\centering
\includegraphics[bb = 0 0 100 100, trim = 11cm 3.5cm 0cm 17.5cm, clip=true, width=0.9\textwidth]{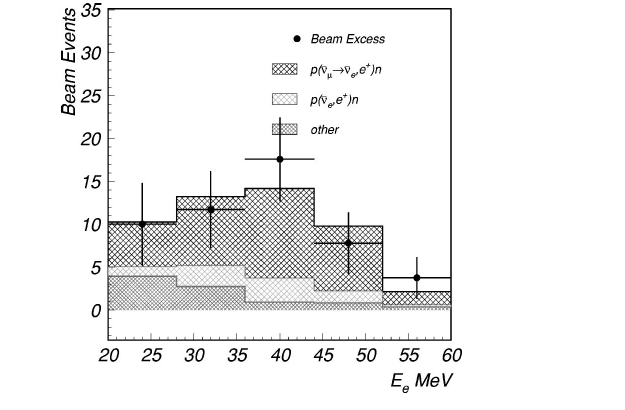}
\caption[The reconstructed positron energy distribution for candidate oscillation events in LSND.]{Distribution of reconstructed positron energy for selected oscillated candidate events. 
The top shaded region shows the expectation from neutrino oscillation at low $\Delta m^2$. Bottom two shared histograms represent the expected neutrino backgrounds. Taken from Fig. 16 of Ref. \cite{LSND_full}}
\label{fig:LSND}
\end{figure}

The detection of $\overline{\nu}_{e}$ in LSND is through inverse beta decay: $$\overline{\nu}_{e} + p \rightarrow e^{+} + n $$ which could be identified by detecting the Cherenkov light of the prompt $e^+$ and a correlated 2.2 MeV $\gamma$ from neutron capture. LSND took data from 1993-1998 and a $\overline{\nu}_{\mu} \rightarrow \overline{\nu}_{e}$ oscillation search was performed with the combined dataset. 
A total excess of $87.9 \pm 22.4 \pm 6.0$ events with positron energy between 20 and 60 MeV was observed above the expected background as highlighted in Fig.~\ref{fig:LSND}. An oscillation fit with the entire sample (events with $20 < E_{e} < 200$ MeV) including both $\overline{\nu}_{\mu} \rightarrow \overline{\nu}_{e}$ and $\nu_{\mu} \rightarrow \nu_{e}$ oscillation yields a best fit at $\Delta m^2 = 1.2 \text{eV}^2$. This is much higher than the mass-squared differences between the three mass eigenstates $v_1, v_2, v_3$, which give magnitudes of mass splittings of $\Delta m^2_{21} \sim 10^{-5} \text{eV}^2$ and $\Delta m^2_{32} \sim 10^{-3} \text{eV}^2$ \cite{ParticleDataGroup:2020ssz}, and seems to suggest additional, heavier ``sterile'' neutrino states\footnote{The additional neutrinos should not couple to $W$ and $Z$ bosons, as limited by the number of active neutrino flavors allowed by $Z$ boson decay width. Thus they don't interact weakly, thus are called ``sterile'' neutrinos.}. 

\subsection{The MiniBooNE Experiment}
The MiniBooNE experiment is a Cherenkov detector at the Fermi National Accelerator Laboratory (Fermilab). MiniBooNE was proposed to follow up on the signal observed by LSND with similar $L/E$, while utilizing different detector technology and being exposed to a different neutrino flux. 

MiniBooNE makes use of the Booster Neutrino Beam (BNB) at Fermilab, which is produced from 8 GeV protons incident on a beryllium target. The secondary mesons produced in the process decay in flight (instead of at rest) and produce (anti)neutrino beam. When running in neutrino (antineutrino) mode, MiniBooNE received an intense beam of highly pure muon neutrinos (anti-neutrinos). 
The $\nu_{\mu}$ and $\overline{\nu}_{\mu}$ flux peak at $\sim 600$ and $400$ MeV respectively and extend to $\sim 3000$ MeV \cite{BNB_flux_MiniBooNE}. The MiniBooNE detector is located 541m from the beryllium target, thus satisfying $L/E \sim 1 \text{m}/\text{MeV}$. It's a 12.2 m in diameter spherical tank filled with pure mineral oil (CH$_2$) and records neutrino interactions by detecting primarily the directional Cherenkov light, and isotropic scintillation light generated by charged particles from the neutrino interaction.  

To measure neutrino oscillation, MiniBooNE searches for charge-current quasi-elastic (CCQE) scattering of $\nu_{e}$ and $\overline{\nu}_{e}$: $$\overline{\nu}_{e} + p \rightarrow e^+ + n, \; \nu_{e} + n \rightarrow e^- + p$$ While the KE of heavy nucleons usually is not enough to pass the Cherenkov threshold in mineral oil ($\sim 350$ MeV KE threshold for protons) and thus can't be observed via Cherenkov radiation, light particles such as $e^{\pm}$ and $\mu^{\pm}$ can usually produce Cherenkov rings, thus be distinguished by the distinct pattern of the ring, and classified as electron- or muon-like. 

MiniBooNE collected data between 2002 and 2019 \cite{MiniBooNE_antinue_osc}.
Combining the data collected in both neutrino mode and antineutrino mode, an excess in data of $4.8 \sigma$ significance was observed in the observed electron-like events 
\cite{MiniBooNE_combine_osc}. 
While the excess in antineutrino mode is consistent with $\overline{\nu_{\mu}} \rightarrow \overline{\nu_{e}}$ oscillation with $0.01<\Delta m^{2}<1.0 \text{eV}^2$ and overlaps with what's observed in LSND \cite{MiniBooNE_antinue_osc}, the excess in neutrino mode is marginally compatible with such neutrino oscillation assumption (assuming two neutrinos). 

\begin{figure}[h]
\centering
\begin{subfigure}[b]{0.48\textwidth}
         \centering
         \includegraphics[width=\textwidth, trim = 1.5cm 11cm 11cm 11cm, clip=true]{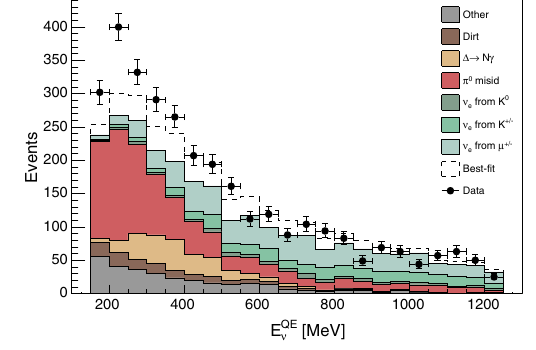}
         \caption{Data-MC comparison in $E_{\nu}^{\text{QE}}$}
         \label{fig:MiniBooNE_data_mc_energy}
\end{subfigure}
\hfill
\begin{subfigure}[b]{0.48\textwidth}
\centering
         \includegraphics[width=1.13\textwidth, trim = 11cm 20.3cm 0.5cm 1.5cm, clip=true]{figures/chapter1/MiniBooNE_Combine_OSC.pdf}
         \caption{Data-MC comparison in $\cos{\theta}$}
         \label{fig:MiniBooNE_data_mc_angle}
\end{subfigure}
\caption[Distributions of neutrino-mode data with reconstructed neutrino energy $150 < E_{\nu}^{\text{QE}} < 1250$ MeV observed in MiniBooNE.]{Distribution of MiniBooNE data collected in neutrino mode with reconstructed energy $150 < E_{\nu}^{\text{QE}} < 1250$ MeV. The stacked color histograms represent the MC prediction, with green histograms represent contributions from intrinsic $\nu_{e}$s in the beam and belgian and red histograms denotes the dominant photon contribution from $\Delta$ radiative decay and $\pi^0$ background. Fig.~\ref{fig:MiniBooNE_data_mc_energy} shows the comparison between data and MC simulation as a function of reconstructed neutrino energy $E_{\nu}^{\text{QE}}$. Fig.~\ref{fig:MiniBooNE_data_mc_angle} shows the comparison in the cosine of the measured angle of outgoing reconstructed candidate electron. The dashed line represents the best-fit two-neutrino oscillation to neutrino-mode data. Taken from Figs.~17 and 18 from Ref. \cite{MiniBooNE_combine_osc}. }
\label{fig:MiniBooNE_data_mc}
\end{figure}

\begin{figure}[h]
\centering

\includegraphics[width=0.8\textwidth, trim = 11.5cm 10.8cm 0cm 10cm, clip=true]{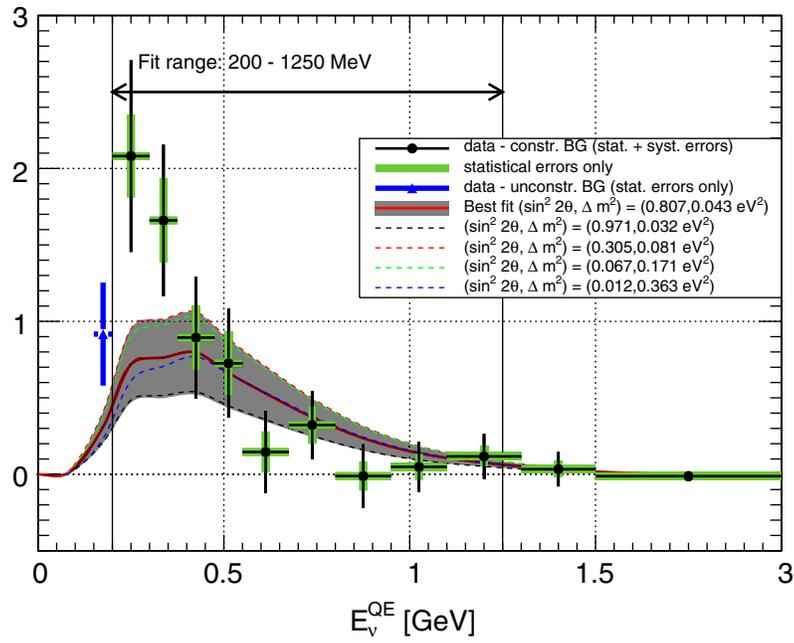}
\caption[Distribution of the MiniBooNE event excess in reconstructed neutrino energy $E_{\nu}^{\text{QE}}$ in neutrino mode data]{Distribution of the MiniBooNE event excess in reconstructed neutrino energy $E_{\nu}^{\text{QE}}$ in neutrino mode data. The solid red line shows prediction at best-fit to neutrino-mode and antineutrino-mode data assuming two-neutrino oscillation. Taken from Fig.~19 from Ref.~\cite{MiniBooNE_combine_osc}. }
\label{fig:MiniBooNE_data_excess}
\end{figure}

Figure \ref{fig:MiniBooNE_data_mc} shows the comparison between data collected during neutrino mode and Monte Carlo (MC) prediction as a function of reconstructed neutrino energy $E_{\nu}^{\text{QE}}$, assuming CCQE kinematics, in Fig.~\ref{fig:MiniBooNE_data_mc_energy} and the cosine of the angle of reconstructed lepton in Fig. \ref{fig:MiniBooNE_data_mc_angle}, highlighting a data excess focused in the low energy region and relatively forward direction. Figure \ref{fig:MiniBooNE_data_excess} shows the excess in data over the predicted background, with predictions of oscillated spectra with different oscillation parameters assuming two-neutrino oscillations overlaid. The data excess in the low energy region ($200 < E_{\nu}^{\text{QE}} < 475$ MeV) can not be explained by two-neutrino oscillation with $\Delta m^{2} \sim \mathcal{O}(1) \text{eV}^2$. This anomaly in the low energy region is referred to as the ``MiniBooNE Low Energy Excess'' (LEE).

Due to the identical Cherenkov signatures of electrons and photons in the detector, MiniBooNE is not able to distinguish electrons from photons, as highlighted by the photon contributions in Fig.~\ref{fig:MiniBooNE_data_mc}. A plethora of interpretations have been proposed to explain the LEE~\cite{whitepaper_sterile}, such as misestimated SM background in the selection, beyond-SM interpretations like more complex scenarios with sterile neutrinos, or more exotic explanations that would generate a similar Cherenkov ring signatures in the MiniBooNE detector. 

The MicroBooNE experiment was proposed to investigate the true nature of the MiniBooNE LEE~\cite{MicroBooNE_proposal}. With exceptional performance offered by liquid argon time projection chamber technology, MicroBooNE has the ability to unambiguously differentiate electrons from photons. Some proposed LEE interpretations have been investigated or are currently being pursued with MicroBooNE. One of the proposed interpretations is that of anomalously large rate of the SM process NC $\Delta \rightarrow \text{N}\gamma$~\cite{MiniBooNE_combine_osc}. This has been investigated with MicroBooNE's first three years of data~\cite{uboone_delta_LEE} and is described in more detail in Ch.~\ref{ch3:ncdelta_LEE} due to its close relation with the main thesis work. More discussions on the MiniBooNE LEE interpretations can be found in the App.~\ref{app:miniboone_lee_interpretations}.

%% file: chapter2.tex
\chapter{The MicroBooNE Experiment}\label{chap:2}
The Micro Booster Neutrino Experiment (MicroBooNE) is an experiment located at Fermilab with the primary physics goal of investigating the MiniBooNE LEE. The MicroBooNE experiment is exposed to the same on-axis Booster Neutrino Beam (BNB) as MiniBooNE at a slightly different baseline, and makes use of the state-of-the-art liquid argon time projection chamber (LArTPC) technology for its superior spatial and calorimetric resolution. This chapter gives an overview of the MicroBooNE experiment, covering the BNB in Sec.~\ref{ch2:uB_nu_beam}, the MicroBooNE LArTPC detector in Sec.~\ref{ch2:uB_detector} and at last MicroBooNE readout and triggering system in Sec.~\ref{ch2:uB_readout_trigger}. 

\section{The Neutrino Beam}\label{ch2:uB_nu_beam}
\subsection{The Booster Neutrino Beam Line}
MicroBooNE is exposed to two independent neutrino beams at different energies: on-axis neutrino flux from the BNB and off-axis neutrino flux from Neutrinos at the Main Injector (NuMI) beam. Since analysis in this thesis uses only data collected with the BNB, we will only discuss the BNB beam here. 

The neutrino beam is produced by the decay of secondary mesons from bombarding high-energy protons onto a beryllium target. It comprises predominantly $\nu_{\mu}$ or $\overline{\nu_{\mu}}$ with an average energy of $\sim 800$ MeV~\cite{BNB_flux_MiniBooNE}. The protons are produced by the Fermilab accelerator complex~\cite{fermilab_accelerator}. First, $\text{H}^{-}$ hydrogen ions are accelerated in the linear accelerator (Linac) to energy up to 400 MeV and then pass through a carbon foil, which removes electrons from the ions, producing a beam of protons. Then protons are injected into the Booster, which is a 474-meter circumference synchrotron, travel around the Booster about 20,000 times and get accelerated to a momentum of 8.9 GeV$/c$ before they are extracted to the BNB. 

\begin{figure}[h!]
\centering
\includegraphics[width=\textwidth, trim = 2cm 17cm 2cm 2cm, clip=true]{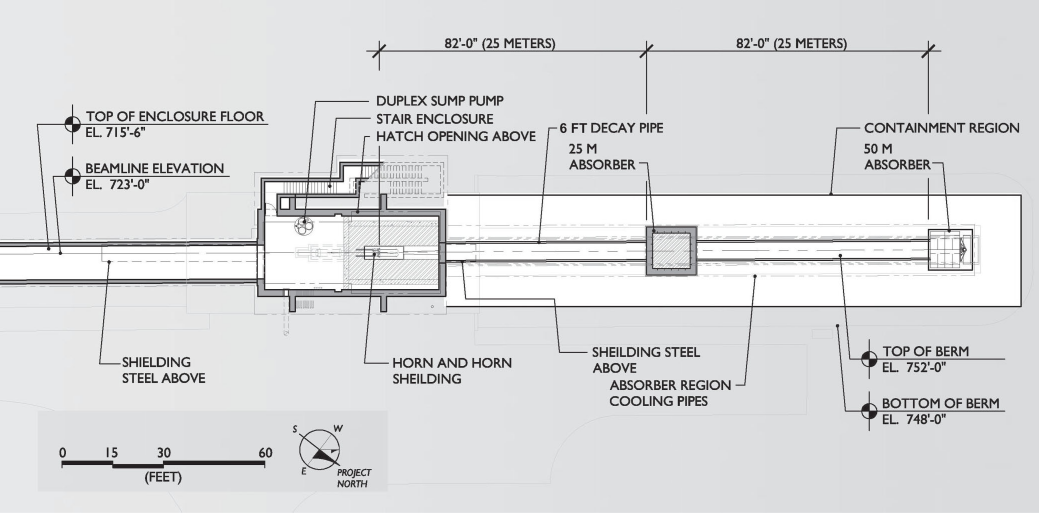}
\caption[Overall layout of the BNB.]{Overall layout of the BNB. The proton beam extracted from the booster enters the target hall from
the left. After leaving the target hall, particles enter a 50-meter-long decay region which terminates on the right at the beam stop. Taken from Fig.~1 in Ref.~\cite{BNB_flux_MiniBooNE}.}
\label{fig:bnb_layout}
\end{figure}

Figure~\ref{fig:bnb_layout} shows the layout of the BNB. The 8.9 GeV/$c$ momentum protons strike the beryllium target embedded within a ``horn'' electromagnet. The monitoring systems located upstream of the target monitor the time and intensity of the proton beam before they are incident on the target. The delivery rate of proton pulses to the BNB is 5 Hz, and approximately $4\times 10^{12}$ protons-per-pulse. Each pulse is called a beam-spill, and lasts $\sim$1.6 $\mu$s with 82 bunches of protons each $\sim$2 ns wide and 19 ns apart~\cite{BNB_flux_MiniBooNE, uboone_WC_eLEE}. 

The horn receives a pulsed current coinciding with the arrival of the proton beam at the target, producing a magnetic field inside. When running in neutrino mode, the primary particles from $p$-Be interaction and secondary particles further produced from the interaction of primary particles with the surrounding materials such as $\pi^{\pm}$, $K^{\pm}$ and $K^{0}$ are then (de)focused by the magnetic field. The flow of the current can be changed such that charged particles with different signs can be focused in neutrino versus antineutrino mode.

After the target-horn assembly is a concrete collimator followed by the decay pipe. The collimator is 214 cm long with an aperture of increasing radius from upstream to downstream. It reduces radiation elsewhere in the beamline by absorbing particles that would not contribute to the neutrino flux. The decay region extends for 45 meters, inside which the secondary mesons decay to produce an intense (anti)neutrino beam. The beam stop sits 50 meters downstream from the upstream end of the beryllium target and absorbs particles apart from the neutrinos. There is another set of absorbing plates located above the decay pipe at 25 meters. This absorber is only used to study systematic uncertainty and was not deployed during MicroBooNE neutrino running.

\subsection{The Neutrino Flux Prediction}\label{ch2:nu_beam_sim}
\begin{figure}[h!]
\centering
\includegraphics[width=0.7\textwidth, trim = 1.8cm 20cm 11cm 2.1cm, clip=true]{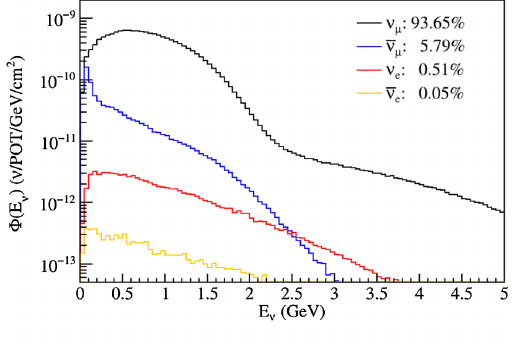}
\caption[The simulated neutrino flux prediction at the MicroBooNE detector.]{The simulated neutrino flux prediction at the MicroBooNE detector. Fluxes shown averaged through the TPC detector volume. Taken from Fig.~1 in Ref.~\cite{uboone_WC_eLEE}.}
\label{fig:bnb_beam_composition}
\end{figure}

MicroBooNE has only taken data during neutrino mode, when positively charged particles are focused and eventually produce a neutrino-enhanced beam. In neutrino mode, the neutrino beam is predominantly $\nu_{\mu}$ produced by pion decay $\pi^{+}\rightarrow \mu^{+} + \nu_{\mu}$ and kaon decay $K^{+}\rightarrow \mu^{+} + \nu_{\mu}$; the later becomes the main mechanism to generate neutrinos with energy higher than 2.3 GeV~\cite{uboone_WC_eLEE}. The contamination from wrong-sign $\overline{\nu_{\mu}}$ neutrino and electron neutrino originates from the decay of $\mu^{+}$ produced in meson two-body decay, $\mu^{+} \rightarrow e^{+} + \nu_{e} + \overline{\nu_{\mu}}$, and from kaon decay, $K^{+}\rightarrow e^{+} + \nu_{e} + \pi^{0}$. The simulation of the MicroBooNE neutrino flux is based on the \textsc{geant4} framework~\cite{geant4}, adapted from earlier work by the MiniBooNE Collaboration~\cite{BNB_flux_MiniBooNE,uB_flux_note}. Figure~\ref{fig:bnb_beam_composition} shows the composition of the beam in neutrino mode as a function of neutrino energy; 93.6\% of the beam is made up of $\nu_{\mu}$'s with 5.8\% $\overline{\nu_{\mu}}$ and $<0.6\%$ $\nu_{e}/\overline{\nu_{e}}$ contamination~\cite{uB_flux_note}. 

\section{The Detector}\label{ch2:uB_detector}
The MicroBooNE detector is situated on-axis in the BNB 468.5~meters downstream from the target and consists mainly of the LArTPC, the light collection system and the cryogenic system~\cite{uB_detector}. The LArTPC is a rectangular field cage of dimension 10.6 meters long, 2.6 meters wide, and 2.3 meters high, filled with about 85 metric tons of highly purified liquid argon. Inside the LArTPC are the cathod plane and anode wire planes connected to different voltages, creating an electric field inside the detector. The light-collection system comprises 32 photomultiplier tubes (PMTs) located behind the LArTPC for the scintillation light collection. The LArTPC and the light-collection system are both housed inside a foam-insulated cryostat vessel containing 170 tons of liquid argon, as depicted in Fig~\ref{fig:ub_detector_cartoon}.

\subsection{The LArTPC}
\begin{figure}
\centering
\begin{subfigure}{.5\textwidth}
  \centering
  \includegraphics[width=\textwidth, page=23, trim = 4cm 16cm 4cm 2.1cm, clip=true]{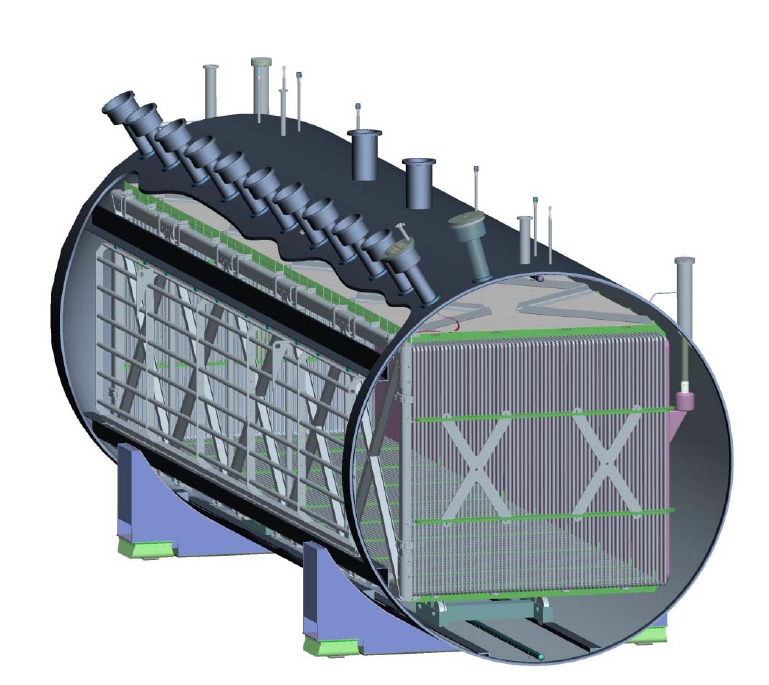}
  \caption{}
  \label{fig:ub_detector_cartoon}
\end{subfigure}%
\begin{subfigure}{.5\textwidth}
  \centering
  \includegraphics[width=\textwidth, page=24, trim = 5cm 10cm 5cm 11cm, clip=true]{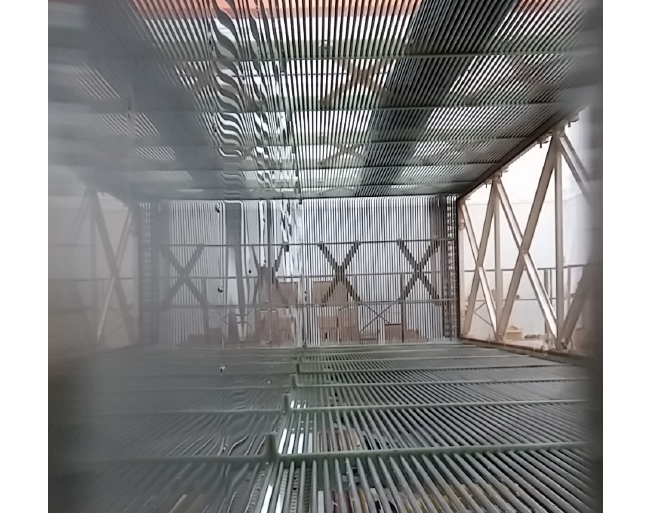}
  \caption{}
  \label{fig:ub_detector_pic}
\end{subfigure}
\caption[A diagram of the MicroBooNE LArTPC and a photo of its interior view.]{(a): Diagram of the MicroBooNE rectangular LArTPC, housed inside the cylindrical cryostat. (b) interior view of the cathode plane (the shiny wall on the left) as viewed from the upstream end of the LArTPC. Also visible are the field rings on the top, bottom and downstream end. Taken from Figs.~9 and~10 in Ref.~\cite{uB_detector}.}
\label{fig:ub_detector}
\end{figure}

\begin{figure}[h!]
\centering
\includegraphics[width=0.9\textwidth, page=9, trim = 2cm 18cm 2cm 2.1cm, clip=true]{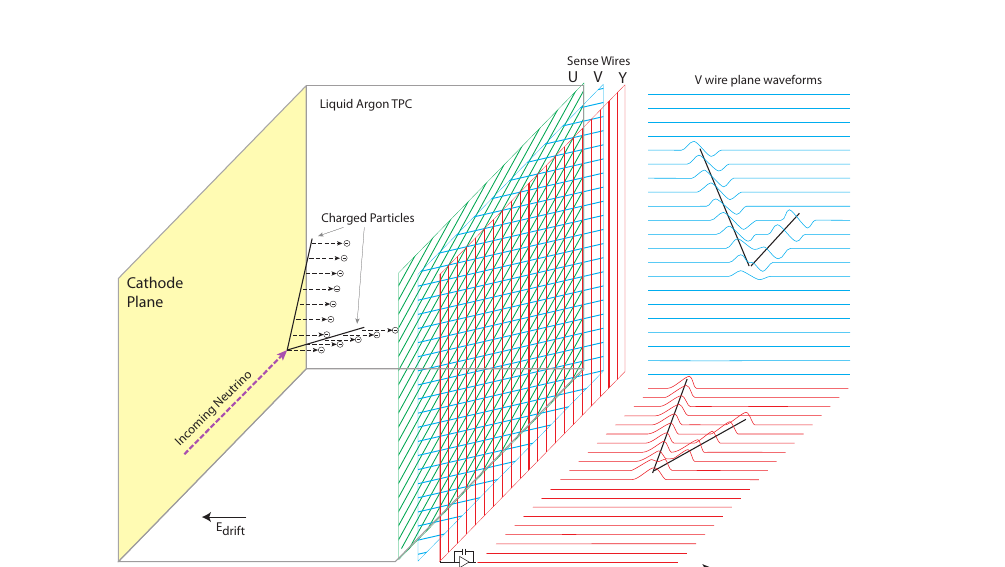}
\caption[Cartoon of the MicroBooNE LArTPC showing its working principle.]{Cartoon of the MicroBooNE LArTPC showing arrangement of the anode and cathode planes, and the detector's working principle. Taken from Fig.~2 in Ref.~\cite{uB_detector}.}
\label{fig:lartpc_principle}
\end{figure}

The MicroBooNE LArTPC is situated along the neutrino beam direction, with the cathode and anode planes sitting at the beam-left and beam-right sides of the detector, parallel to the beam direction. Figure~\ref{fig:lartpc_principle} shows the arrangement of the LArTPC.

The cathode plane is assembled by nine individual stainless steel sheets and kept at a high negative voltage of -70kV during normal operation. A uniform electric field of strength 273 V/cm is established by a series of field rings enclosing the volume between the cathode and anode planes. The field rings are connected by a voltage divider chain such that each ring is kept at a different electrical potential, which maintains a uniform electric field between the cathode and anode planes. A picture of the field rings and the cathode plane is shown in Fig.~\ref{fig:ub_detector_pic}.

At the anode, there are three sense wire planes: two induction planes with wires oriented at $\pm 60^{\circ}$ from vertical (also called ``U'' and ``V'' plane, or plane 0 and plane 1), and one collection plane with vertically oriented wires (referred to as ``Y'' plane, or plane 2). The U, V, and Y wire planes are spaced 3 mm apart and have 2368, 2368, and 3456 wires respectively, each with a wire pitch of 3 mm. The bias voltages set on the U, V and Y planes are $-110$ V, $0$ V and $230$ V.

\begin{figure}[h!]
\centering
\includegraphics[width=0.5\textwidth, trim = 3cm 18.3cm 11cm 2.1cm, clip=true]{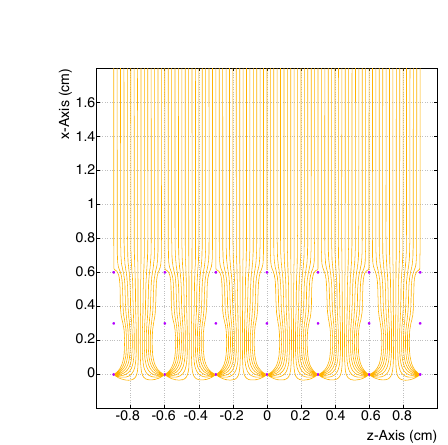}
\caption[The electron drift paths in the applied electric field simulated for a portion of MicroBooNE LArTPC.]{The electron drift paths in the applied electric field (in yellow) for a 2D model of a portion of MicroBooNE LArTPC near a subset of wires, simulated with the Garfield software~\cite{garfield}. The x-axis is the direction of the drifting electric field, and the z-axis is in the beam direction. Taken from Fig. 2 in Ref.~\cite{tpc_signal_processing}.}
\label{fig:drift_path}
\end{figure}

Figure~\ref{fig:lartpc_principle} shows the working principle of the LArTPC. When charged particles from a neutrino interaction traverse through liquid argon, they ionize argon atoms along their path, leaving free ionization electrons and creating prompt scintillation photons. Under the uniform electric field, the ionization electrons drift toward the anode planes. Thanks to the highly-purified liquid argon, the ionization electrons can drift over distances of $\mathcal{O}(m)$ with minimal attenuation until they reach the anode planes. Under 273 V/cm electric field in the detector, electrons drift at a velocity of 1.1 mm$/\mu$s~\cite{electron_drift_argon} leading to a maximum drift time of 2.3 ms from the cathode plane to the anode plane. As shown in Fig.~\ref{fig:drift_path}, due to different bias voltages on three planes, the drifting electrons pass by the two induction planes and eventually get collected by wires on the collection plane. This creates ``bipolar'' induced signals on the induction planes and ``unipolar'' signals on the collection plane. 

\subsection{Light Detection System}
Besides the ionization electrons, light is also produced in this process. Liquid argon is an excellent scintillator and is transparent to its own scintillation light. When excited argon atoms decay, they emit vacuum ultraviolet (VUV) scintillation light with a wavelength of 128 nm isotropically~\cite{uB_detector}. The emitted scintillation photons have a fast component of 6.8 $ns$; thus, once detected in coincidence with a BNB beam spill, they serve as a neutrino interaction candidate trigger, and they are also used to determine the time of neutrino interaction $t_{0}$. This is needed to determine the location of the interaction in the drifting direction, in combination with the measured arrival time of corresponding ionization charge on the anode plane.  

\begin{figure}[h!]
\centering
\includegraphics[width=0.5\textwidth, page=41, trim = 4cm 22cm 11cm 2.1cm, clip=true]{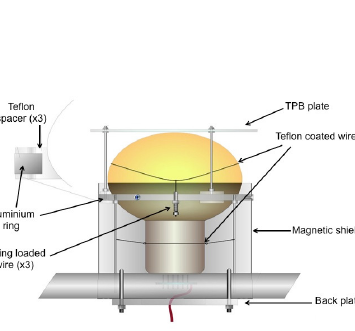}
\caption[Diagram of one optical unit in MicroBooNE.]{Diagram of one optical unit: a PMT situated behind a TPB-plate and surrounded by a mu-metal shield. Taken from Fig.~29 in Ref.~\cite{uB_detector}.}
\label{fig:uB_PMT_pic}
\end{figure}

\begin{figure}[h!]
\begin{subfigure}{.99\textwidth}
  \centering
  \includegraphics[width=\textwidth, page=38, trim = 2cm 21cm 2cm 2.1cm, clip=true]{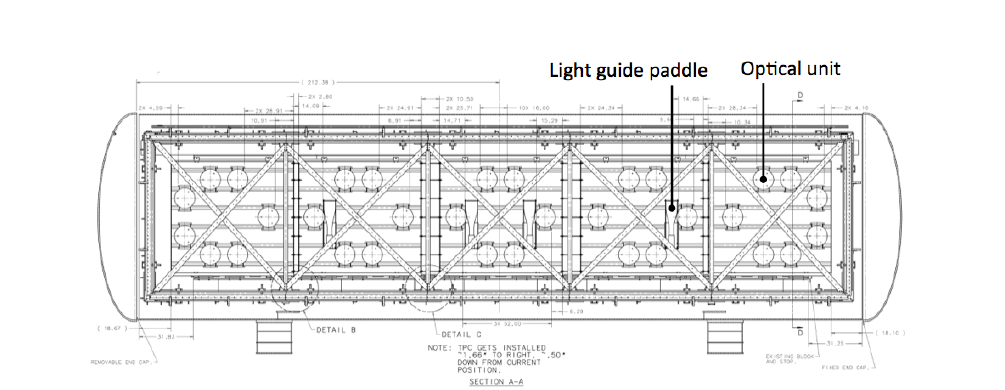}
  \caption{Sketch of MicroBooNE light-collection system}
  \label{fig:ub_light_sys}
\end{subfigure}
\begin{subfigure}{.99\textwidth}
  \centering
  \includegraphics[width=0.9\textwidth, page=40, trim = 2cm 20cm 2cm 2.1cm, clip=true]{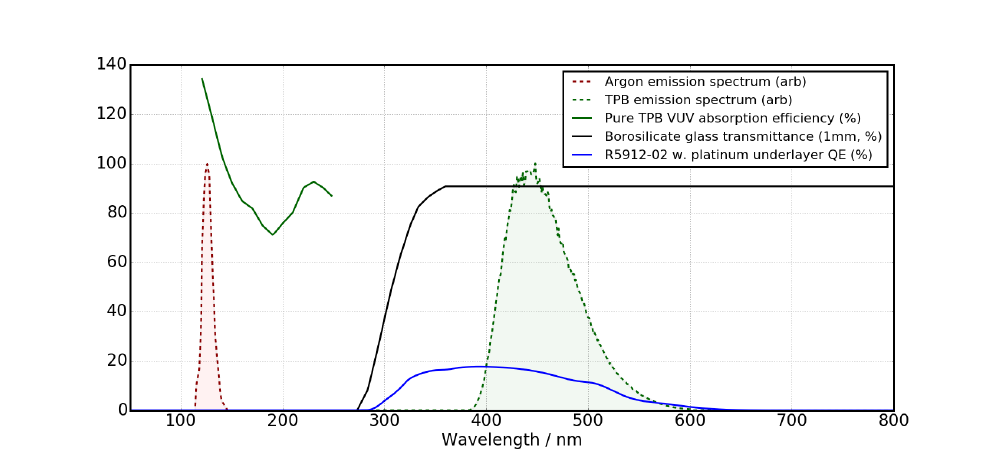}
  \caption{Scintillation light emission and detection spectrum}
  \label{fig:uB_PMT_eff}
\end{subfigure}
\caption[MicroBooNE's light-collection system and detection efficiency of the PMTs.]{(a): Diagram showing the orientation of the 32 PMTs (labeled as ``optical units'') in MicroBooNE detector. (b): This figure highlights the challenge in detecting the scintillation light from argon. The red dashed curve shows the spectrum of scintillation light emitted by argon; the solid blue curve represents the efficiency of the R5912-02MOD PMT employed by MicroBooNE; the dashed green curve represents the re-emission spectrum of the TPB. Taken from Figs.~26 and~28 in Ref.~\cite{uB_detector}.}
\end{figure}

The light-collection system is situated behind the anode plane facing into the detector volume and primarily comprises 32 cryogenic R5912-02MOD Hamamatsu PMTs. The placement of the PMTs are shown in Fig.~\ref{fig:ub_light_sys}, yielding 0.9\% photocathode coverage. As highlighted in Fig.~\ref{fig:uB_PMT_eff}, the MicroBooNE PMTs are sensitive to photons with wavelengths of 300$\sim$600 nm. Thus, in order to detect the scintillation light, tetraphenyl-butadiene (TPB) is used to convert the VUV light to visible wavelengths peaking at 425 detectable by the PMTs. Figure~\ref{fig:uB_PMT_pic} shows one unit of the PMT placed behind an acrylic plate coated with a TPB-rich layer.

\section{Detector Readout and Triggering}\label{ch2:uB_readout_trigger}
\begin{figure}[h!]
\centering
\includegraphics[width=1.1\textwidth, trim = 4.8cm 18cm 3cm 2.1cm, clip=true]{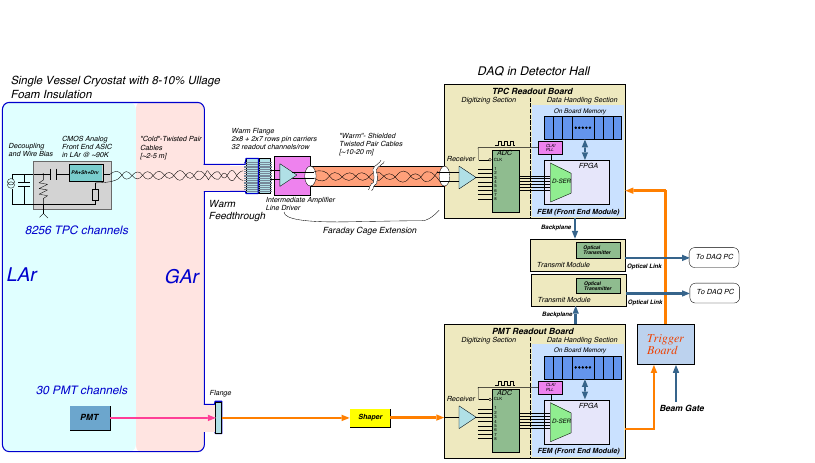}
\caption[Scheme for MicroBooNE's LArTPC and PMT signal processing and readout.]{Scheme for MicroBooNE's LArTPC and PMT signal processing and readout. Taken from Fig.~45 in Ref.~\cite{uB_detector}.}
\label{fig:uB_readout}
\end{figure}

The signals from the LArTPC and light collection system need to be amplified, digitized, and organized by the Data Acquisition (DAQ) systems before being written to disk for analysis. The LArTPC and PMT readout systems use a different designs for analog front-end and digitization and share the same back-end design for organizing and packaging data\footnote{The Columbia neutrino group at Nevis Labs led the design and construction of the warm readout electronics for the MicroBooNE detector.} before sending them to the DAQ system~\cite{uB_detector}, as shown in Fig.~\ref{fig:uB_readout}. This section will briefly describe the readout system for the LArTPC and PMT, as well as the trigger system employed in MicroBooNE. 

\subsection{TPC Readout System}
To minimize the electronics noise, MicroBooNE uses cryogenic front-end electronics situated inside the Cryostat to read out the LArTPC. For this purpose, the analog front-end application-specific integrated circuits (ASICs) are integrated on a cold motherboard directly attached to wire carrier boards on the LArTPC. The signals from 8256 LArTPC wires are independently pre-amplified and shaped by the front-end ASICs before being transmitted to the warm interface electronics through cold cables. The warm interface electronics are installed on top of signal-feedthrough flange onto the Cryostat. Here, the signals are further amplified by intermediate amplifiers shielded in a Faraday cage in order to prepare for $\sim20$ meter-long transmission to the warm digitizing and processing readout electronics in the detector hall.   

The readout of 8256 LArTPC wires is distributed among 9 readout crates, each consisting of $16-18$ Front End readout Modules (FEMs) and connected to a dedicated DAQ server. A custom, BNL-designed analog-to-digital conversion (ADC) module on the FEM handles signals from 64 wires through 8 octal-channel 12-bit ADCs and digitizes the signals continuously at a frequency of 16 MHz. The digitized data then get downsampled to 2 MHz by the field-programmable gate array (FPGA) on the FEM. The FPGA is responsible for processing the signals, reducing the data rate, and preparing the data for readout by the DAQ system. 

MicroBooNE has two separate data streams for readout for different purposes: one stream contains losslessly compressed LArTPC data recorded in coincidence with the event Level-1 trigger (discussed in Sec.~\ref{ch2:uB_trigger}), which is referred to as the ``NU'' data stream; the second stream is a continuous, lossily compressed readout of the LArTPC data. The continuous data is used for beam-unrelated physics analysis, such as the detection of potential supernova bursts, and is referred to as the ''SN'' stream. 

\begin{figure}[h!]
\centering
\includegraphics[width=0.9\textwidth, page=67, trim = 4.5cm 18.5cm 4.5cm 8cm, clip=true]{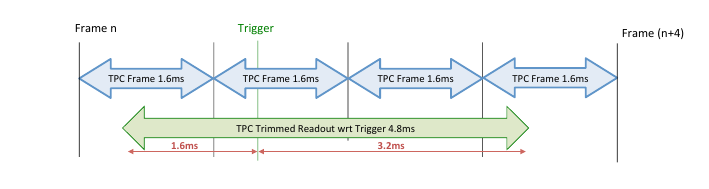}
\caption[MicroBooNE readout for NU stream.]{MicroBooNE readout for NU stream. Taken from Fig.~56 in Ref.~\cite{uB_detector}.}
\label{fig:ub_readout_window}
\end{figure}

In MicroBooNE, the frame size for readout is set to 1.6 ms. For the NU stream, when an event trigger is received, 4.8 ms of data spanning three or four frames are trimmed, as shown in Fig.~\ref{fig:ub_readout_window}. Then, the trimmed data is run through the Huffman encoding~\cite{Huffman}, which losslessly reduces the data rate by a factor of $\sim$5. For the continuous SN stream, further reduction is required. The FPGA applies two sequential data reduction algorithms to the SN data~\cite{uB_SN_stream}. Zero suppression is applied first, where the ADC samples not meeting a configurable threshold (for each channel) with reference to a configured baseline are discarded, followed by the Huffman compression, yielding an overall compression factor of $\approx 20-80$. 

From the readout electronics, the compressed data is then sent to a dedicated sub-event buffer (SEB) DAQ server, which creates sub-event fragments. For the NU stream, these fragments are sent to an event-building machine (EVB) for full-event building, and an additional high-level software trigger utilizing light information is applied to further reduce the event rate before the events are written to the local disk on the EVB and later sent offline for further processing. For the SN stream, data sits on the SEB, and is not transferred to offline storage unless a SuperNova Early Warning System (SNEWS)~\cite{SNEWS} alert is received. If no SNEWS alert is issued and disk occupancy reaches 80\%, the oldest SN data is permanently deleted until the occupancy falls below 70\%~\cite{uB_SN_stream}.

\subsection{Light Readout System}
Light produced by neutrino interactions in the detector is an important complement of the charge information and provides critical information for neutrino event selection and reconstruction. For NU stream, requiring light collected to be in coincidence with the beam spill can help reject cosmic-ray background significantly; for SN stream, the light information can provide event time $t_0$ when matched correctly to corresponding ionization activity in the TPC.

\begin{figure}[h!]
\centering
\includegraphics[width=0.9\textwidth, page=69, trim = 4cm 20cm 3cm 2.1cm, clip=true]{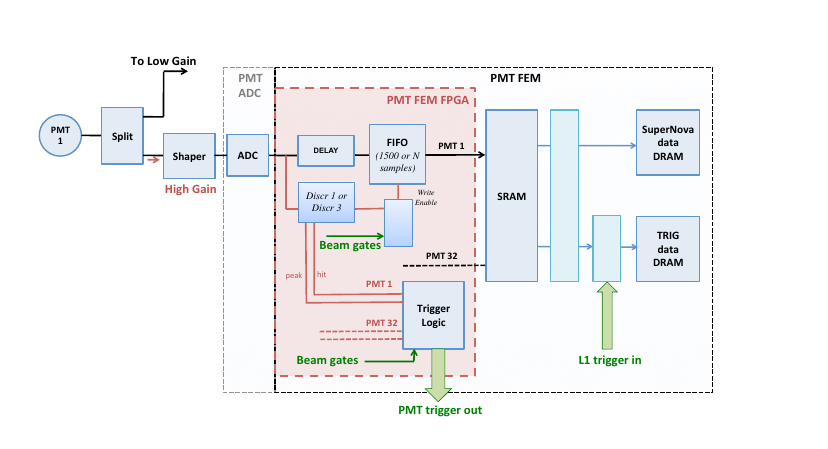}
\caption[Cartoon illustrating the processing of PMT signal in MicroBooNE.]{Cartoon illustrating the processing of PMT signal in MicroBooNE. Taken from Fig.~58 in Ref.~\cite{uB_detector}.}
\label{fig:ub_PMT_readout}
\end{figure}

The PMT readout system processes signals from the PMTs and identifies light coincident with the beam spills for triggering. Figure~\ref{fig:ub_PMT_readout} shows the stages of PMT signal processing. The signal from each PMT is split into two different gains: the high-gain (HG) channel carries 18\% of the signal, and the low-gain (LG) channel carries 1.8\% of the signal~\cite{uB_detector}. Each gain channel is split again into HG1, HG2, and LG1 and LG2. These separate copies allow the PMT readout system to separately process beam-related and beam-unrelated PMT signals with ease. 

After the HV/signal splitter, the PMT signals are pre-amplified and shaped into unipolar signals by preamp/shaper boards and then passed to PMT readout modules, which digitize the signals at 64 MHz and further process them. In the case of a BNB or other external trigger, PMT signals spanning four 1.6 ms frames (frame containing the trigger, one/two frames preceding/following the trigger frame) are recorded for the NU stream. Zero suppression is applied after the digitization in order to reduce the data rate except for data surrounding the beam spill. When a BNB-coincident beam-gate signal is received, 4 $\mu$s earlier than the expected arrival of neutrinos, 1500 consecutive PMT samples (corresponding to 23.4 $\mu$s) surrounding and including the beam-spill period are read out continuously with no compression. After signal processing, PMT data is sent to the designated DAQ in the same way as for LArTPC data. 

While processing the PMT signals, the PMT readout module can also generate two different PMT-based triggers when certain criteria are met: a cosmic PMT trigger or a beam-gate-coincident PMT trigger~\cite{uB_detector}. The formation of the cosmic PMT trigger makes use of PMT signals outside the beam-spill-surrounding period, while the beam-gate-coincident PMT trigger requires PMT signals inside the beam spill period, and the construction of these two triggers has different requirements on the PMT hit amplitude and hit multiplicity. The PMT trigger information is available for the events in the NU stream at the event-building stage and offline for future analysis. 

\subsection{MicroBooNE Trigger System}\label{ch2:uB_trigger}
MicroBooNE employs two-level triggering to control the data rate to a maintainable level. The first trigger is the \textit{hardware trigger}, or ``Level-1'' trigger. The MicroBooNE Trigger Board (TB) takes charge of issuing the ``Level-1'' trigger. The TB takes input from the BNB, NUMI, a configurable fake beam (strobe) trigger, and two calibration signals as trigger inputs. These trigger inputs can be independently pre-scaled, masked, and mixed together to generate a Level-1 trigger. Once a hardware trigger is received, the readout of the NU data stream from both the LArTPC and the PMT systems is recorded.

If every triggered event were to be recorded, running with a 5 Hz beam trigger rate would lead to a data volume of $\sim$13 TB per day with Huffman compression. Since, on average, MicroBooNE expects to see one neutrino interaction inside the detector active
volume per $\sim$600 beam-spills, to reduce the data rate further, MicroBooNE applies a high-level \textit{software trigger}, run within the DAQ in the event-building stage. For all Level-1 triggered events, the software trigger looks at the PMT signal during the beam-spill, and if there is a significant amount of PMT light coincident with the arrival of the neutrino beam, the event will be saved. This software trigger reduces the data rate by a factor of $\sim22$ to 500 GB per day with negligible efficiency loss for the neutrino interactions~\cite{uboone_WC_eLEE}. This is lower than the factor of $\sim 600$ because of random coincident cosmics since MicroBooNE detector operates on-surface. 

\section{Detector Operations}
\begin{figure}[h!]
\centering
\includegraphics[width=\textwidth, trim = 1cm 16cm 3cm 2.1cm, clip=true]{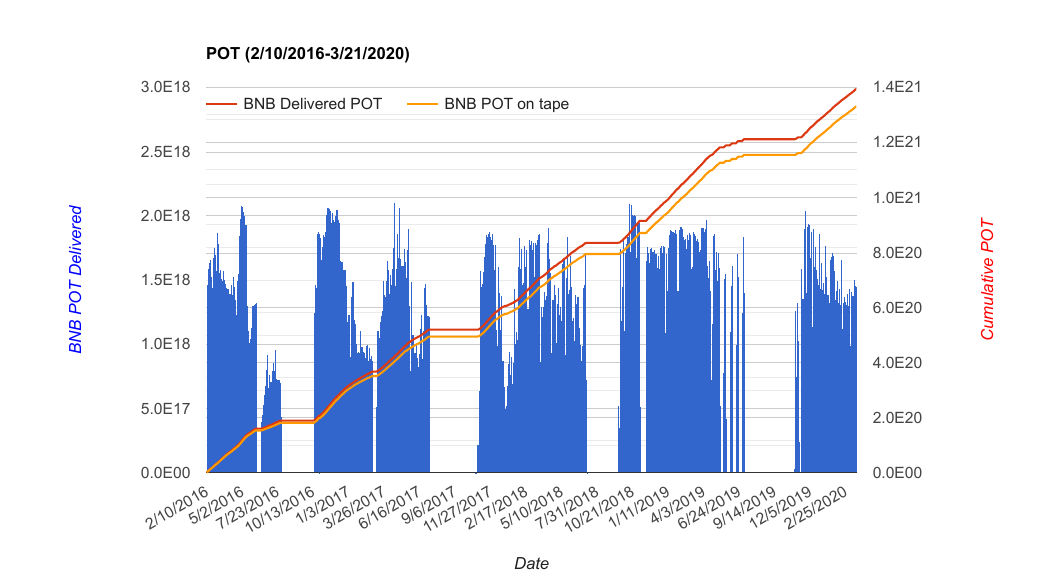}
\caption[Accumulated POT delivered to MicroBooNE during data-taking period.]{POT delivered to MicroBooNE over five-year data-taking period.}
\label{fig:ub_runs15_POT}
\end{figure}

MicroBooNE ran for 5 years between 2015 and 2020, with a total exposure of $1.3\times 10^{21}$ protons-on-target (POT). Data collected each year marks an individual run, starting with first-year data considered as ``Run 1'', etc. Figure~\ref{fig:ub_runs15_POT} shows the accumulated POT delivered to MicroBooNE during Runs 1-5 data taking. The analysis presented in this thesis utilizes $6.868\times 10^{20}$ POT of data collected during the first three years of running (referred to as ``Runs 123''). 

%% file: chapter2.1.tex
\chapter{Simulation and Reconstruction in MicroBooNE}\label{chap:2.1}
This chapter discusses the simulation and reconstruction of events in MicroBooNE. Section~\ref{ch2.1:MC} describes general event simulation and background modeling in MicroBooNE, followed by Sec.~\ref{ch2.1:ub_event_reco} where the signal processing and pattern-recognition reconstruction are briefly discussed, and Sec.~\ref{ch2.1:high_level_reco} focusing on high-level reconstruction specific to the analyses presented in this thesis. 

\section{Neutrino Interaction Simulation and Cosmic Modeling}\label{ch2.1:MC}
Any new process search or precise cross-section measurement typically requires an accurate MC simulation of the experiment for a reliable estimate of the contribution from different types of backgrounds. MicroBooNE uses both MC simulation and data-driven techniques for background evaluation. 

\subsection{Cosmic Modeling}\label{ch2.1:cosmic_model}
As a detector operating on the Earth's surface, without significant overburden, the MicroBooNE LArTPC sees a high rate of cosmic-ray muons at 5 kHz; this leads to $\mathcal{O}(10)$ cosmic muons entering the detector volume during each 2.3 ms drift time window~\cite{uB_cosmic_corsika}. Thus, it is crucial to have a precise model of the cosmic ray activity in the detector. MicroBooNE utilizes recorded data collected when the neutrino beam is turned off to estimate the cosmic activity in the detector. Two sets of beam-off data are recorded for different purposes. One data stream records cosmic activity without any optical requirements, and is overlaid with simulated neutrino activity to account for cosmic activity in the detector during the TPC readout window. The other data stream contains events that are triggered by the optical signals from cosmic activity during the fake beam spill window and saved in the same manner as the NU stream. This data stream is used to estimate recorded events that are triggered by cosmic activity when no neutrino interacts in the detector during the BNB beam-spill. 

In MicroBooNE, the cosmic-ray rate can also be simulated using CORSIKA \cite{corsika} \texttt{\detokenize{v7.4003}}. CORSIKA simulates a flux of the cosmic nuclei and the hadronic interactions in the air showers initiated by the cosmic nuclei~\cite{uB_cosmic_corsika}; then the primary particles from the interaction and their interaction/decay products are propagated through the MicroBooNE LArTPC using GEANT4 and subsequent simulation tools in MicroBooNE.

\subsection{Neutrino Interaction Simulation}
In general simulation of neutrino interactions in MicroBooNE can be broken down into four steps. First, the flux of the neutrino beam is simulated, as described in Sec.~\ref{ch2:nu_beam_sim}. Then, the simulated neutrino flux profile is fed into the \textsc{genie}~\cite{genie} neutrino event generator. MicroBooNE adopted \textsc{genie} \texttt{\detokenize{v3.0.6}} as the core event generator with a customized tune~\cite{uB_genie_tune} of the CCQE and CC two-particle two-hole (CC2p2h) models to T2K CC$0\pi$ data~\cite{T2K_cc0pi}. The \textsc{genie} simulation takes into account the energy of the neutrino sampled, the materials the neutrino encounters, and the cross-section of the interaction, and simulates neutrino interactions inside and outside the MicroBooNE cryostat.

The final state particles from \textsc{genie} interactions are passed to LArSoft~\cite{larsoft} framework, which is a toolkit for the full simulation in the detector, reconstruction as well as analysis of the LArTPC events. The propagation of the final state particles in the detector is simulated by the GEANT4~\cite{geant4} toolkit \texttt{\detokenize{v4_10_3_03c}} implemented in LArSoft. This includes the energy deposition, ionization electrons, and optical photons produced along the path of the particle trajectory. LArSoft simulates the propagation of scintillation light in the detector medium to the PMTs and the drift of the ionization charge to the wires while taking into account different detector effects, such as Rayleigh
scattering, reflection, and partial absorption of the light in the medium, electron recombination with argon ions~\cite{electron_recomb}, and space charge effects~\cite{SCE,spacecharge,ub_SCE_effect,uB_Efield_measurement}. 

Then, a dedicated detector simulation is used to simulate the electric field response\footnote{Field response function is defined by the induced current on one wire due to a single electron charge~\cite{tpc_signal_processing}.} and electronics response of the detector~\cite{tpc_signal_processing}, while taking into account the diffusion~\cite{uB_elec_diffusion} and absorption of the ionization charge during the drift. The field response functions of wires are calculated based on the electron drift paths (example shown in Fig.~\ref{fig:drift_path}) and the weighting fields of wires simulated with the Garfield program~\cite{garfield}. The electronics response includes two parts and models the behavior of the pre-amplifier that amplifies and shapes the analog signals from the TPC, and the behavior of resistor-capacitor (RC) circuits employed to remove the baseline from the pre-amplifier and the intermediate amplifier~\cite{uB_noise_filter}. 

\begin{figure}[h!]
\centering
\includegraphics[width=0.6\textwidth, trim = 3cm 12.5cm 10.5cm 2.1cm, clip=true]{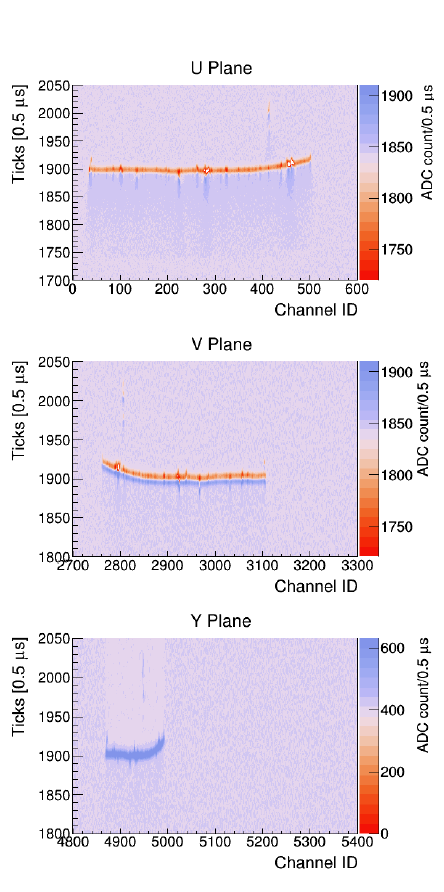}
\caption[Simulated waveforms from a GEANT4-based 5-GeV muon track in liquid argon.]{Simulated waveforms from a GEANT4-based 5-GeV muon track in liquid argon. The x-axis represents the channel (wire) of the three wire planes, the y-axis represents the sampled time tick, and the z-axis shows the ADC value of the simulated waveform proportional to the energy deposition per unit length. Taken from Fig.~22 in Ref.~\cite{tpc_signal_processing}.}
\label{fig:uB_simulated_muon_waveform}
\end{figure}

Eventually, the TPC signal for each channel is simulated as~\cite{tpc_signal_processing}:
\begin{equation}\label{eq:ch2.1:TPC_signal}
    M = (\text{\textbf{I}}^{\text{\textbf{depo}}}\otimes\text{\textbf{T}}^{\text{\textbf{drift}}}\otimes \text{\textbf{R}}^{\text{\textbf{duct}}} + \text{\textbf{N}})\otimes \text{\textbf{R}}^{\text{\textbf{digit}}}
\end{equation}
where 
\begin{itemize}
    \item $\text{\textbf{I}}^{\text{\textbf{depo}}}$ represents the initial distribution of the ionization electrons due to energy deposition of charged particles.
    \item $\text{\textbf{T}}^{\text{\textbf{drift}}}$ represents a function that transforms initial ionization charges to distributions of electrons arriving at the wires, which includes various detector effects.
    \item $\text{\textbf{R}}^{\text{\textbf{duct}}}$ represents the response of the wires, which is the convolution of field response and electronics response. 
    \item $\text{\textbf{N}}$ represents simulated inherent electronics noise~\cite{tpc_signal_processing}.
    \item $\text{\textbf{R}}^{\text{\textbf{digit}}}$ models the waveform digitization process. 
\end{itemize}
Figure~\ref{fig:uB_simulated_muon_waveform} shows the simulated waveform for a 5-GeV muon. 

As discussed in Sec.~\ref{ch2.1:cosmic_model}, the MicroBooNE detector is exposed to a constant cosmic-ray rate. Thus, to account for cosmic activities during the readout window, at last, the simulated waveforms are overlaid with the data waveform collected for beam-off events; the resulting simulation sample is referred to as Monte Carlo ``overlay''. It's worth-noting that the NC coherent $1 \gamma$ events, signal targeted by one of the two analyses presented in this thesis, is simulated in a standalone \textsc{genie}, not the default MicroBooNE \textsc{genie} tune. Therefore all steps remain the same for the simulation of NC coherent $1 \gamma$ events, except the \textsc{genie} simulation step. This is discussed in more details in Ch.~\ref{ch:signal_model}.

\clearpage
\section{Event Reconstruction}\label{ch2.1:ub_event_reco}
Event reconstruction is a sequential process, which includes processing of the TPC signals and PMT light signals applied to all recorded data events, and high-level reconstructions leading to reconstructed particle tracks and showers. Section~\ref{ch2.1:light_reco} and~\ref{ch2.1:tpc_reco} describes the TPC and optical reconstruction respectively, and Sec.~\ref{ch2.1:pandora_reco} briefly describes the Pandora~\cite{pandora_reco} event reconstruction used in the analyses presented in this thesis. 
\subsection{Light Signal Reconstruction}\label{ch2.1:light_reco}
The optical reconstruction processes the raw waveform from all PMTs, and identifies PMT signals close in time to form ``flashes'' which are usually caused by the neutrino or cosmic-ray interactions in the detector. First, a baseline is estimated for the PMT ADC waveform, and PMT pulses are formed from the waveform above a threshold relative to the baseline~\cite{WireCell_light_charge_matching, DelTutto_thesis}. Then, a flash is formed if a group of pulses across several PMTs within the same time interval are identified. The starting time of the flash is defined as the time of the maximal summed photoelectrons (PEs) from all PMTs, and each flash window is 7.8 $\mu$s long, in order to collect the slow component of the scintillation light and exclude noise. Each flash is characterized by the time of flash with respect to the trigger and the total PEs of the flash which are integrated across all PMTs over the entire window. Flashes are used in charge-light matching later in the reconstruction chain to identify neutrino interactions. 

\subsection{TPC Signal Reconstruction}\label{ch2.1:tpc_reco}
The TPC signal processing starts with the removal of inherent noise from electronics, which includes inherent noise from the cold ASICs, the intermediate amplifier, and the ADC. During the initial operation of the detector from October 2015 through July 2016, excess noise was identified in the TPC readout, and offline noise filtering was developed, which removes most of the excess noise while preserving the signal~\cite{uB_noise_filter}. 
Hardware upgrades were performed in the summer of 2016, after which the noise level was suppressed by a factor of 3 for the induction planes and a factor of 2 for the collection plane. The offline noise filter further suppresses the noise by an additional 10$-$20\%. 

\begin{figure}[h!]
\centering
\includegraphics[width=0.8\textwidth, trim = 3cm 16cm 3cm 2.1cm, clip=true]{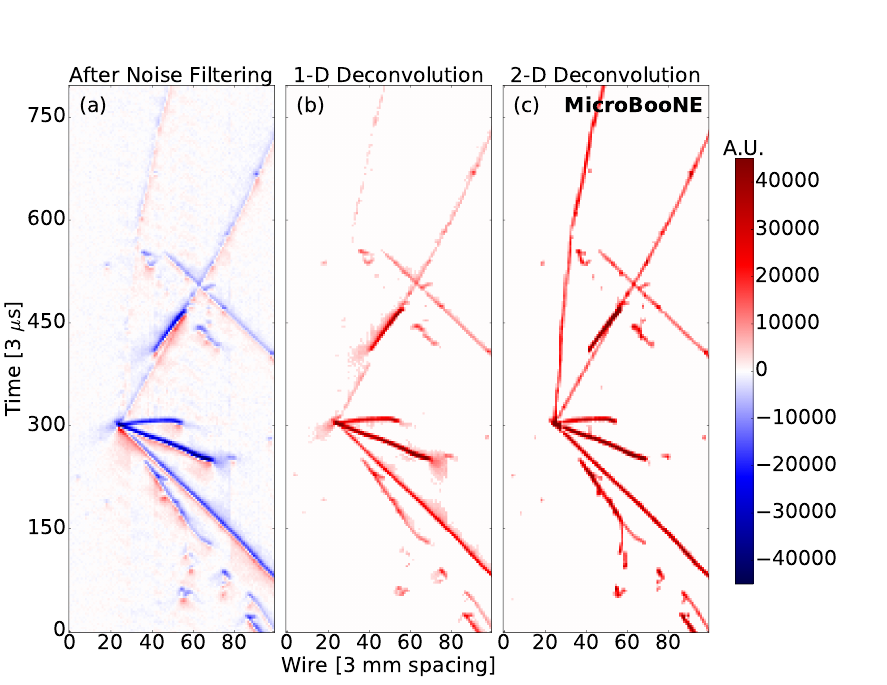}
\caption[The U plane waveforms for a recorded neutrino candidate in MicroBooNE demonstrating the outcome of deconvolution.]{The U plane waveforms for a recorded neutrino candidate in MicroBooNE: the x-axis represents the wire and the y-axis represents the time in units of 3 $\mu$s (corresponding to 6 sampling ticks). (a) Raw waveform after noise filtering, shown in unit of baseline subtracted ADC scaled by 250. (b) Charge spectrum in units of electrons after 1D deconvolution. (c) Charge spectrum in units of electrons after 2D deconvolution.  Taken from Fig.~11 in Ref.~\cite{tpc_signal_processing}.}
\label{fig:uB_2D_deconvolv}
\end{figure}

The second step of TPC signal processing is to reconstruct the ionization charge distribution. As shown in Eq.~\ref{eq:ch2.1:TPC_signal}, the raw wire signal is a result of convolution between the ionization electron distribution and the detector response. MicroBooNE employs two-dimensional (2D) deconvolution~\cite{tpc_signal_processing, tpc_signal_processing_2} to deconvolve the detector response and recover the ionization distribution. Unlike traditional 1D deconvolution~\cite{1d_deconvolution}, which considers current induced by ionization electrons when passing by the sense wire, 2D deconvolution also takes into account the contribution from ionization electrons drifting in neighboring wire regions and thus involves both time and wire dimensions. During the deconvolution, an additional software filter inspired by the Wiener filter~\cite{wiener_filter} is applied to attenuate high-frequency noise while excluding the suppression at low-frequency in the Wiener filter. The 2D deconvolution method yields a more accurate recovery of the ionization charge distribution, as highlighted in Fig.~\ref{fig:uB_2D_deconvolv}. 

In order to reduce the data size, the region-of-interest (ROIs) technique~\cite{1d_deconvolution} is applied on the deconvolved signal distribution. For each channel, the root mean square (RMS) noise is calculated from the deconvolved signal over the entire readout window (4.8 ms) and used to set the threshold for ROI identification. Once ROIs are identified on the deconvolved signals, the entire TPC readout is replaced by a set of ROIs\footnote{For the induction plane, additional linear baseline subtraction is performed in the ROI window, in order to suppress the low-frequency noise arising from 2D deconvolution~\cite{tpc_signal_processing}. The baseline is calculated as the interpolation of baselines at the start and end of the ROI window.}. Each ROI can contain more than one (typically less than five) Gaussian pulses (referred to as ``hits''), and a hit-finding algorithm~\cite{hit_finding_algo} is utilized to identify these Gaussian hits with associated hit time, width and peak amplitude information. These reconstructed Gaussian hits serve as inputs to high-level reconstruction algorithms later in the reconstruction chain.

\subsection{Pandora High-level Reconstruction}\label{ch2.1:pandora_reco}
MicroBooNE has explored multiple high-level reconstruction paradigms, including Pandora reconstruction~\cite{pandora_reco}, Wire-Cell tomographic reconstruction~\cite{wirecell_reco}, and Deep-learning based reconstruction~\cite{DL_reco}. The wire cell reconstruction leverages the sparsity of neutrino interaction and utilizes the charge, light, and time information to identify energy deposition in 3D first, prior to forming particle trajectories. The deep-learning based reconstruction takes advantage of MicroBooNE's 2D readout from each plane and uses convolutional neural networks to associate pixels with different particle activities. The Pandora reconstruction technique is used by the analyses discussed in this thesis and is discussed in more detail below. 

The Pandora reconstruction is integrated into the LArSoft framework, and it takes a multi-algorithm approach to pattern recognition. Given reconstructed hits with wire number and waveform information as inputs, Pandora runs two reconstruction paths on them: ``PandoraCosmic'' and ``PandoraNu'', each involving multiple algorithms. 

``PandoraCosmic'' is run first, aiming to reconstruct cosmic-ray particles and associated delta rays. The clustering algorithms are run on hits to identify 2D clusters, which are groups of continuous hits on each plane. Then, all combinations of clusters from three planes (one from each plane) are examined, and the time and consistency of each group are checked to identify track-like trajectories in 3D. The clustering algorithms and 3D track reconstruction algorithms are run back and forth until there is no ambiguity in the clusters formed, and the final 3D tracks are taken as the reconstructed cosmic-ray muons. After the track reconstruction, any hits that are not associated with reconstructed tracks are regrouped in 2D and matched in 3D to reconstruct the delta rays and associate them with parent muon tracks. For any reconstructed particle, 3D hits (called ``spacepoints'') are created from the 2D hits across planes.

``PandoraNu'' then takes hits that are \textit{not} reconstructed as cosmic activity by ``PandoraCosmic'' and reconstructs neutrino interactions out of them. The same algorithms are run as in ``PandoraCosmic'', and the original 2D input hits are split into different groups (called ``\textit{slices}'') based on the geometric information of associated 3D hits~\cite{pandora_reco}. The 2D Gaussian hits in each slice are then processed individually to reconstruct neutrino interactions. 

In neutrino interaction reconstruction, a 3D vertex that defines the interaction point or the first energy deposit is first identified. Vertex candidates are formed from pairs of 2D clusters across different planes, scored based on cluster energy, distribution of hits, and the position of the candidate vertex, and the vertex candidate with the highest score is picked. 

Following the identification of interaction vertex, shower and track reconstruction starts. 2D clusters are classified as shower-like or track-like based on the linearity, hit dispersion, and cluster extent. On each plane, a long shower-like cluster that typically points back to the vertex is identified and represents the shower spine, and small shower-like clusters are then added recursively to the shower spine, eventually yielding a large, complete 2D shower cluster. The 2D shower clusters on three planes are matched to form 3D shower particles. Like the track reconstruction case, 2D clustering and 3D matching processes are run recursively until unambiguous shower particles can be formed. Track-like clusters are examined to form 3D track reconstruction in the same way as discussed for ``PandoraCosmic'' path.

\section{High-Level Energy Reconstruction}\label{ch2.1:high_level_reco}
The output from Pandora is processed further to extract additional high-level information for physics analysis. This section focuses on the two main calorimetric quantities associated with reconstructed showers most relevant to analyses in this thesis: the total shower energy, and energy loss per unit length ($dE/dx$) at the shower start. 

\subsection{Shower Energy Reconstruction}
The energy of the reconstructed shower starts with the energy reconstruction of the hits making up the shower, and involves two steps: 1) translating the integral of hit ADC waveforms to ionization charge; and 2) using the charge to recover the deposited energy of the particle. To convert from the hit integral (in ADC) to the charge ($Q$ in $e^{-}$), the conversion constant is calculated for MC simulation, and extracted for data using the $dE/dx$ profile of stopping muons in the detector~\cite{ub_SCE_effect, adc_charge_conversion_TN}. The conversion constants measured for data and MC for each readout plane are summarized in Tab.~\ref{tab:ch2.1_adc_conversion}~\cite{MCC9_dEdx_cali}.

\begin{table}
    \centering
    \begin{tabular}{|c|c|c|c|}
        \hline 
        Conversion Constant & Plane 0 & Plane 1 & Plane 2\\\hline
        Monte Carlo Overlay & 235.5 &249.7& 237.6\\\hline
        Data & 230.3 & 237.6 & 243.7\\\hline
    \end{tabular}
    \caption[The ADC-charge conversion constants measured for each plane for data and MC.]{The ADC-charge conversion constants measured for each plane for data and MC, in units of $e^{-}/ADC$.}
    \label{tab:ch2.1_adc_conversion}
\end{table}

Then to convert from charge to the energy of the particle, a simplified conversion is applied: 
\begin{equation}\label{eq:Q2E}
    E = \frac{W_f}{R_C} \times Q
\end{equation}
where $W_f = 23.6 \text{eV}/e^{-}$ is the ionization work function of argon, and $R_C = 0.62$ is the recombination factor, accounting for the fact that the dispersed drift electrons may quickly thermalize and recombine with nearby ions. Using a fixed recombination factor is a simplification because the recombination factor is a function of energy deposition. Larger energy deposition leads to a higher density of positive ions, thus the probability that one of the liberated electrons can recombine with an ion is higher and the recombination factor $R_C$ decreases. Here a simplified assumption of fixed $R_C$ is adopted, and its value is calculated using the so-called modified box model with parameter values measured by the ArgoNeuT experiment~\cite{electron_recomb_model} at MicroBooNE's electric field strength of 273 V/cm.

The reconstructed energy of the entire shower is found by summing up the energy $E$ of all hits on a plane-by-plane basis, and taking the max of all planes. This is due to the fact that clumps of energy are often not associated to the shower on all planes, due to dead wires or incorrect clustering, and choosing the maximum provides the most reliable information. 

The shower $dE/dx$ at the shower start is a powerful tool for distinguishing showers originating from photons from those originating from electrons. A photon doesn't ionize the argon until it pair-produces an $e^{+}$ and $e^{-}$ and the electromagnetic shower cascade happens. So while the initial $dE/dx$ of electron showers corresponds to the $dE/dx$ of minimum ionizing particle (MIP), showers induced from photons will have higher initial $dE/dx$ corresponding to twice that of MIPs. To evaluate the $dE/dx$ at the shower start, a Kalman filter~\cite{kalman_filter} based procedure is applied to the reconstructed shower to identify the main trunk of the shower and reject transversely or longitudinally displaced hits. The values of $dE/dx$ measured for the first 4-cm of the shower are extracted, and the median $dE/dx$ is used as the reconstructed $dE/dx$ at the shower start.

\subsection{Shower Energy Correction}\label{ch2.1:energy_corr}
\begin{figure}[h!]
    \centering
    \begin{subfigure}{0.45\textwidth}
        \includegraphics[width=\linewidth]{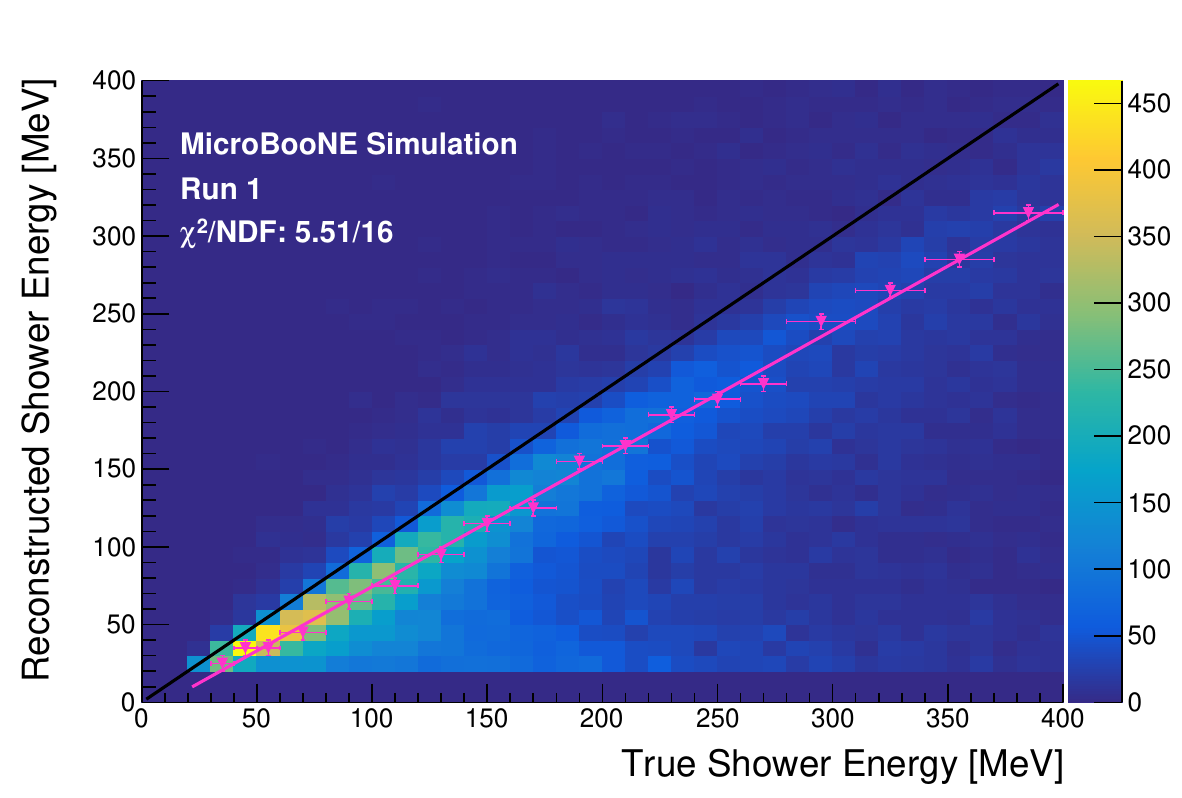}
        \caption{Uncorrected}
        \label{fig:shower_energy_corr_uncorrected}
    \end{subfigure} %
    \begin{subfigure}{0.45\textwidth}
        \includegraphics[width=\linewidth]{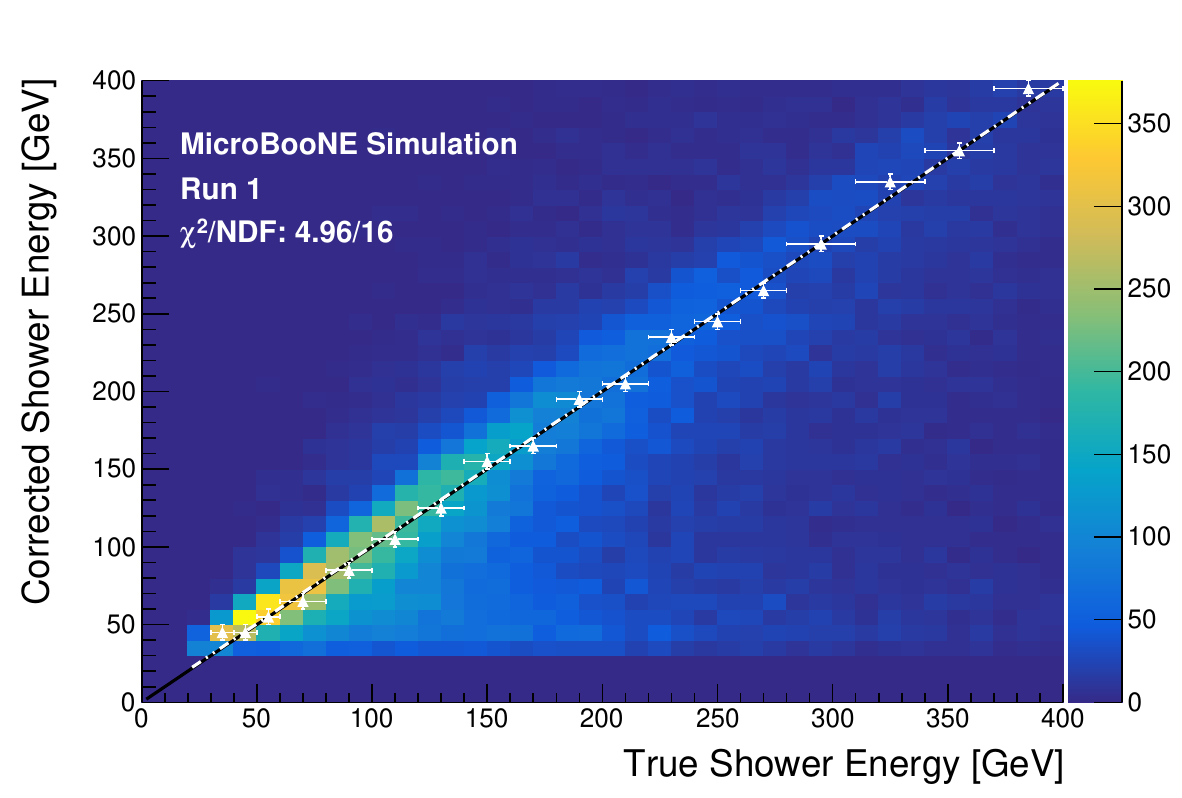}
        \caption{Corrected}
        \label{fig:shower_energy_corr_corrected}
    \end{subfigure} %
    \caption[2D distributions of reconstructed shower energy versus the true photon energy.]{(a) 2D distribution of reconstructed shower energy (y-axis) v.s. true energy (x-axis) for reconstructed showers with at least 30 MeV, evaluated with $\sim$100,000 simulated NC $\pi^0$ events. The points represent the most probable value in each true energy bin. (b) 2D distribution of corrected shower energy (y-axis) v.s. true shower. Taken from Fig.~10 in Ref.~\cite{gLEE_internal_note}
    }
    \label{fig:shower_energy_corr}
\end{figure}

As highlighted in Fig.~\ref{fig:shower_energy_corr_uncorrected}, the resulting reconstructed shower energy turns out to be systematically below the true shower energy. This is due to lossy effects such as thresholding, where small local energy depositions created during the shower cascade may go undetected because they fall below the detection threshold, and mis-clustering, which refers to hits being missed during clustering in the reconstruction leading to a partially reconstructed shower. On average, there is $\sim$20\% energy loss due to these effects~\cite{Caratelli_thesis}.

To account for reconstructed energy bias, a shower energy correction factor is extracted using simulated NC $\pi^0$ sample with Run 1 overlay. Reconstructed shower energy is plotted against the true photon energy for all showers with at least 30 MeV of simulated and reconstructed energy, as shown in Fig. \ref{fig:shower_energy_corr_uncorrected}. Then a linear fit is performed to the most probable value (MPV) in bins of true shower energy. The various-sized bins are hand-tuned to account for lower statistics at high energies and to provide a reasonable fit to the underlying 2D distribution. The correction equation derived from linear fit result is:
\begin{equation}\label{eqn:ncpi0_energy_corr_factor}
    E_{\textrm{corr}} = (1.21 \pm 0.03)E_{\textrm{reco}} + (9.88 \pm 4.86) \ \textrm{MeV}
\end{equation}
where $E_{\textrm{reco}}$ and $E_{\textrm{corr}}$ correspond to the reconstructed shower energy before and after correction, respectively. This represents an approximately 20\% correction, consistent with a previous MicroBooNE study~\cite{Caratelli_thesis}. Figure \ref{fig:shower_energy_corr_corrected} shows the 2D distribution of the corrected shower energy and true photon energy which are much closer after the correction. The validity of the reconstruction energy correction for Runs 2 and 3 is cross-checked through the reconstructed $\pi^0$ mass distributions across different runs. Good data-MC agreements are seen, and fitting the $\pi^0$ mass distributions with Gaussian-plus-linear functions yields consistent $\pi^0$ mass and width parameters across runs that are consistent to the expected values ~\cite{gLEE_internal_note}.

%% file: chapter3.tex
\chapter{MicroBooNE Neutral Current Delta Radiative Decay Measurement}
\label{ch3:ncdelta_LEE}
Single photons from NC $\Delta$ radiative decay is the second largest contribution to the photon background at low energy in MiniBooNE's $\nu_{e}$ measurement, as shown in Fig.~\ref{fig:MiniBooNE_data_mc_energy}. NC $\Delta$ radiative decay becomes a background because, while the nucleon from $\Delta$ decay is too heavy to generate Cherenkov light in the MiniBooNE detector, the decay photon creates Cherenkov light which cannot be separated from Cherenkov rings produced by electrons in MiniBooNE. On the other hand, the extraordinary spatial and energy resolution offered by the MicroBooNE LArTPC enables MicroBooNE to distinguish photons from electrons, making it ideal for investigating the root cause of the MiniBooNE LEE. 

This chapter gives a summary of the investigation of one of the leading interpretations of the MiniBooNE LEE by MicroBooNE: an anomalous rate of neutrino-induced NC $\Delta$ radiative decay, targeting specifically a 3.18 times enhancement of the process. The NC $\Delta$ radiative decay search is the first search performed by MicroBooNE under the photon hypothesis of the MiniBooNE LEE. Section~\ref{ch3:gLEE_motivation} briefly discusses the motivation and analysis overview, Secs.~\ref{ch3:glee_1p}, ~\ref{ch3:glee_2p},~\ref{ch3:glee_2p_rate} and~\ref{ch3:glee_fit} describe different parts of the analysis in more
detail. 
The full details of the analysis can be found in Ref.~\cite{uboone_delta_LEE, gLEE_internal_note, gLEE_ncpi0}.   

\section{Analysis Motivation and Overview}\label{ch3:gLEE_motivation}
\begin{figure}[h!]
    \centering
    \includegraphics[trim = 2cm 22cm 15.5cm 1.5cm, clip=true, width=0.4\textwidth]{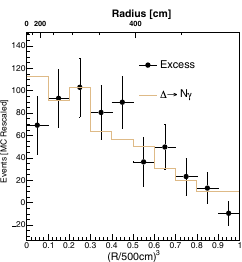}
    \caption[The vertex radial distribution of the MiniBooNE excess in neutrino-mode data.]{The vertex radial distribution of the excess in neutrino-mode data in MiniBooNE's $\nu_{\mu} \rightarrow \nu_{e}$ oscillation measurement, overlaid with the prediction for NC $\Delta$ radiative decay in MiniBooNE scaled to explain the excess. This distribution shows the good shape agreement between the observed excess and scaled NC $\Delta$ radiative decay prediction. Taken from Fig. 23 in Ref.~\cite{MiniBooNE_combine_osc}. } 
    \label{fig:miniboone_NCDelta_fit}
\end{figure}

NC $\Delta$ radiative decay has never been directly measured in MiniBooNE; it was only indirectly constrained through \textit{in-situ} NC $\pi^0$ measurements~\cite{PhysRevD.81.013005}. MiniBooNE's latest data release shows that a normalization scaling of 3.18 of the predicted NC $\Delta \rightarrow N+\gamma$ background provides the best fit to the observed excess~\cite{MiniBooNE_combine_osc}. More specifically, different processes are tested through a shape-only fit to the vertex radial distribution of the excess with only statistical uncertainty, where each process is normalized to explain the size of the excess. It was found that an enhancement of the NC $\Delta \rightarrow N\gamma$ process (as predicted by the \textsc{nuance}~\cite{nuance} neutrino generator used by MiniBooNE) by a factor of 3.18 fits the radial distribution the best, as highlighted in Fig.~\ref{fig:miniboone_NCDelta_fit}. 

A previous search of neutrino-induced NC $1\gamma$ processes in the $\mathcal{O}(1)$ GeV neutrino energy regime was performed by the T2K experiment~\cite{T2K_NC1gamma}, using a neutrino beam with $\langle E_{\nu}\rangle  \sim 0.6$ GeV. The search had limited sensitivity and no positive evidence of NC $1 \gamma$ was observed by T2K. 
If all NC $1\gamma$ are assumed to be solely from NC $\Delta$ radiative decay, T2K's result leads to an upper limit on the NC $\Delta$ radiative decay rate at $\sim100$ times the SM predicted rate at 90\% confidence level (CL). Thus in order to test whether there was a factor of 3.18 enhancement of the $\Delta$ radiative decay process with respect to the SM prediction sufficient to explain the MiniBooNE LEE, a highly sensitive research for this process was warranted.   

MicroBooNE tested the hypothesis of a factor of 3.18 enhancement of the $\Delta$ radiative decay rate using the first three years of data (referred to as ``Runs 1-3'') through four exclusive, independent selections. First, in order to yield a pure and efficient selection of the NC $\Delta$ radiative decay events, two selections targeting exclusive topologies were developed to select events that have the same topologies as and similar kinematics with the NC $\Delta$ radiative decay event in MicroBooNE detector. Second, the NC 1$\pi^0$ process was anticipated to be a dominant background for the NC $\Delta$ radiative decay search. Thus, in order to ensure the correct modeling of this important background, two mutually exclusive, high-statistic NC 1$\pi^0$ selections were developed. Moreover, due to the high correlation of the NC 1$\pi^0$ selection with both the NC 1$\pi^0$ background in the NC $\Delta$ single photon selections and the NC $\Delta$ radiative decay signal, the high-statistic NC 1$\pi^0$ selections can constrain the background prediction and systematic uncertainty in the NC $\Delta$ single photon selections. This was achieved through a combined fit performed with four selections simultaneously and was critical in yielding a sensitive search of NC $\Delta$ single photon events. The result came out in 2021 and was published in Physical Review Letters with editors' suggestion~\cite{uboone_delta_LEE}. My direct contributions to this analysis are the rate validation of NC $\pi^0$ background, and the exploration of methods and tool development for the final combined fit, which extracts an ultimate limit on the NC $\Delta$ radiative decay process; these are discussed in Sec.~\ref{ch3:glee_2p_rate} and~\ref{ch3:glee_fit}.

\section{NC $\Delta$ Radiative Decay Targeted Selections}\label{ch3:glee_1p}
\begin{figure}[h!]
\begin{subfigure}{0.49\textwidth}
   \includegraphics[width = \textwidth, trim=0.3cm 4.8cm 12.9cm 1.5cm, clip]{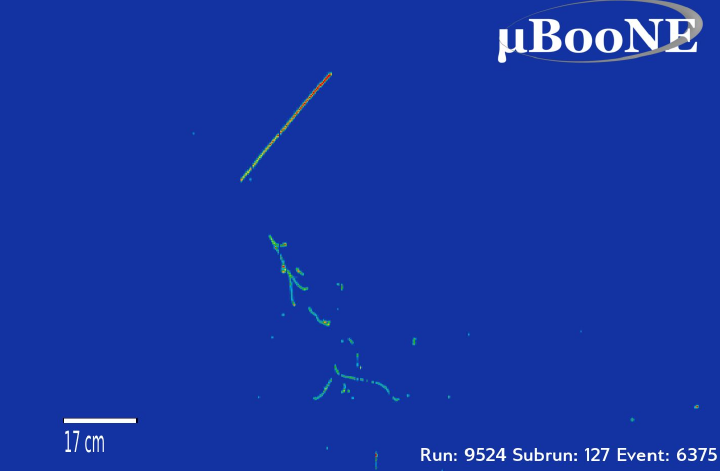}
  \caption{Candidate data event in $1\gamma 1p$ selection}
  \label{fig:delta_1g1p_events}
  \end{subfigure}
  \begin{subfigure}{0.49\textwidth}
   \includegraphics[width = \textwidth, trim=0.1cm 4.5cm 12.8cm 1.5cm, clip]{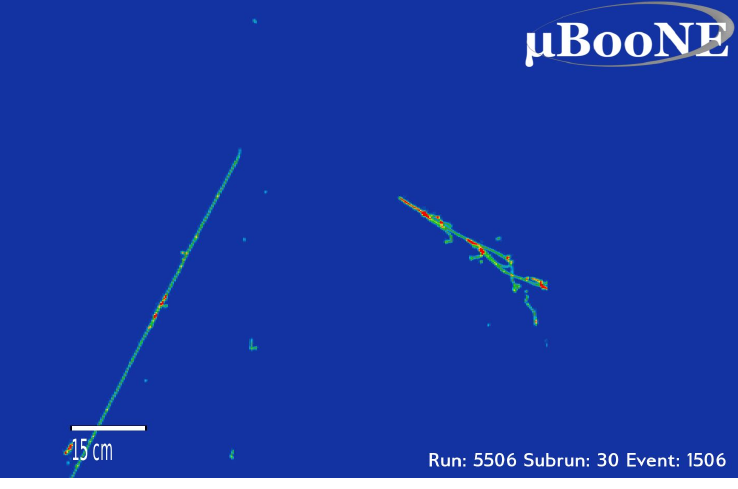}
  \caption{Candidate data event in $1\gamma 0p$ selection}
  \label{fig:delta_1g0p_events}
  \end{subfigure}
  \caption[Event displays of selected NC $\Delta$ radiative decay candidate events in $1\gamma 1p$ and $1\gamma 0p$ selections.]{Event displays of selected candidate NC $\Delta$ radiative decay event in $1\gamma 1p$ selection (left) and $1\gamma 0p$ selection (right). The horizontal axis represents wires on a plane while the vertical axis represents the time of the induced wire signal. Colored pixel represents the deposited energy in each hit; the red pixels indicate higher energy deposition than the green pixels. In the selected $1\gamma 1p$ event, we can see clean proton and photon candidates: the reconstructed track shows calorimetry in good agreement with a proton-like Bragg peak, and the shower is displaced from the track start. In the selected $1\gamma 0p$ event, we see a single shower, and a track nearby. The track has an energy deposition profile of a MIP, which is correctly identified to be cosmic-ray and \textit{not} associated with the reconstructed neutrino interaction.}
  \label{fig:delta_events}
\end{figure}

NC $\Delta \rightarrow N+\gamma$ creates two different topologies in the MicroBooNE detector, depending on the type of the nucleon that excites: 
\begin{enumerate}
    \item If nucleon is proton, the corresponding $\Delta$ resonance decay is 
    $\Delta^{+} \rightarrow p + \gamma$. Since the proton is a charged particle, it will ionize the argon atoms along its path, leading to an identifiable ``track'' in the detector. The photon from the decay will cascade and create ``shower'' some distance away from the proton. An example of such topology is shown in Fig.~\ref{fig:delta_1g1p_events}.
    \item If nucleon is neutron, the corresponding $\Delta$ resonance decay is 
    $\Delta^{0} \rightarrow n + \gamma$. Being a charge-neutral particle, neutrons will not ionize argon atoms and will be invisible in the detector. So in this case, only one ``shower'' from the decay photon is expected in the detector, with no other associated vertex acitivities. An example of such topology is shown in Fig.~\ref{fig:delta_1g0p_events}.
\end{enumerate}

Thus, two selections targeting different and exclusive topologies are developed: $1\gamma 1p$ ($1\gamma 0p$), selecting events with a single shower and exactly one (zero) protons exiting the nucleus with no charged leptons in the final states. Note that particles from the interaction might experience final state interactions on their way out of the nucleus. Thus, the actual particles exiting the nucleus might be different from daughter particles from the decay. Hence the $1\gamma 1p$ and $1\gamma 0p$ selections will end up with events from both decay modes.

With the specific topologies in mind, the 1$\gamma$ selections require events to have exactly one reconstructed shower and one (or zero) reconstructed tracks, depending on whether it is the $1p$ or $0p$ selection. Preselection quality cuts are subsequently applied to the reconstructed objects, to reject obvious muon background and ensure good reconstruction of the shower, and that the events are well contained within the TPC active volume. Machine learning methods, specifically boosted decision trees (BDTs) using the XGBoost library \cite{xgboost}, are then used and optimized to separate the NC $\Delta$ radiative decay signals from various neutrino-induced and cosmic backgrounds. There are five BDTs developed for the selection:

\begin{itemize}
    \item Cosmic BDT rejects cosmogenic backgrounds and is trained on cosmic ray data events collected when no neutrino beam was present. Track calorimetry is used to reject cosmic muons, with track and shower directionality-based variables proving powerful discriminators.
    \item NC $\pi^0$ BDT compares the relationship of the reconstructed shower and track to those expected from $\pi^{0}$ decay kinematics to separate true single-photon events from $\pi^{0}$ background where a second photon from $\pi^{0}$ decay is not reconstructed.
    \item CC $\nu_e$ BDT targets the intrinsic $\nu_e$ background events, which leverages the power of the photon conversion distance and shower calorimetry variables.
    \item A dedicated ``second-shower'' BDT vetos events in which a second shower from a $\pi^0$ decay deposits some charge in the detector, but fails 3D shower reconstruction. Such events can result in 2D charge hits near the neutrino interaction that are not associated with a 3D object. A plane-by-plane clustering algorithm,  DBSCAN \cite{dbscan}, is used to group these unassociated hits, and properties including direction, shape and energy of the cluster are used to determine consistency with a second shower from a $\pi^{0}$ decay. The clustering procedure for the second shower clusters is described in more detail in Sec.~\ref{ch5:hit_clustering}.
    \item CC $\nu_\mu$-focused BDT removes any remaining backgrounds, primarily targeting the muon track through track calorimetry variables.  
\end{itemize}
The $1\gamma0p$ sample is limited to shower-only variables due to the absence of a track. As such, it uses variations of the cosmic and NC$\pi^0$ BDTs, and a third BDT merging the functionality of the CC $\nu_e$ and CC $\nu_\mu$-focused BDTs, targeting all remaining backgrounds. The BDTs for the $1\gamma1p$ and $1\gamma0p$ selections are trained and optimized independently for each selection towards the highest statistical significance of the NC $\Delta\rightarrow N\gamma$ signal over background in each sample. The topological, pre-selection, BDT selection, and combined signal efficiencies are summarized in Table~\ref{tab:efficiencies}. 
\begin{table}[t!]
    \centering
    \begin{tabular}{|l|r|r|}
    \hline
        \multicolumn{3}{|c|}{Selection Efficiencies of NC $\Delta$ $1\gamma1p$ and $1\gamma0p$ selections}\\\hline\hline
        Selection Stage &  $1\gamma1p$ eff. & $1\gamma0p$ eff. \\ \hline
        Topological & 19.4\% & 13.5\%  \\
        Pre-selection & 63.9\% & 98.4 \% \\
        BDT Selection & 32.1\% & 39.8\%  \\ \hline
        Combined & 3.99\% & 5.29\%  \\ \hline
    \end{tabular}
    \caption[Signal efficiencies for the $1\gamma1p$ and $1\gamma0p$ selections.]{Signal efficiencies for the $1\gamma1p$ and $1\gamma0p$ selections. The topological and combined efficiencies  are evaluated relative to all true NC $\Delta\rightarrow N\gamma$ events inside the active TPC in the simulation (124.1 events expected for $6.80 \times 10^{20}$ POT). The pre-selection and BDT selection efficiencies are evaluated relative to their respective preceding selection stage.Taken from Tab.~I in Ref.~\cite{uboone_delta_LEE}.}
    \label{tab:efficiencies}
\end{table}

\begin{table}[t!]
    \centering
    \begin{tabular}{|l|c|c|}
    \hline 
    & $1\gamma1p$ & $1\gamma0p$ \\ \hline\hline
    Unconstr. bkgd. & 27.0 $ \pm $ 8.1    & 165.4 $ \pm $ 31.7 \\
    Constr. bkgd. & 20.5 $ \pm $ 3.6   & 145.1 $ \pm $ 13.8\\
    \hline
    NC $\Delta\rightarrow N\gamma$ & 4.88 & 6.55 \\
    LEE ($x_\text{MB}=3.18$) & 15.5 & 20.1 \\
    \hline 
    Data & 16 &153 \\
    \hline
    \end{tabular}
    \caption[Number of predicted signal, background, and observed data events for the $1\gamma1p$ and $1\gamma0p$ selections in NC $\Delta$ radiative decay search.]{Number of predicted background, predicted signal, and observed data events for the $1\gamma1p$ and $1\gamma0p$ samples, with background systematic uncertainties. ``Unconstr. bkgd.'' represents the nominal MC prediction for the backgrounds, while ``Constr. bkgd.'' represents the prediction for background after constraint from the NC $\pi^0$ selections is applied. The constraint is discussed later in Sec.~\ref{ch3:glee_fit}. The table is taken from Tab.~IV in Ref.~\cite{uboone_delta_LEE}.}
    \label{tab:delta_results}
\end{table}

For MicroBooNE's Runs 1-3 data, with an exposure equivalent to $6.8\times 10^{20}$ POT, a total of 16 data events are selected in the $1\gamma 1p$ channel and 153 in the $1\gamma 0p$ channel~\cite{uboone_delta_LEE}, compared to an expected constrained\footnote{Constraint is discussed later in Sec.~\ref{ch3:glee_fit}.} background of $20.5 \pm 3.6 \text{(sys.)} \pm 4.5 \text{(stat.)}$ events in the $1\gamma1p$ sample and of $145.1  \pm 13.8 \text{(sys.)} \pm 12.0 \text{(stat.)}$ events in the $1\gamma0p$ sample. The expected signal, overall background predictions, and observed data in both $1\gamma$ selections are summarized in Tab.~\ref{tab:delta_results}.
The small number of events selected arises from the fact that strict cuts on the BDTs are required to improve the signal-to-background ratio due to the sheer amount of (cosmic-ray) background present and the difficulty of cleanly separating signals from background events thanks to the overlapping kinematic phase space. 

\begin{figure}[h!]
    \centering
    \includegraphics[page=6, trim = 1cm 20.2cm 11cm 1.6cm, clip=true, width=0.49\textwidth]{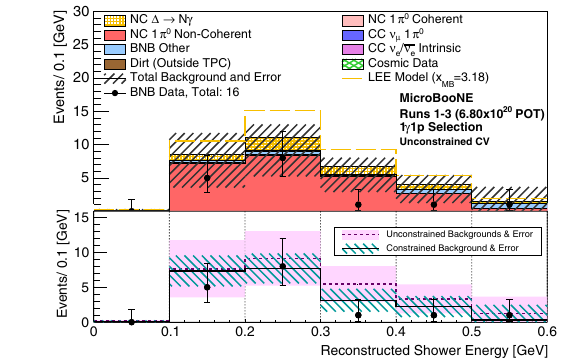}
     \includegraphics[page=6, trim = 1cm 13.4cm 11cm 8.5cm, clip=true, width=0.49\textwidth]{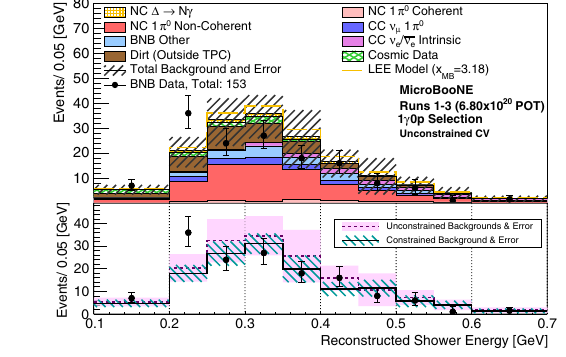}
    \caption[Reconstructed shower energy distributions of the final selected samples in MicroBooNE NC $\Delta$ radiative decay search.]{Distribution of reconstructed shower energy of the final selected sample in $1\gamma1p$ (left) and $1\gamma0p$ (right) analysis. The upper panel of both plots show the observed data overlaid on top of the MC prediction before any constraint. Prediction of only backgrounds is shown as the stacked histogram beneath the predicted signal in yellow; central value spectrum including the background and nominal NC $\Delta$ radiative decay prediction is shown as the stack of all colored histograms; overall prediction with enhanced NC $\Delta$ radiative decay rate at a factor of 3.18 is highlighted in yellow dashed line. The error band represents the systematic uncertainty evaluated on the predicted background only. The bottom panel shows the comparison between data and overall predictions before (dashed line) and after (solid line) the constraint from $2\gamma$ selection is applied. Taken from Fig. 2 in Ref.~\cite{uboone_delta_LEE}}
    \label{fig:delta_1g_final}
\end{figure}

The distribution of the final selected events in the $1\gamma 1p$ and $1\gamma 0p$ samples with MicroBooNE Runs 1-3 data are shown in Fig.~\ref{fig:delta_1g_final}. Compared to the $0p$ channel, the $1p$ selection leverages the presence of the reconstructed track to aid in the rejection of CC background and yield higher purity and less diverse backgrounds. In both $1\gamma 1p$ and $1\gamma 0p$ selections, the observed data agree better with the central value prediction than the prediction with enhanced NC $\Delta$ radiative decay rate. 

\section{NC 1$\pi^0$ Targeted Selection}\label{ch3:glee_2p}
As shown in Fig.~\ref{fig:delta_1g_final}, NC $\pi^0$ by far dominates the background in the $1\gamma$ selections. 
This occurs when one of the decay photons from $\pi^0$ is missed by the reconstruction either due to the low photon energy below the detection threshold, or because the photon pairproduces outside the detector active volume. For BNB with $\langle E_{\nu}\rangle  \sim 0.8$ GeV, NC $\pi^0$ events are mainly from the NC $\Delta$ resonance production followed by $\Delta \rightarrow N+ \pi^0$ decay. The $\Delta$ resonance decay to $\pi^0$ has a branching ratio of 99.4\%, compared to the $\Delta$ radiative decay branching ratio of $0.55\sim0.65$\% ~\cite{particle_phys_review}. Thus NC $\pi^0$ events are much more abundant than NC $\Delta$ radiative decay events in MicroBooNE. Moreover, a high correlation exists between $\pi^0$s and single photons originating from the same type of parent resonances. 
Therefore acquiring a high-statistic NC $\pi^0$ sample not only allows us to validate the modeling of this key background in $1\gamma$ selections, but also provides us a constraining sample to help constrain the systematic uncertainty in the $1\gamma$ channels through correlations. 

In order to get a pure, high-statistic sample of NC $\pi^0$ events, exactly two reconstructed showers are required. In a similar vein as the $1\gamma$ analyses, exclusive $2\gamma 1p$ and $2\gamma 0p$ selections are developed to select NC 1$\pi^0$ events with one or zero protons exiting the nucleus. The $2\gamma$ selections follow the same selection scheme (topological, preselection quality cut, followed by BDT selection) as the $1\gamma$ selections, and yield an NC 1$\pi^0$ purity of 63.4\% and 59.6\% in the $2\gamma1p$ and $2\gamma0p$ selections respectively.

\begin{figure}[h!]
    \centering
    \includegraphics[page=4, trim = 11cm 11.5cm 1cm 10cm, clip=true, width=0.49\textwidth]{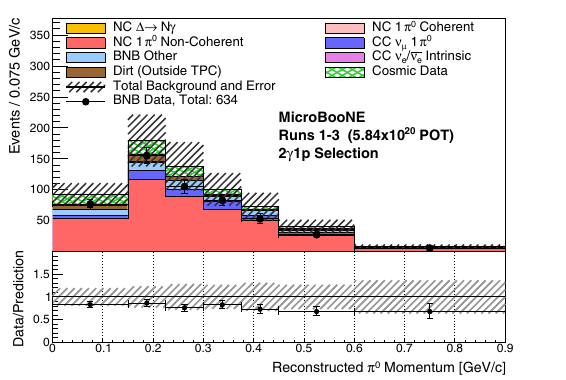}
     \includegraphics[page=4, trim = 11cm 4.5cm 1cm 17cm, clip=true, width=0.49\textwidth]{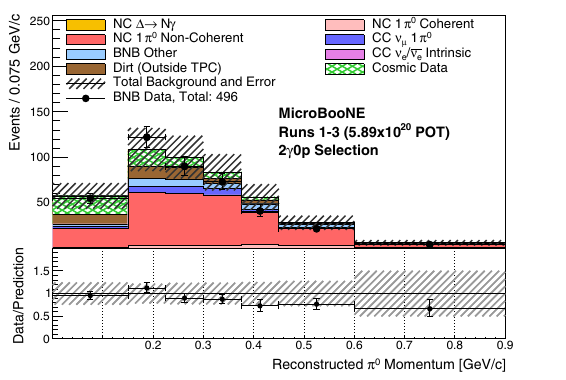}
    \caption[The reconstructed $\pi^0$ momentum distributions of the final selected samples in MicroBooNE NC !$\pi^0$ measurements.]{Distribution of reconstructed $\pi^0$ momentum of the final selected sample in $2\gamma1p$ (left) and $2\gamma0p$ (right) analysis. The upper panel of both plots shows the observed data compared with the central value prediction; the bottom panel shows the data-MC simulation ratio in each bin. The error band shows the size of systematic uncertainties. Taken from Fig. 1 in Ref. \cite{uboone_delta_LEE}}
    \label{fig:delta_2g_final}
\end{figure}

The distributions of the selected $2\gamma 1p$ and $2\gamma 0p$ sample with MicroBooNE's Runs 1-3 data are shown in Fig.~\ref{fig:delta_2g_final}, as a function of the reconstructed $\pi^0$ momentum~\cite{gLEE_ncpi0}. Slight deficits in data are observed in both selections, with the data-to-MC simulation ratio of $0.80 \pm 0.22$ (stat $\oplus$ sys) in $2\gamma1p$ and $0.91\pm 0.19$ (stat $\oplus$ sys) in $2\gamma0p$, suggesting a correction to NC $\pi^0$ background in the $1\gamma$ selections may be warranted. 

\section{NC 1$\pi^0$ Rate Validation}\label{ch3:glee_2p_rate}
The high-statistic NC 1$\pi^0$ samples is used to validate the observed overall rate of this process as currently modeled with \textsc{genie}, which is of critical importance to the NC $\Delta$ radiative decay search. Assuming \textsc{genie} provides a sufficient description of the observed data, we can use NC 1$\pi^0$ samples to provide an \textit{in situ} constraint on NC $1\pi^0$ mis-identified backgrounds for NC $\Delta$ radiative decay search. Alternatively, these measurements can be used to increase our understanding of our current NC $\pi^0$ modelling and potentially motivate \textsc{genie} tuning.

As shown in Fig.~\ref{fig:delta_2g_final}, both the $2\gamma1p$ and the $2\gamma0p$ selections see an overall deficit in data relative to the MC prediction. The deficit is more pronounced in the $2\gamma1p$ selection with an indication of dependence on the $\pi^0$ momentum. As it is also known that the \textsc{genie} branching fraction of coherent NC $1\pi^{0}$ production on argon is significantly lower than expectation extrapolated from MiniBooNE's $\pi^{0}$ measurement on mineral oil~\cite{MiniBooNE:2008mmr}, the possibility of a correction to \textsc{genie} predictions on both non-coherent and coherent NC $1\pi^0$ production is explicitly examined. The MC predictions are fitted to data allowing both coherent and non-coherent NC $1\pi^{0}$ rates to vary. Both normalization-only and normalization plus shape variations to the coherent and non-coherent rates are explored; all yield similar conclusions. Here, I focus on the normalization plus shape variation fit.  

The normalization plus shape variation fit is performed as a function of reconstructed $\pi^0$ momentum for both 2$\gamma1p$ and $2\gamma0p$ selections, using [0, 0.075, 0.15, 0.225, 0.3, 0.375, 0.45, 0.525, 0.6, 0.675, and 0.9] GeV/c bin limits. In the fit, the predicted NC coherent $1\pi^0$ events are scaled by a normalization factor $N_{coh}$, and predicted NC non-coherent $1\pi^0$ events are scaled on an event-by-event basis depending on their corresponding true $\pi^0$ momentum according to formula $(a+b|\vec{p}^{~true}_{\pi^{0}}|)$, where the true $\pi^0$ momentum is given in [GeV/c]. This linear scaling as a function of $\pi^0$ momentum was chosen because it was the simplest implementation that was consistent with the observed data-to-MC deficit, as observed in Fig.~\ref{fig:delta_2g_final}.

At each set of fitting parameters ($N_{coh}$, $a$, $b$), the consistency between the scaled prediction for this parameter set and the observed data is evaluated by the Combined-Neyman-Pearson (CNP) $\chi^2$~\cite{CNP}:
\begin{align}\label{ch3:cnp_chi}
    \chi^2 & = \sum_{i,j}(P^{\text{scaled}}_{i} - D_{i})(M_{\text{stat}}+M_{\text{syst}})^{-1}_{i,j}(P^{\text{scaled}}_{j} - D_{j}) \\
    M_{\text{stat}\; i,j} & = \frac{3}{1/D_{i}+2/MC_{i}}\delta_{ij}
\end{align}
where $D$ is the observed data spectrum; $P^{\text{scaled}}$ is the MC spectrum scaled according to the set of scaling factors ($N_{coh}$, $a$, $b$); $M_{\text{stat}}$ is a diagonal matrix with CNP statistical uncertainty terms; and $M_{\text{syst}}$ is the covariance matrix incorporating systematic uncertainties and correlations. Flux, cross section, detector and hadronic reinteraction systematic uncertainties are included in the fit including bin-to-bin systematic correlations. As the goal of the fit is to extract the normalization and scaling parameters of the coherent and non-coherent NC $1\pi^{0}$ rates, the cross-section normalization uncertainties of NC coherent and non-coherent $1\pi^{0}$ are not included\footnote{Note that the cross-section normalization uncertainties of coherent and non-coherent NC $1\pi^{0}$ are only removed for the purposes of this fit and not for the constraint discussed in the next section.}. The fit is performed in an iterative approach, where the $M_{\text{stat}}$ and $M_{\text{syst}}$ matrices are evaluated with the MC prediction at the best-fit parameters from previous interaction and kept fixed during each iteration, and stops until the difference in $\chi^2$ between two iterations converge.

The data-extracted best-fit parameters correspond to $a=0.98$ and $b=-1.0$~[c/GeV] for the scaling parameters of the NC non-coherent $1\pi^0$ events, and $N_{coh}=2.6$ for the NC coherent 1$\pi^0$ normalization factor with no enhancement, $N_{coh}=1$, being allowed within the $1\sigma$ error bands. This best-fit gives a $\chi^2$ per degree of freedom ($dof$) of 8.46/17. The $\chi^2/dof$ at the \textsc{genie} central value (CV) prediction is 13.74/20 yielding a $\Delta\chi^2$ between the \textsc{genie} CV and the best-fit point of 5.28 for 3 $dof$. Although the goodness-of-fit $\chi^2/dof$ values for both scenarios are acceptable due to the generally large uncertainties, the momentum-dependent shift is preferred over the \textsc{genie} CV at the 1.43$\sigma$ level. 

\begin{figure}[h!]
\centering
  \begin{subfigure}{0.7\textwidth}
  \includegraphics[width = \textwidth]{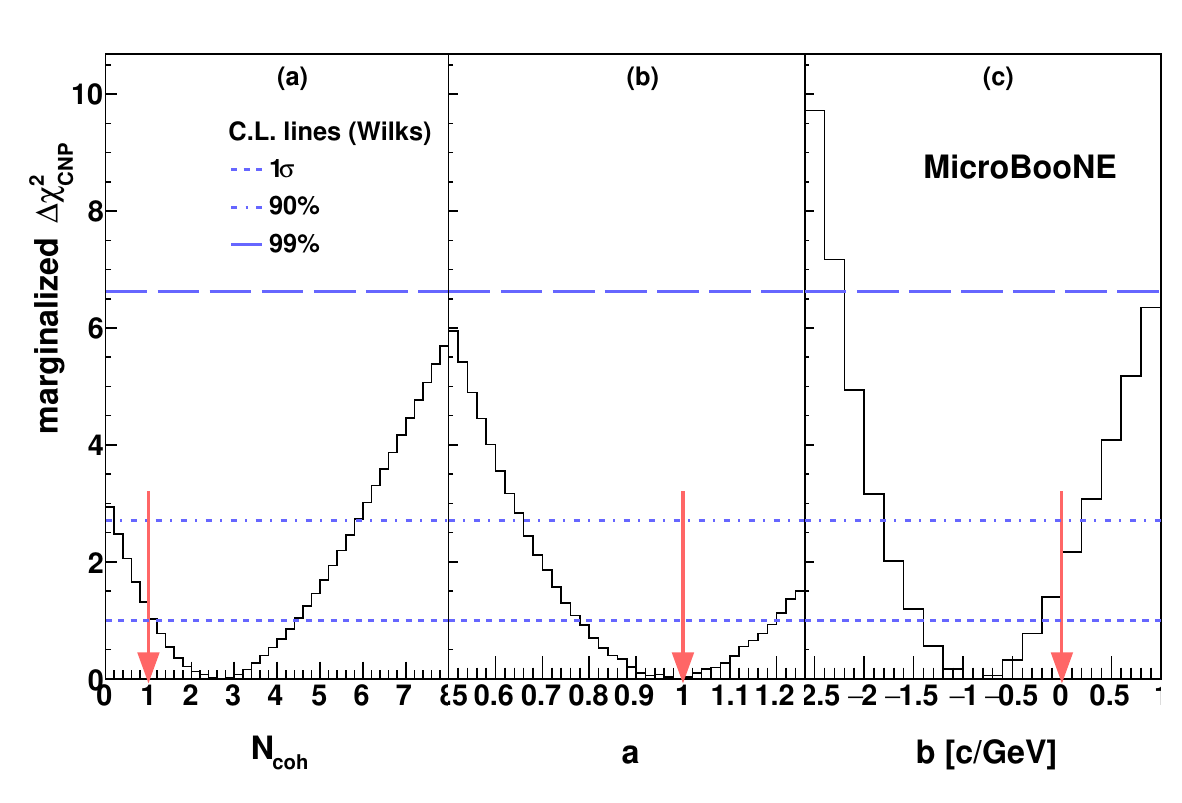}
  \end{subfigure}
\caption[The distribution of marginalized $\Delta\chi^2$ from the NC 1$\pi^0$ rate fit.]{The distribution of marginalized $\Delta\chi^2$, as a function of flat normalization factor for (a) coherent NC 1$\pi^0$ momentum-independent scaling factor, (b) non-coherent NC 1$\pi^0$ momentum-independent scaling factor, and (c) coefficient of momentum-dependent scaling factor for NC non-coherent 1$\pi^0$, marginalized over the other two parameters. The red arrows indicate parameter values expected for the \textsc{genie} central value prediction. The 1$\sigma$, 90\% and 99\% C.L.~lines are based on the assumption that the distribution follows a $\chi^2$ distribution with 1 degree of freedom. Taken from Fig.~10 in Ref.~\cite{gLEE_ncpi0}.
}
\label{fig:marginalized_chi}
\end{figure}

\begin{figure}[h!]
  \begin{subfigure}{0.49\textwidth}
   \includegraphics[width = \textwidth]{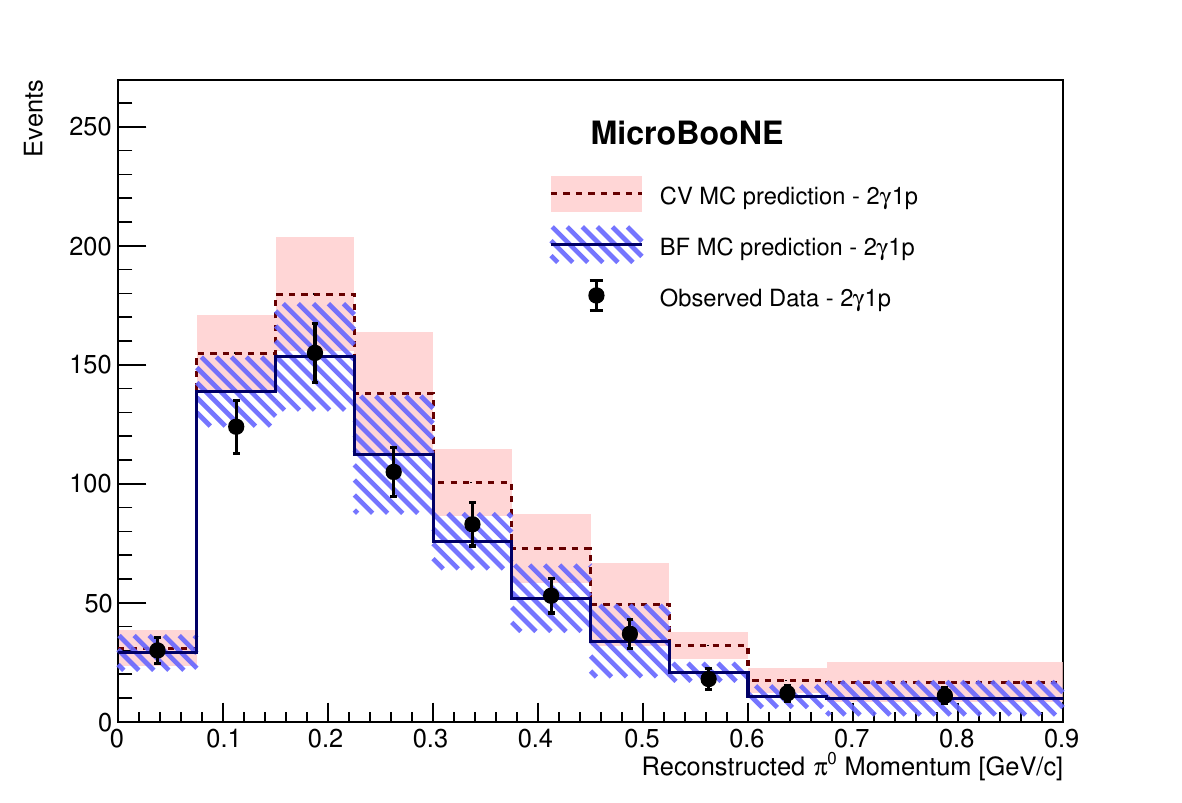}
  \caption{$2\gamma1p$}
  \end{subfigure} 
  \begin{subfigure}{0.49\textwidth}
   \includegraphics[width = \textwidth]{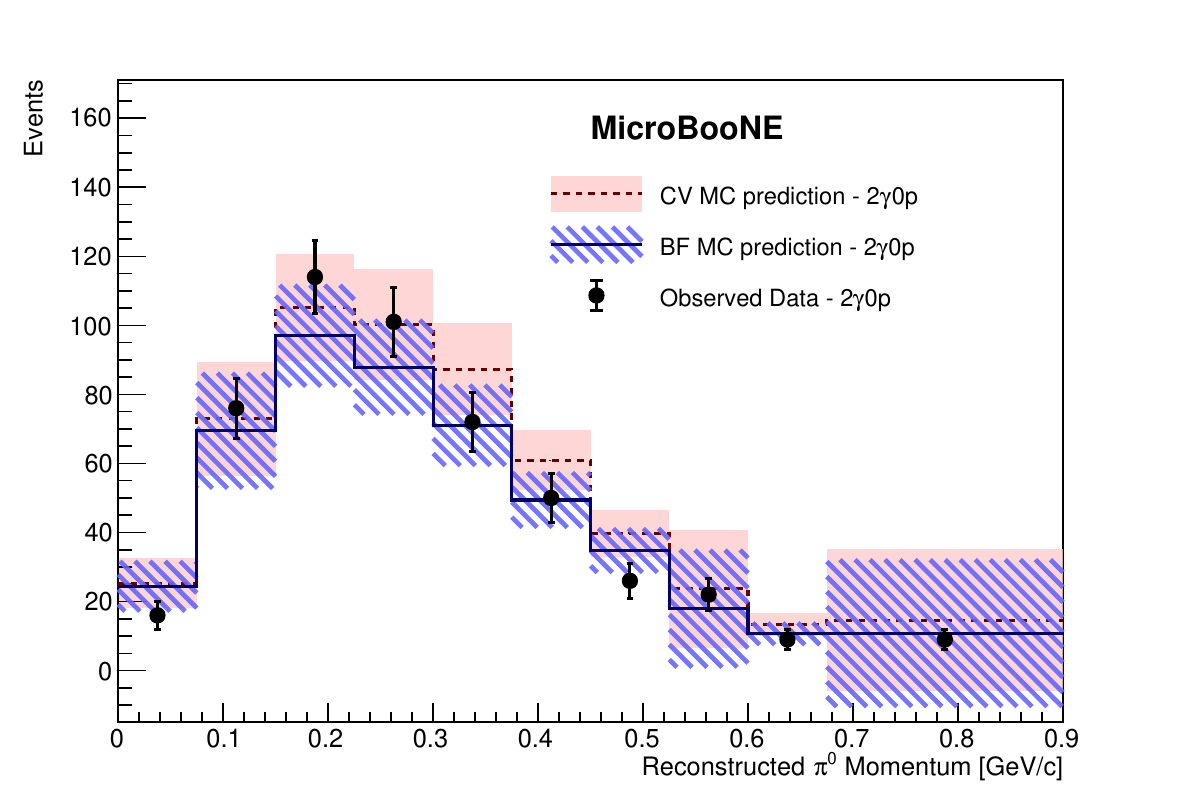}
  \caption{$2\gamma0p$}
  \end{subfigure}
  \caption[The data-MC comparison for $2\gamma1p$ and $2\gamma0p$ selections at CV and the best-fit parameters.]{The data-MC comparison for (a) $2\gamma1p$ and  (b) $2\gamma0p$ selections, as a function of reconstructed $\pi^0$ momentum. Monte Carlo predictions at the central value and at the best-fit point ($\text{N}_{coh} = 2.6$, $a=0.98$, $b=-1.0$~[c/GeV]) are both shown, with prediction and corresponding systematic error evaluated at the \textsc{genie} central value in salmon, and at the best-fit in blue. Note that the systematic uncertainties on the plot include MC intrinsic statistical error and all the systematic errors (flux, cross-section and detector), with the exception of cross section normalization uncertainties on coherent and non-coherent NC 1$\pi^0$. Taken from Fig.~11 in Ref.~\cite{gLEE_ncpi0}.}
  \label{fig:momentum_BF}
\end{figure}
The 1D marginalized $\Delta\chi^2$ distributions in Fig.~\ref{fig:marginalized_chi} also confirm that the \textsc{genie} CV prediction agrees with data within uncertainty. The data and MC comparisons of the reconstructed $\pi^0$ momentum distributions scaled to the best-fit parameters are provided in Fig.~\ref{fig:momentum_BF} and, compared to those corresponding to the \textsc{genie} CV, show better agreement with data after the fit.

While the data suggest that \textsc{genie} may over-estimate NC $1\pi^0$ production, the results demonstrate that the \textsc{genie} prediction of NC $1\pi^{0}$s is accurate within uncertainty. This validates the approach of using the measured NC $1\pi^{0}$ event rate as a powerful \textit{in situ} constraint of \textsc{genie}-predicted NC $1\pi^0$ backgrounds in the NC $\Delta$ radiative decay search.  It should be stressed that while this result motivates a momentum dependent shift in how NC $\pi^0$ events are modeled, this change is not applied to our modeling, relying instead on the fact that the assigned uncertainty covers the observed discrepancy.

\section{NC 1$\pi^0$ Constraint and Fit Results}\label{ch3:glee_fit}
\begin{figure}[h!]
    \centering
    \includegraphics[trim = 11cm 12.5cm 1cm 10cm, clip=true, width=0.7\textwidth]{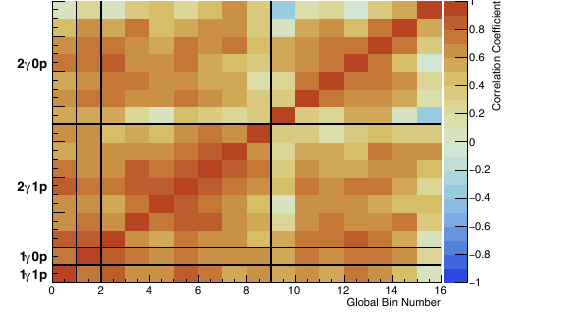}
    \caption[The systematic correlation between the final selections of all four channels in MicroBooNE NC $\Delta$ radiative decay search.]{The systematic correlation between the final selected $1\gamma1p$, $1\gamma0p$, $2\gamma1p$ and $2\gamma0p$ samples. The $1\gamma$ selections are shown in single bins due to low statistics, while the $2\gamma$ selections are shown in the same binning as in Fig.~\ref{fig:delta_2g_final}. The binning shown here is the same as the ones adopted in the fit to extract the normalization scaling of NC $\Delta$ radiative decay branching ratio. Taken from Fig. 2 in the supplementary material of Ref.~\cite{uboone_delta_LEE}.}
    \label{fig:1g_2g_corr}
\end{figure}

As mentioned in the previous section, the selected $2\gamma$ samples are expected to be highly correlated with the $1\gamma$ samples. Figure~\ref{fig:1g_2g_corr} highlights the strong correlation between the final selected $1\gamma1p$, $1\gamma0p$, $2\gamma1p$ and $2\gamma0p$ samples. Through the correlation, the information of the observed data deficit in $2\gamma$ selection can be translated to the $1\gamma$ selections through a fit involving a full systematic covariance matrix.

To illustrate the power of the NC 1$\pi^0$ constraint on the 1$\gamma$, the conditional constraint procedure~\cite{conditional_constraint} is performed. The bin-to-bin covariance matrix for $1\gamma$ and $2\gamma$ selections is shown in Eq.~\ref{eq:1g_2g_matrix}. It can be divided into block matrices: correlation matrix within $1\gamma$ selections $\Sigma^{1\gamma}$, correlation matrix within $2\gamma$ selections $\Sigma^{2\gamma}$, and the covariance matrix between $1\gamma$ and $2\gamma$ selections $\Sigma^{1\gamma, 2\gamma}$ and $\Sigma^{2\gamma, 1\gamma}$. 
\begin{equation}\label{eq:1g_2g_matrix}
    \Sigma = \begin{pmatrix}
\Sigma^{1\gamma} & \Sigma^{1\gamma, 2\gamma} \\
\Sigma^{2\gamma, 1\gamma} & \Sigma^{2\gamma}
\end{pmatrix}
\end{equation}
Given the bin-to-bin covariance matrix Eq.~\ref{eq:1g_2g_matrix}, the predicted spectrum for $1\gamma$ selection $P^{1\gamma}$, the prediction for $2\gamma$ selection $P^{2\gamma}$, and the observation for $2\gamma$ selection $O^{2\gamma}$, the constrained prediction for the $1\gamma$ selections can be calculated via formula:
\begin{equation}\label{}
    P^{1\gamma}_{constrained} = P^{1\gamma} + \Sigma^{1\gamma, 2\gamma}(\Sigma^{2\gamma})^{-1}(O^{2\gamma} - P^{2\gamma})
\end{equation}
and the constrained systematic covariance matrix for the $1\gamma$ selections follows:
\begin{equation}
    \Sigma^{1\gamma, 1\gamma}_{constrained} = \Sigma^{1\gamma} - \Sigma^{1\gamma, 2\gamma}(\Sigma^{2\gamma})^{-1}\Sigma^{2\gamma, 1\gamma}
\end{equation}

The result of the constraint is shown in the bottom panels of Fig.~\ref{fig:delta_1g_final}, where the predicted spectra of the $1\gamma$ selections before and after the $2\gamma$ constraint are shown and compared to the observed data. The impact of the $2\gamma$ constraint on the overall normalizations of $1\gamma$ selections is shown in Tab.~\ref{tab:delta_results}. The constraint from $2\gamma$ channels reduces the total background prediction by an amount consistent with the data-MC ratio observed in the $2\gamma$ selections, and significantly suppresses the systematic uncertainties by 40\% and 50\% respectively for the $1\gamma1p$ and $1\gamma0p$ selections.

A single parameter $x_{\Delta}$, the normalization factor on the nominal rate of NC $\Delta$ radiative decay, is extracted from the observation through a fit, and a simultaneous fit strategy with all four selected samples ($1\gamma1p$, $1\gamma0p$, $2\gamma1p$ and $2\gamma0p$) is adopted. The extracted $x_{\Delta}$ is compared with the 3.18 scaling factor required to explain the MiniBooNE LEE, to provide a test for this specific photon hypothesis. Unlike the illustration above where the NC $\pi^0$ constraint is directly applied on the $1\gamma$ selection, in the fit $2\gamma$ selections implicitly constrain the systematic uncertainty and background prediction in the $1\gamma$ selections through correlation. 

The CNP $\chi^2$ test statistic in Eq.~\ref{ch3:cnp_chi} is used for the fit. The $M_{\text{syst}}$ matrix encapsulates the full systematic uncertainties except the uncertainty on the braching ratio of $\Delta \rightarrow N \gamma$ decay, since the branching ratio is closely related to the $x_{\Delta}$ parameter of interest. The fit is performed once, and the $M_{\text{stat}}$ and $M_{\text{syst}}$ matrices are reevaluated at every $x_{\Delta}$ value for $\chi^2$ calculation across the grid. During the fit, the $1\gamma1p$, $1\gamma0p$ samples each are binned in one bin due to low statistics, and the binning of $2\gamma1p$ and $2\gamma0p$ samples is the same as shown in Fig.~\ref{fig:delta_2g_final}. 
\begin{figure}[h!]
    \centering
    \includegraphics[page=7, trim = 1.8cm 20.3cm 11cm 1.5cm, clip=true, width=0.7\textwidth]{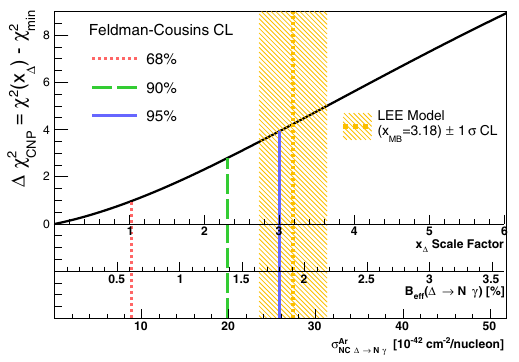}

    \caption[Resulting $\Delta \chi^2$ distribution as a function of $x_{\Delta}$ using the Feldman-Cousins procedure in MicroBooNE NC $\Delta$ radiative decay search.]{The resulting $\Delta \chi^2$ from the fit to $x_{\Delta}$ using the Feldman-Cousins procedure, showing extracted confidence intervals. The best fit is found to be at $x_{\Delta} = 0$ with a $\chi^2_{bf}  =  5.53$. Shown also is the reinterpretation of this scaling factor as both an effective branching fraction and a cross section. Taken from Fig.~4 in Ref.~\cite{uboone_delta_LEE}.}
    \label{fig:gLEE_delta_fit}
\end{figure}

The result of the fit is shown in Fig.~\ref{fig:gLEE_delta_fit} with a best-fit value at $x_{\Delta}=0$ and corresponding $\chi^2/ndof$ of 5.53/15. The confidence level of the analysis is evaluated with the Feldman-Cousins frequentist method~\cite{feldman_cousins}, and a one-sided bound is placed at $x_{\Delta} < 2.3$ with 90\% CL. Under
a two-hypothesis test, the interpretation of the MiniBooNE LEE as a factor of 3.18 enhancement to the $\Delta$ radiative decay rate is ruled out by the data while the nominal prediction is favored at 94.8\% CL~\cite{uboone_delta_LEE}.

%% file: chapter4.tex
\chapter{Neutrino-induced NC Coherent Single Photon Production in MicroBooNE}\label{ch:signal_model}

In Ch.~\ref{ch3:ncdelta_LEE}, MicroBooNE's search for NC $\Delta$ resonance production followed by radiative decay $\Delta \rightarrow N\gamma$ ($N = p, n$) is discussed, and a tight limit on the $\Delta$ radiative decay branching ratio is reported. Besides the incoherent NC $1\gamma$ emission, there are also contributions from coherent channels that are expected to be roughly ten times smaller~\cite{coh_gamma_model}. NC coherent $1\gamma$ is interesting to search for because, currently, only theoretical predictions exist; the NC coherent $1\gamma$ process has never been directly measured or constrained. An experimental search for this process could allow evaluation of its cross-section and serve as a check of the theoretical models. 

The last chapter of the thesis focuses on the search for NC coherent $1\gamma$ events in MicroBooNE with Runs 1-3 data. Given current theoretical predictions~\cite{coh_gamma_pred, Alvarez-Ruso:2018fdm}, the amount of statistics MicroBooNE collected is likely insufficient to discover this process experimentally. Nevertheless, the methodology and tools developed can be applied in future experiments with higher statistics. One example of such experiments is the SBND experiment~\cite{SBND}: it will start to take data very soon and expects to collect approximately $20\times$ more data than MicroBooNE, thus can observe this process experimentally. This chapter discusses the state-of-the-art theoretical modeling of the NC coherent $1\gamma$ process and its simulation in MicroBooNE, while the following chapter details the selection chain and results with MicroBooNE data.

\section{Theoretical Model of Neutral Current Coherent 1$\gamma$ Process}\label{ch4:coh_model}
The neutrino-induced NC coherent $1 \gamma$ process studied in this thesis concerns the scattering of $\mathcal{O}(1)$ GeV neutrinos on argon atoms that specifically leads to a weak photon production and leaves the nucleus in its ground state: 
\begin{align}
    \nu + \text{Ar}_{gs} & \rightarrow \nu + \text{Ar}_{gs} + \gamma  \\
    \overline{\nu} + \text{Ar}_{gs} & \rightarrow \overline{\nu} + \text{Ar}_{gs} + \gamma
\end{align}
The theoretical model of this process was developed by L. Alvarez-Ruso \textit{et al.}~\cite{coh_gamma_model}, following the same procedure as neutrino-induced coherent pion modeling in~\cite{coh_pion_model}. It is built upon the coherent sum of photon production amplitudes of a neutrino on single nucleons\cite{coh_gamma_model}. The nucleon wave function remains unchanged during the scattering; thus, after summing over all nucleons in the nucleus, we obtain the nuclear densities. 

\begin{figure}[h!]
    \centering
    \includegraphics[trim = 2cm 22.2cm 10.5cm 3.5cm, clip=true, width=0.8\textwidth]{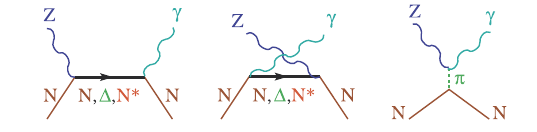}
    \caption[Feynman diagrams for the hadronic weak couplings considered in the calculation of NC photon emission off nucleon.]{Feynman diagrams for the hadronic weak couplings considered in the calculation of NC photon emission off nucleon. (Left) direct nucleon and resonance pole terms; (Middle) crossed nucleon and resonance pole terms; (Right) t-channel pion exchange term. $N^{\ast}$ represents ($N^{\ast}(1440)$, $N^{\ast}(1520)$, and $N^{\ast}(1535)$) resonances. Taken from Fig. 1 in Ref.~\cite{coh_gamma_pred}}
    \label{fig:coh_nucleon_contribution}
\end{figure}

The NC coherent photon emission off of a nucleon takes into account contributions from direct and crossed baryon pole terms with the nucleon and other resonances; this includes the resonant contributions from $\Delta (1232)$, $N^{\ast}(1440)$, $N^{\ast}(1520)$, and $N^{\ast}(1535)$, as well as non-resonant contributions from nucleon-pole terms, and contribution from t-channel pion exchange $\pi Ex$, shown in Feynman diagrams in Fig. \ref{fig:coh_nucleon_contribution}. 

Kinematic variables of all particles involved in the coherent photon emission off of a nucleon are labeled as:
\begin{align}
    \nu(k) + N(p) & \rightarrow \nu(k^{'}) + N(p^{'}) + \gamma (k_{\gamma}) \\
    \overline{\nu}(k) + N(p) & \rightarrow \overline{\nu}(k^{'}) +  N(p^{'}) + \gamma (k_{\gamma}) 
\end{align} where $\vec{k_{\gamma}}$, $\vec{k}$ and $\vec{k^{'}}$ are the outgoing photon, incoming neutrino, and outgoing neutrino momenta; $\vec{p}$ and $\vec{p^{'}}$ are the nucleon momenta before and after the scattering, all in the laboratory frame. The differential cross section derived from this model with respect to the photon kinematics in the laboratory frame is given by: 
\begin{align}
    \frac{d^2\sigma(\nu, \overline{\nu})}{dE_{\gamma}d\Omega(\hat{k_{\gamma}})} = \frac{E_{\gamma}}{|\vec{k}|}\frac{G^2}{16\pi^2}\int \frac{d^3 k^{'}}{|\vec{k^{'}}|} L^{(\nu, \overline{\nu})}_{\mu, \sigma}W^{\mu, \sigma}_{\text{NC}, \gamma}
\end{align}
where $E_{\gamma}=|\vec{k_{\gamma}}|$ is the energy of the outgoing photon, and neutrinos are assumed massless. $G= 1.1664 \times 10^{-11}$~$\text{MeV}^{-2}$ is the Fermi constant, and $L^{(\nu, \overline{\nu})}$ and $W_{\text{NC}, \gamma}$ represent the leptonic and hadronic tensors respectively. 
The leptonic tensor is
\begin{equation}
    L^{(\nu, \overline{\nu})}_{\mu, \sigma} = k^{'}_{\mu}k_{\sigma} + k^{'}_{\sigma}k_{\mu} + g_{\mu\sigma}\frac{q^2}{2}\pm i\epsilon_{\mu \sigma \alpha \beta}k^{'\alpha}k^{\beta} 
    \;\;
    (+ \rightarrow \nu, - \rightarrow \overline{\nu})
\end{equation}
where $q=k-k^{'}$ is the momentum-transfer and $q^{2}=-2k\cdot k^{'}$. The hadronic tensor includes nonleptonic vertices and is evaluated to be: 
\begin{align}
    W^{\mu, \sigma}_{\text{NC}, \gamma} &= - \frac{\delta(E_{\gamma} - q^0)}{64\pi^3 M^2}\mathcal{A}^{\mu, \rho}(q, k_{\gamma})(\mathcal{A}^{\sigma}_{.\rho}(q, k_{\gamma}))^{\ast} \\
    \mathcal{A}^{\mu, \rho}(q, k_{\gamma}) &= \int d^3re^{i(\vec{q} - \vec{k_{\gamma}})\cdot \vec{r}}\{ \rho_{p}(r)\hat{\Gamma}^{\mu, \rho}_{p}(r; q, k_{\gamma}) + \rho_{n}(r)\hat{\Gamma}^{\mu, \rho}_{n}(r; q, k_{\gamma}) \}
\end{align}
where $M$ is the mass of the nucleon; $\rho_{p(n)}(r)$ is the local density of proton (neutron) normalized to the number of protons (neutrons). $\delta(E_{\gamma} - q^0)$ term imposes $q^0 = E_{\gamma}$ for energy conservation which is justified by the large nucleus mass. Term $\hat{\Gamma}^{\mu, \rho}_{N}$ represents the nucleon helicity averaged photon production amplitudes off nucleon, where contributions from diagrams in Fig. \ref{fig:coh_nucleon_contribution} are taken into account.
\begin{align}
    \hat{\Gamma}^{\mu, \rho}_{N}(r;q,k_{\gamma}) &= \sum_{i} \hat{\Gamma}^{\mu, \rho}_{i;N}(r;q,k_{\gamma})\\
    &= \sum_{i}\frac{1}{2}\text{Tr}[(\slashed{p} + M)\gamma^{0}\Gamma^{\mu, \rho}_{i;N}]\times \frac{M}{p^{0}}\\
    i &= NP, CNP, \pi Ex, RP, CRP\;\; [R = \Delta (1232), N^{\ast}(1440), N^{\ast}(1520), N^{\ast}(1535)]
\end{align}
Here, four-vector matrices $\Gamma^{\mu, \rho}_{i; N}$ are the amputated photon production amplitude off a nucleon for different channels shown in Fig.~\ref{fig:coh_nucleon_contribution} (See Eq.~10-12 in Ref.~\cite{coh_pion_model} for details). During the evaluation, the momentum transfer is accommodated by the nucleon wave function and taken to be equally shared between the initial and final nucleon: 
\begin{align}
    p^{\mu} &= (\sqrt{M^2 + \frac{1}{4}(\vec{k_{\gamma}} - \vec{q})^2}, \frac{\vec{k_{\gamma}} - \vec{q}}{2}) \\
    p^{'\mu} &= q - k_{\gamma} + p = (\sqrt{M^2 + \frac{1}{4}(\vec{k_{\gamma}} - \vec{q})^2}, -\frac{\vec{k_{\gamma}} - \vec{q}}{2})
\end{align} which greatly simplifies the calculation. Nuclear medium corrections, including the Pauli blocking, Fermi motion, and the $\Delta$ resonance broadening in the medium, are also considered during the calculation. 

Infrared divergence arises from the nucleon-pole terms (NP and CNP)  when the energy of the outgoing photon $E_{\gamma} \rightarrow 0$. This divergence should be canceled by others in the electromagnetic radiative corrections to neutrino elastic scattering~\cite{coh_gamma_model}. Driven by the MiniBooNE detection threshold on photons of $E_{\gamma} \gtrsim 140$ MeV~\cite{miniboone_nue_appear}, an artificial cut-off is introduced within the \textsc{genie} implementation to the outgoing photon energy $E_{\gamma}$ at 140 MeV. This is a caveat for the MicroBooNE search since the MicroBooNE detector has a much lower detection threshold for photons.

\begin{figure}[h!]
    \centering
    \includegraphics[page=16, trim = 1cm 3.5cm 13.5cm 17.1cm, clip=true, width=0.5\textwidth]{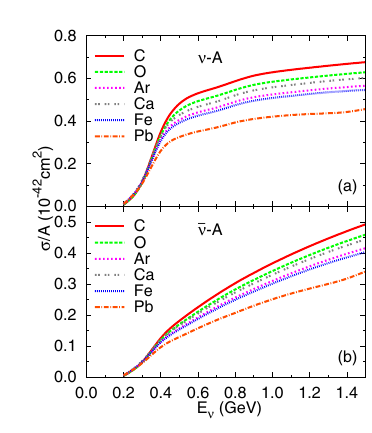}
    \caption[The predicted total cross section as a function of incoming neutrino energy for NC coherent $1\gamma$ process.]{The predicted total cross section as a function of incoming neutrino energy for neutrinos (top) and antineutrinos (bottom), on different nuclei targets. Taken from Fig. 12 in Ref.~\cite{coh_gamma_model}.}
    \label{fig:coh_total_xs}
\end{figure}

\begin{figure}[h!]
    \centering
    \includegraphics[page=16, trim = 7.5cm 3.5cm 1cm 17.1cm, clip=true, width=0.9\textwidth]{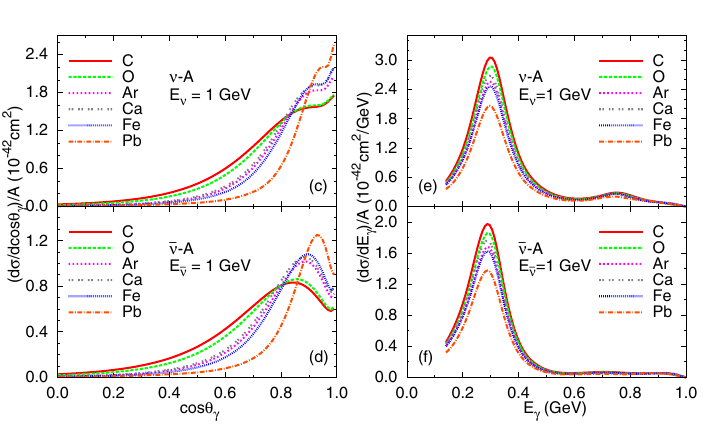}
    \caption[The predicted differential cross-section as a function of the outgoing photon angle and energy for incoming (anti)neutrino of 1 GeV for NC coherent $1\gamma$ process.]{The predicted differential cross-section as a function of the outgoing photon angle (left) and outgoing photon energy for incoming (anti)neutrino of 1 GeV on different nuclei targets. Taken from Fig. 12 (c)-(f) in Ref.~\cite{coh_gamma_model}.}
    \label{fig:coh_differential_xs}
\end{figure}

The leptonic and hadronic tensors are independent of neutrino flavors, thus the cross section is the same for the electron, muon, or tau (anti)neutrinos. The predicted total cross sections for neutrinos and antineutrinos on different nucleus targets are shown in Fig.~\ref{fig:coh_total_xs}, generally on the order of $\frac{\sigma}{A} \sim \mathcal{O}(10^{-43} \text{cm}^2)$. Figure \ref{fig:coh_differential_xs} shows the differential cross section as a function of the outgoing photon angle with respect to the incoming neutrino beam and energy for incident neutrino energy $E_{\nu} = 1$ GeV. The peak at $\sim 0.3$ GeV in the photon energy distribution is due to the dominant $\Delta$ contribution and the second small peak at $\sim 0.8$ GeV that can be observed for neutrinos corresponds to the $N^{\ast}(1520)$ resonance.

\section{Simulation of NC Coherent $1\gamma$ in MicroBooNE}

The model of NC coherent 1$\gamma$ process is not implemented in the \textsc{genie} \texttt{\detokenize{v3.0.6}} generator incorporated in MicroBooNE software. Instead it has been successfully implemented and validated~\cite{coh_gamma_validation} in an updated \textsc{genie} version~\cite{genie_v3.2}\footnote{More specifically the signal simulation used in this analysis is produced with a development branch (\textit{NCGammaFix}) of \textsc{genie} with source code at git repository: \href{https://github.com/mroda88/Generator.git}{GENIE generator}.}. Thus, differently from background events, which are simulated following the simulation chain discussed in Sec.~\ref{ch2.1:MC} with default \textsc{genie}, the NC coherent 1$\gamma$ signal events for this analysis are simulated in a standalone \textsc{genie} generator.

The \textsc{genie} generator takes as input the BNB $\nu_{\mu}$ flux profile expected in the MicroBooNE detector and simulates the production of neutrino-induced NC coherent $1\gamma$ emission on argon. Then, the relevant truth information from \textsc{genie} output is saved in plain text-based \textsc{hepevt}~\cite{hepevt} format. For computing efficiency, the simulated signal event is assumed to happen within the 1.6$\mu$s beam spill with the interaction vertex randomly scattered in the detector active volume. The \textsc{hepevt} file is then fed into MicroBooNE software for a full simulation of the final states from this process in the detector, including the particle propagation in the detector as well as the detector response discussed in Sec.~\ref{ch2.1:MC}. 

\begin{figure}[h!]
\centering
    \includegraphics[width=0.49\textwidth]{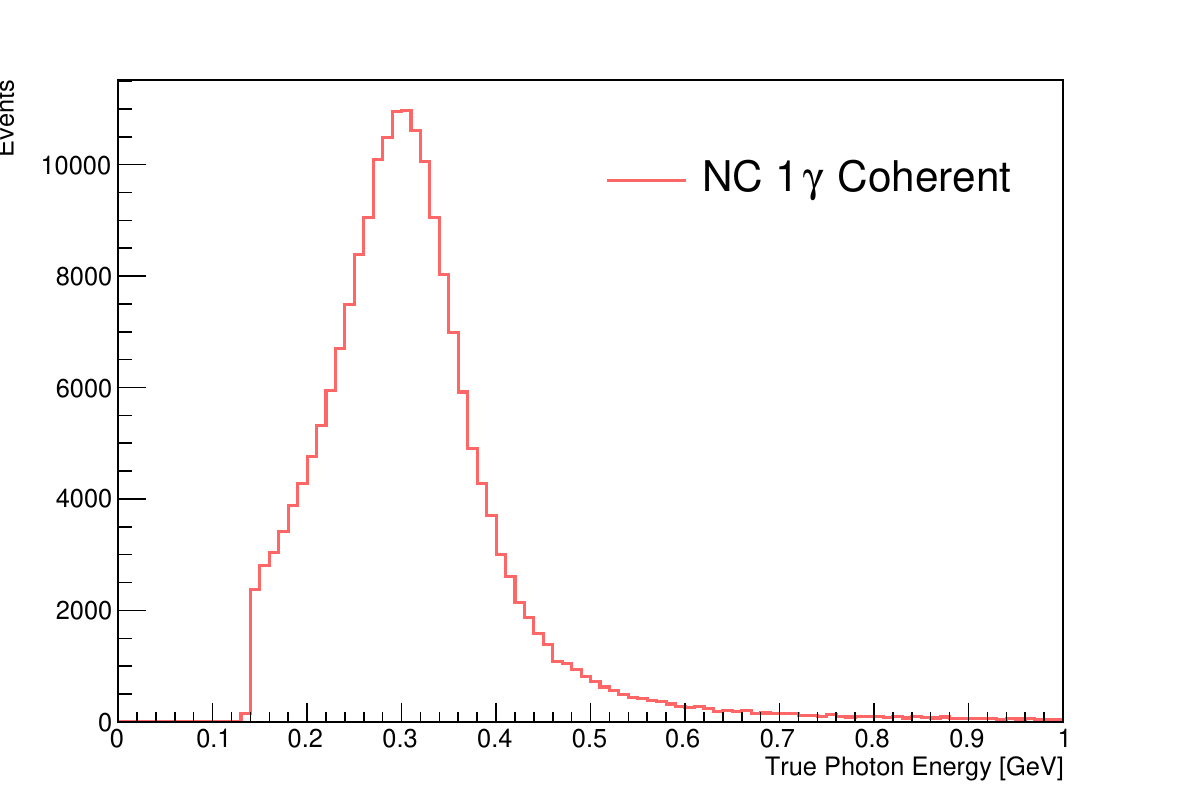}
    \includegraphics[width=0.49\textwidth]{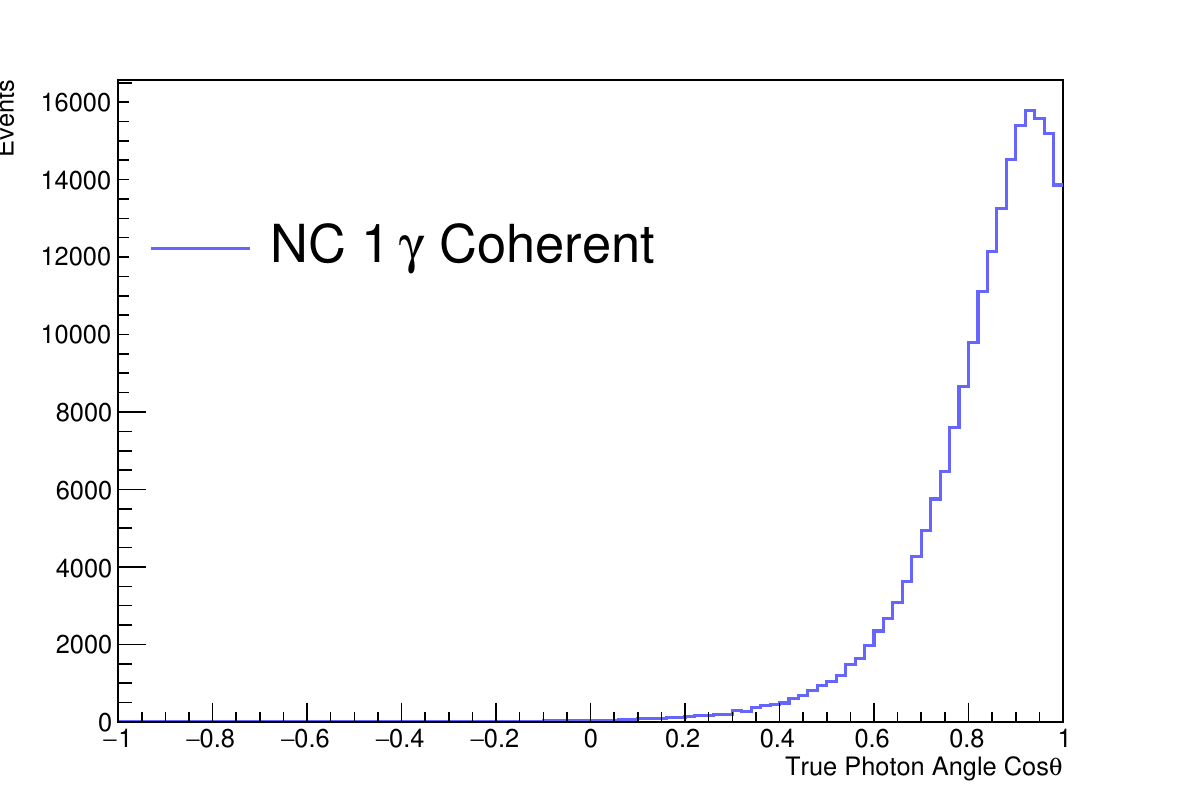}
    \caption[Distribution of true photon energy and angle with respect to the incoming neutrino beam of simulated NC coherent 1$\gamma$ events in the MicroBooNE detector.]{Distribution of true photon energy (left) and photon angle with respect to the incoming neutrino beam (right) of all simulated NC coherent 1$\gamma$ events in MicroBooNE detector. Note the 140 MeV cut-off introduced in the \textsc{genie} implementation.}
    \label{fig:coh_signal_truth}
\end{figure}

\begin{figure}[h!]
\centering
    \includegraphics[width=0.6\textwidth]{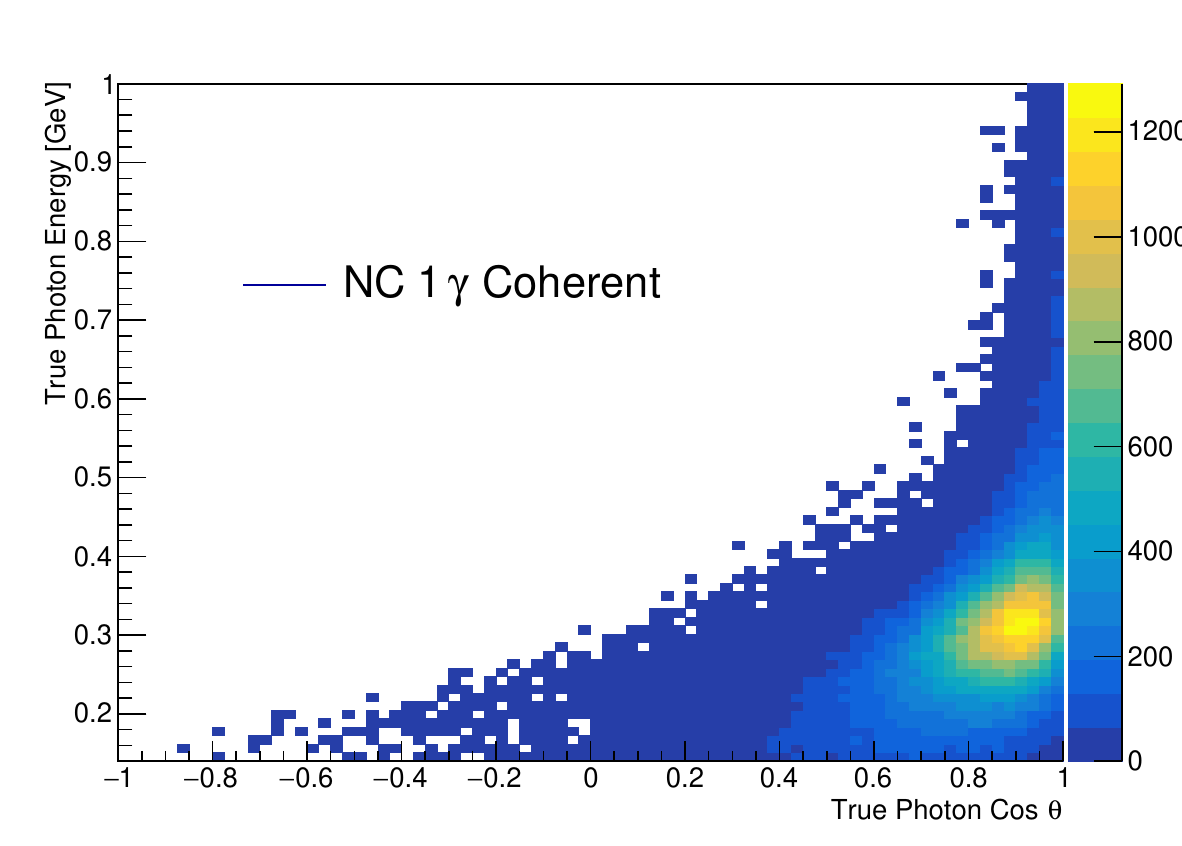}
    \caption{Correlation between true photon energy and angle with respect to the beam of all simulated NC 1$\gamma$ coherent events.}
    \label{fig:signal_2D}
\end{figure}

The distributions of the simulated signal in the MicroBooNE detector are shown in Fig.~\ref{fig:coh_signal_truth} highlighting the energy range and forward nature of the final state photons. Note that compared to Fig.~\ref{fig:coh_differential_xs}, there is no longer a bump at $\sim0.8$ GeV in the photon energy distribution in Fig.~\ref{fig:coh_signal_truth}. This is because, different from the full model discussed in Sec.~\ref{ch4:coh_model}, the current implementation of the model includes only contribution from the \textit{$\Delta$ resonance}. This is a small effect since the BNB neutrino has an average energy of $\langle E_{\nu}\rangle \sim 0.8$ GeV and the higher-order resonance contribution is expected to be smaller than shown in Fig.~\ref{fig:coh_differential_xs}. A 2D distribution of the simulated signals is provided in Fig.~\ref{fig:signal_2D} showing the correlations between the shower angle and energy. 

A caveat for the above signal simulation is that only contribution from the $\nu_{\mu}$ flux is simulated, while for NC processes contributions from all neutrino species are expected. As such, every simulated event is assigned a unique scaling weight based on its associated true $\nu$ energy, to take into account contributions from $\overline{\nu_{\mu}}$, $\nu_{e}$ and $\overline{\nu_{e}}$. For the detailed procedure of how the scaling weight is determined, please see App.~\ref{app:signal_weight}. 

\section{Normalization of NC Coherent $1\gamma$ Simulation}
To evaluate the number of predicted events at a given POT, the simulated NC coherent sample needs to be carefully normalized. More specifically, the number of NC coherent 1$\gamma$ events needs to be mapped to the amount of POT it's equivalent to. This is done based on the \textsc{genie} simulated cross-section splines for the NC coherent 1$\gamma$ process and the simulated neutrino flux profile. 

The cross-section spline $S(E_{\nu})$ from \textsc{genie} provides the cross-section as a function of the true neutrino energy in units of [$\text{cm}^{2}$]. The simulated neutrino flux $F(E_{\nu})$ at MicroBooNE is expressed as a function of true neutrino energy in units of [$\text{POT}^{-1}*\text{bin}^{-1}*\text{cm}^{-2}$], which gives the number of neutrinos predicted in each bin for unit area per POT. Given the number of target atom $N_{target}$, the number of interactions per POT can be calculated as
\begin{equation}\label{eq:pot_norm}
    P_{\text{per POT}} = N_{target}*\int_{0}^{\inf} S(E_{\nu})*F(E_{\nu})dE_{\nu}
\end{equation}
Naturally, the corresponding POT of $n$ predicted events is $n/P_{\text{per POT}}$. The BNB beam contains neutrinos of four flavors: $\nu_{\mu}$, $\overline{\nu_{\mu}}$ , $\nu_{e}$ and $\overline{\nu_{e}}$, thus Eq.~\ref{eq:pot_norm} is updated to: 
\begin{equation}
    P_{\text{per POT}} = N_{target}*\int_{0}^{\inf} \sum_{k}S_{k}(E_{\nu})*F_{k}(E_{\nu})dE_{\nu}
\label{eq_POT}
\end{equation}
where $k$ denotes different neutrino flavors. 

\begin{figure}[h!]
\centering
    \includegraphics[width=0.49\textwidth]{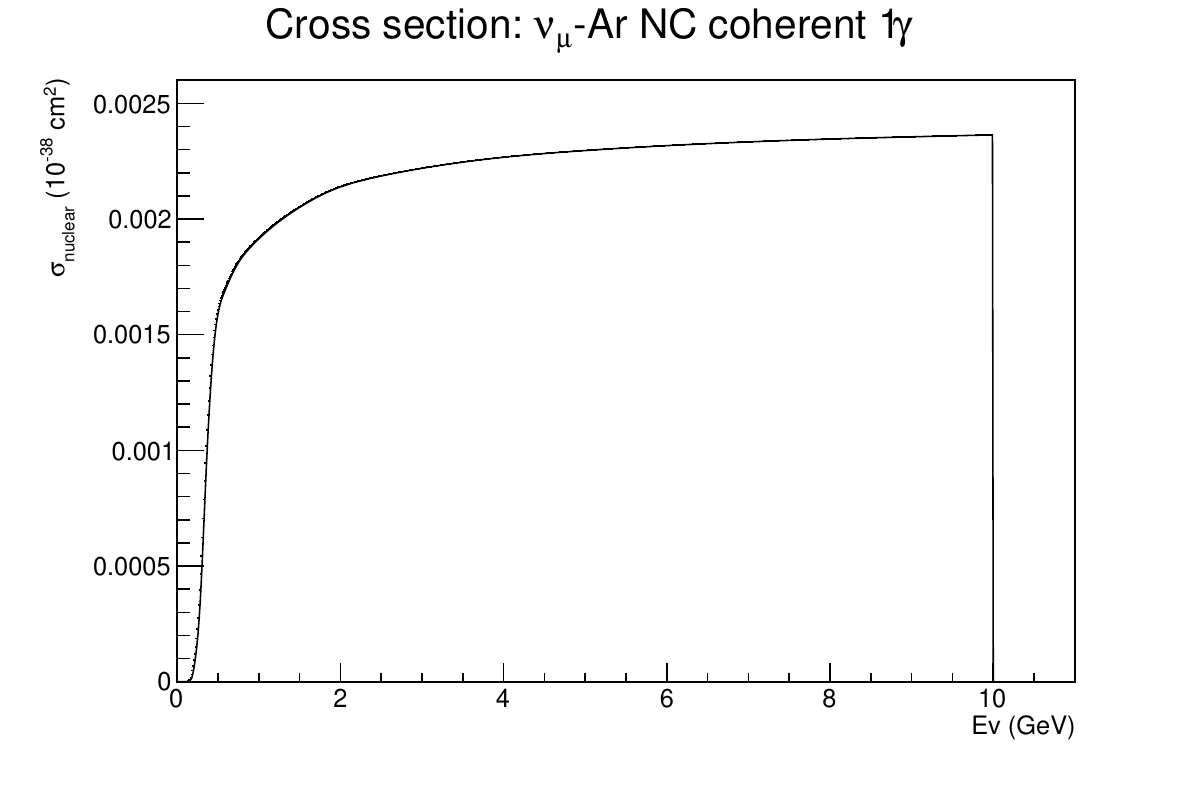}
    \includegraphics[width=0.49\textwidth]{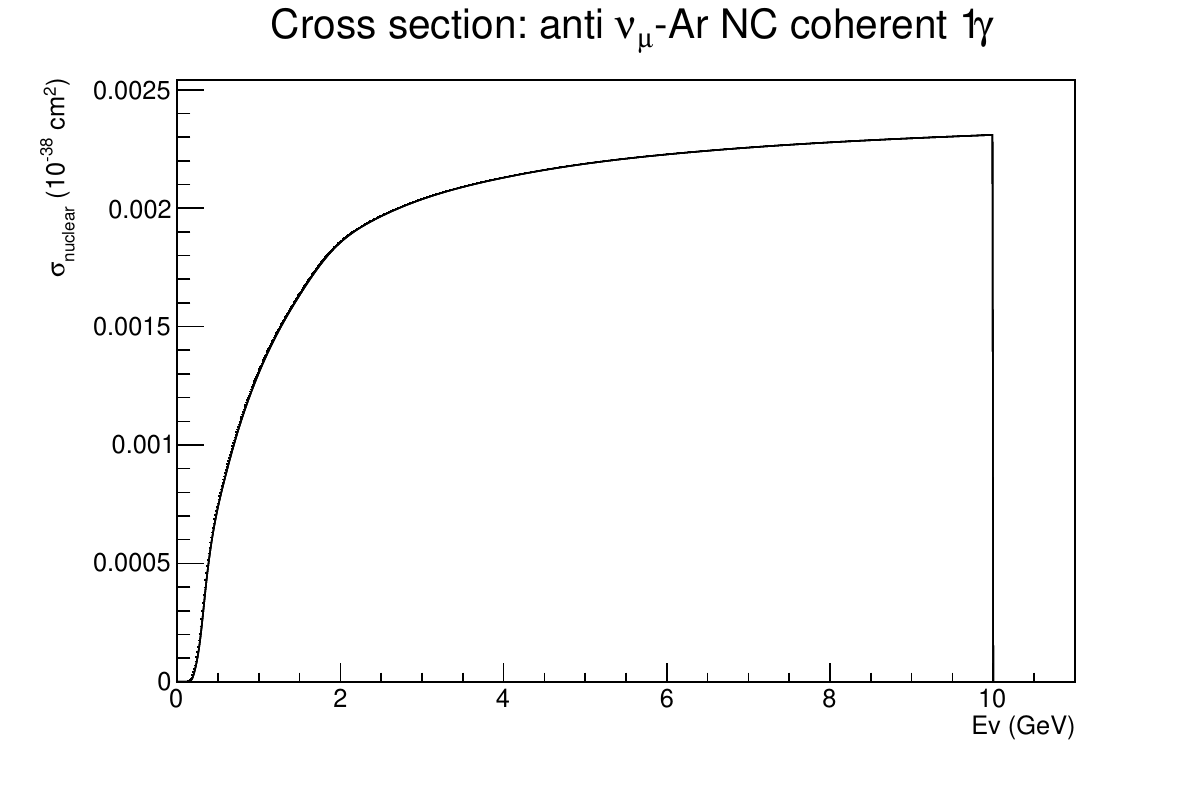}
    \caption[Predicted cross section for neutrino induced NC coherent $1\gamma$ interaction in \textsc{genie}, as a function of true neutrino energy.]{Cross section for neutrino induced neutral current coherent single photon interaction, as a function of true neutrino energy. (left): predicted cross section for neutrinos ($\nu_{\mu}$, $\nu_{e}$ and $\nu_{\tau}$); (right): predicted cross section for antineutrinos ($\overline{\nu_{\mu}}$, $\overline{\nu_{e}}$ and $\overline{\nu_{\tau}}$). The cross-section is shown as per-nucleus cross-section, in a unit of $10^{-38} \text{cm}^{2}$.}
    \label{fig:signal_spline}
\end{figure}
The flux prediction of neutrino beam $F_{k}(E_{\nu})$ at MicroBooNE can be found in Ref.~\cite{BNB_flux_MiniBooNE}. The cross sections $S_{k}(E_{\nu})$ for $\nu$($\overline{\nu}$)-induced NC coherent single photon interaction on argon are shown in Fig.~\ref{fig:signal_spline}, as a function of neutrino energy. For the MicroBooNE detector dimensions, 
the number of argon atoms $N_{target}$ is calculated to be $1.30\times 10^{30}$. This leads to a prediction of 10.43 NC coherent $1\gamma$ events for $6.80\times 10^{20}$ POT.

%% file: chapter5.tex
\chapter{MicroBooNE NC Coherent Single Photon Search}\label{ch:5}
With the expected topology of single shower with no hadronic activity nearby, the search for NC coherent $1\gamma$ focuses on events with a single reconstructed shower and zero reconstructed tracks. The analysis builds upon the framework and tools developed for the NC $\Delta$ radiative decay search discussed in Ch.~\ref{ch3:ncdelta_LEE}, and takes a similar analysis approach. Starting with outputs from Pandora reconstruction~\cite{pandora_reco}, it combines tranditional selection cuts with BDTs to help reject background and utilizes the high-statistic NC $\pi^0$ samples to help constrain background and systematic uncertainty. 

Given the extreme rarity of the NC coherent $1\gamma$ process ($\sim10$ events expected in the MicroBooNE detector over three years of data), a total of six BDT's are developed to maximize the background rejection while preserving the coherent single photon signal events. Though some BDTs share the same names as the ones in the NC $\Delta$ radiative decay search, all BDTs in this analysis are developed and trained independently to target the different signal definition. A major difference from the previous NC $\Delta$ radiative decay search is a new proton-stub veto (PSV) tool developed to reject events with visible proton-candidate activities. The PSV specifically targets events with protons exiting the nucleus that are too low-energy to be reconstructed, through tiny but visible traces of energy deposition the protons leave in the detector. The PSV tool proves to be very efficient in removing NC background with exiting protons.

This analysis makes use of MicroBooNE Runs 1-3 data, and it is designed to be a blind analysis and developed using MC overlay samples without full access to the data. A small subset of ``open data'' that is $\sim 10$\% of the full Runs 1-3 data is accessible during the analysis development, and is used to validate data-MC agreement in BDT training and other important variables. Fake data studies are performed to ensure that the selection is sensitive to the injected signal, and a sideband sample exclusive of the signal region is inspected with full Runs 1-3 data before the signal region is unblinded. The following sections describe the detailed selection, sideband result, and final data result of the analysis. 
Full details of the selection, fake data studies, and sideband studies can be found in Ref.~\cite{coherent_sp_TN}. 


\section{Monte Carlo Prediction Breakdown}
In this analysis, MC simulations are broken down into different categories based on truth information, in a similar manner as the NC $\Delta$ analysis in Ch.~\ref{ch3:ncdelta_LEE}. The definitions of different categories are described below, and their corresponding color scheme can be found in Fig.~\ref{fig:precut}: 
\begin{itemize}
    \item NC coherent $1\gamma$: This is the targeted signal. It includes all NC coherent single-photon production on argon inside the active TPC volume, regardless of incoming neutrino flavor. 
    \item NC 1$\pi^0$: All NC interactions in the active volume that produce one exiting $\pi^0$ regardless of incoming neutrino flavor. This category is further split into two sub-categories, ``NC 1$\pi^0$ Coherent'' and ``NC 1$\pi^0$ Non-Coherent'' contributions, based on their interaction types. Non-Coherent scattering occurs when a neutrino interacts with a nucleon inside the argon nucleus, potentially knocking out one or more nucleons. NC 1$\pi^0$ Coherent, like the NC coherent $1\gamma$ signal, is produced from coherent scattering of the neutrino with the nucleus, and the resulting $\pi^0$ tends to be very forward relative to the incoming neutrino beam.
    \item NC $\Delta \rightarrow N\gamma$: NC single-photon production from the radiative decay of the $\Delta$(1232) baryon in the active volume. This is the largest contribution of NC single photon production relevant to the BNB beam. In this analysis, this is further broken down into two exclusive sets, NC $\Delta \rightarrow N\gamma$ (1+p) and NC $\Delta \rightarrow N\gamma$ (0p), depending on the outgoing nucleon and its kinetic energy (KE). NC $\Delta \rightarrow N\gamma$ (1+p) refers to NC $\Delta$ radiative decay events that have at least one proton exiting the nucleus with KE larger than 50MeV. $\Delta \rightarrow N\gamma$ (0p) is the complementary set and includes NC $\Delta$ radiative decay events with either no protons exiting the nucleus, or protons exiting the nucleus but with low KE ($<50$MeV). 
    \item CC $\nu_\mu\;1\pi^0$: All $\nu_\mu$ CC interactions in the active volume that have one true exiting $\pi^0$.
    \item CC $\nu_e/\overline{\nu}_{e}$ Intrinsic: All CC $\nu_e$ or $\overline{\nu}_{e}$ interactions in the active volume.
    \item BNB Other: All remaining BNB neutrino interactions that take place in the active TPC volume and are not covered by the above five categories, such as multiple $\pi^0$ events and $\eta$ meson decay.
    \item Dirt (Outside TPC): All BNB neutrino interactions that take place outside the MicroBooNE active TPC but have final states that enter and deposit energy inside the active TPC detector. This can originate from scattering off liquid argon in the cryostat vessel outside the active TPC volume or from interactions in the concrete and ``dirt'' surrounding the cryostat itself. 
    \item Cosmic Data: Coincident cosmic ray interactions that take place during a BNB beam-spill but without any true neutrino interaction present.
\end{itemize}


\section{Topological selection and Preselection}\label{ch5:precut}
\begin{figure}[h!]
    \centering
    \includegraphics[width=0.7\linewidth]{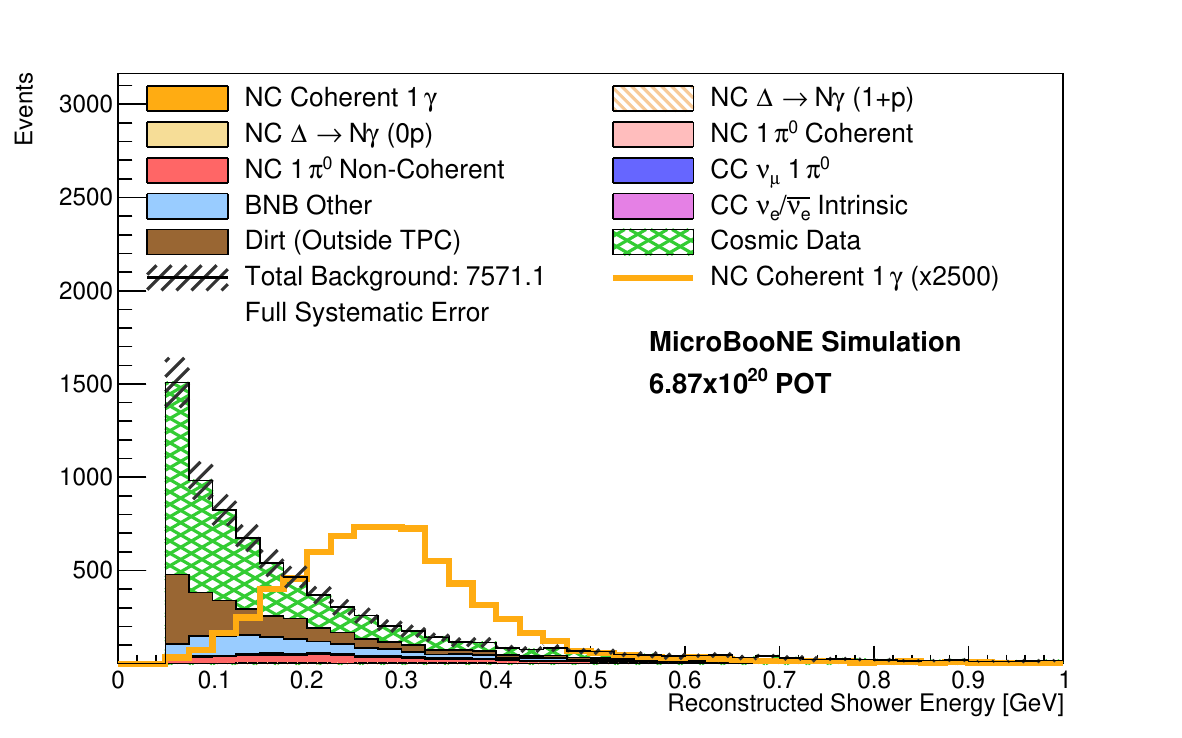}
    \caption[MC predicted background distribution after pre-selection for MicroBooNE NC coherent $1\gamma$ analysis.]{MC predicted background distribution as a function of the reconstructed shower energy after pre-selection. The orange histogram overlaid shows the distribution of NC coherent $1\gamma$ signal scaled by 2500 times for it to be visible. At this stage, the signal and backgrounds populate different regions of phase space in the reconstructed shower energy.}
    \label{fig:precut}
\end{figure}
The event selection starts with a topological selection on outputs from Pandora, specifically requiring one reconstructed shower and zero reconstructed tracks ($1\gamma0p$), which is the expected topology for NC coherent $1\gamma$ signal in the detector. It's worth noting that $71.9$\% of simulated NC coherent $1\gamma$ events are removed by the $1\gamma0p$ topological requirement, because $32.7$\% of the signal events are reconstructed with no track or shower by Pandora, while $39.2$\% are reconstructed with either $\ge 1$ tracks or $\ge 2$ showers. 

Then, loose preselection cuts are applied in order to remove obvious backgrounds. 
The preselection requires that the reconstructed shower energy be larger than 50 MeV to remove Michel electrons from cosmic muon decays. Additionally a fiducial volume cut of at least 2 cm away from the space charge boundary~\cite{spacecharge} is made on the reconstructed shower start, in order to remove showers originating from particles scattering into the TPC. Relative to the topological stage, the preselection cuts further reject 33.9\% of the overall background, mostly dirt and cosmic background, while preserving 98.6\% of the signal. 

After preselection stage, 27.7\% of the NC coherent $1\gamma$ signal remains out of all simulated signal events in the TPC, and the corresponding signal to background ratio is $\sim 1:2696$. Figure~\ref{fig:precut} shows the predicted NC coherent $1\gamma$ signal scaled by a factor of $2500$ and the nominal predicted background after the preselection requirements are applied, highlighting the rarity of the signal and different regions of phase space in reconstructed shower energy populated by the signal and backgrounds.

\section{Event-level Boosted Decision Tree Based Selection}\label{ch5:event_BDT}
After the preselection, BDTs are developed and optimized to further differentiate the signal from background events. There are in total six (6) tailored BDTs developed using the XGBoost library~\cite{xgboost}, each trained with variables derived from reconstructed objects. Every BDT targets a different background and thus is trained with different labeled background events. There are four BDTs trained with event-level variables, which are variables unique in each event, and two BDTs trained with cluster-level variables associated with clusters of reconstructed hits in each event. Simulated MC events are separated into statistically exclusive sets; test sets independent from the BDT training sets are used to obtain the MC prediction, which later is used in signal region selection optimization and data-MC comparison. The event-level BDTs are discussed first in this section, and the cluster formation and cluster-level BDTs are described in Sec.~\ref{ch5:clusterBDT}. 

The largest background contributions remaining after preselection are cosmic-ray background, dirt ($\sim 50$\% of which is $\pi^0$ events), BNB Other, and NC non-coherent $1\pi^0$, ordered by the size of predicted contributions in decreasing order. Three event-level BDTs are employed to target these backgrounds, and another BDT is designed to specifically target the electron background from CC interactions induced by the intrinsic $\nu_{e}/\overline{\nu_{e}}$ in the beam. To enhance the ability of BDTs in learning the differences between NC coherent $1\gamma$ events and backgrounds, all four BDTs are trained with selections of events passing the preselection requirements. The same NC coherent $1\gamma$ 
overlay sample is used as the ``signal'' during training in four BDTs, with the requirements that the reconstructed shower originates from photon activity in truth and $>50$\% of the reconstructed hits in the shower result from energy deposition of the simulated activities, in addition to the preselection requirements. The background definitions used and the key features in training the four BDTs are described below:
\paragraph{Cosmic BDT:} The goal of the cosmic BDT is to differentiate NC coherent $1\gamma$ signal from misidentified cosmic backgrounds. The cosmic BDT trains on BNB external data as background, and makes use of the fact that cosmic rays usually travel from the top of the detector down to Earth vertically, resulting in reconstructed showers that are oriented more vertically and have less extent in the beam direction when compared to that of the signals.
\paragraph{CC $\nu_{e}$ BDT:} CC $\nu_{e}$ BDT aims to remove reconstructed showers originating from electrons instead of photons, and trains on simulated, preselected CC $\nu_{e}/\overline{\nu_{e}}$ events. One key handle for photon/electron separation is the shower $dE/dx$. Most photons in the energy range relevant to this analysis lose energy through $e^{+}e^{-}$ pairproduction; these then further radiate, producing an electromagnetic shower. Compared to the electron shower, whose $dE/dx$ at the shower start is similar to the $dE/dx$ of a MIP around 2 MeV/cm, the photon shower start has $dE/dx \sim $ 4 MeV/cm. 
\paragraph{CC $\nu_{\mu}$ focused BDT:} The CC $\nu_\mu$ focused BDT aims to remove any backgrounds other than the cosmic, NC $\Delta$ radiative decay, CC $\nu_{e}/\overline{\nu_{e}}$ and NC 1$\pi^{0}$. Among these background events, $\sim$76\% are CC $\nu_{\mu}/\overline{\nu_{\mu}}$ events, 60\% of which are CC events without $\pi^{0}$ exiting the nucleus, and for which the muon is mis-reconstructed as a shower. Since muons are MIPs, variables such as shower $dE/dx$, averaged energy per hit in the reconstructed shower, and Pandora score are of importance for the separation of mis-identified muons from the true coherent single photons. The CC $\nu_{\mu}$ focused BDT is trained with simulated $\nu/\bar{\nu}$-interactions other than NC $\Delta$ radiative decay, CC $\nu_{e}/\bar{\nu_{e}}$ and NC 1$\pi^{0}$ as background events. 
\paragraph{NC $1\pi^0$ BDT:}   NC 1$\pi^{0}$ background is harder to remove compared to cosmic and other $\nu$-induced backgrounds because in most cases the reconstructed shower in mis-identified NC 1$\pi^{0}$ is indeed from a true photon (most likely the leading photon from $\pi^{0}$ decay). There are a variety of reasons that could lead to single shower reconstruction in $\pi^{0}$ sample: 
\begin{enumerate}
    \item The second shower from $\pi^{0}$ decay is not visible in the detector; this could happen when the photon is absorbed by the medium, leaves the detector before pair-producing, or is too low energy to be detected. 
    \item The second shower deposits energy in the detector but the 3D reconstruction fails; this could happen when the cluster-matching across planes in Pandora fails. 
    \item The hits from the second photon are reconstructed in 3D but not associated with the neutrino interaction; this happens when cosmic-rays interfere with the energy deposition on one or more planes.
\end{enumerate} 
To mitigate failure case (2), a cluster-level second-shower veto (SSV) BDT is developed following the exact same approach as in the NC $\Delta$ analysis~\cite{uboone_delta_LEE} to identify the possible presence of the second shower that misses reconstruction. The SSV BDT is discussed in more detail in Sec.~\ref{ch5:clusterBDT}, and a selection of the SSV BDT outputs is used to train the NC 1$\pi^{0}$ BDT. In addition to variables involving the missing second shower, variables associated with the primary reconstructed shower that contribute the most in separating the signal from NC $1\pi^0$ backgrounds are the reconstructed shower energy, angle, and projected momentum on the vertical plane (perpendicular to the neutrino beam direction). 
\begin{figure*}[hbt!]
    \centering 
    \begin{subfigure}{0.49\textwidth}
        \includegraphics[width =\textwidth]{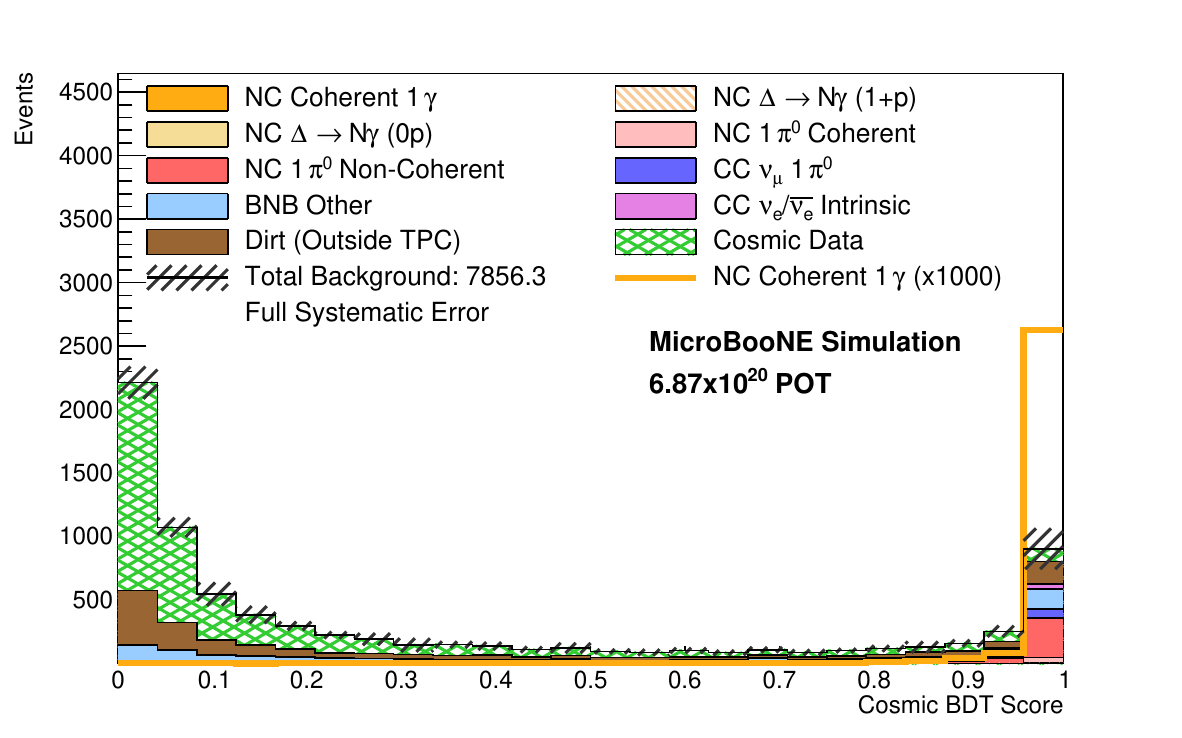}
        \caption{Cosmic BDT score}
        \label{fig:cosmic_bdt}
  \end{subfigure} 
  \begin{subfigure}{0.49\textwidth}
        \includegraphics[width =\textwidth]{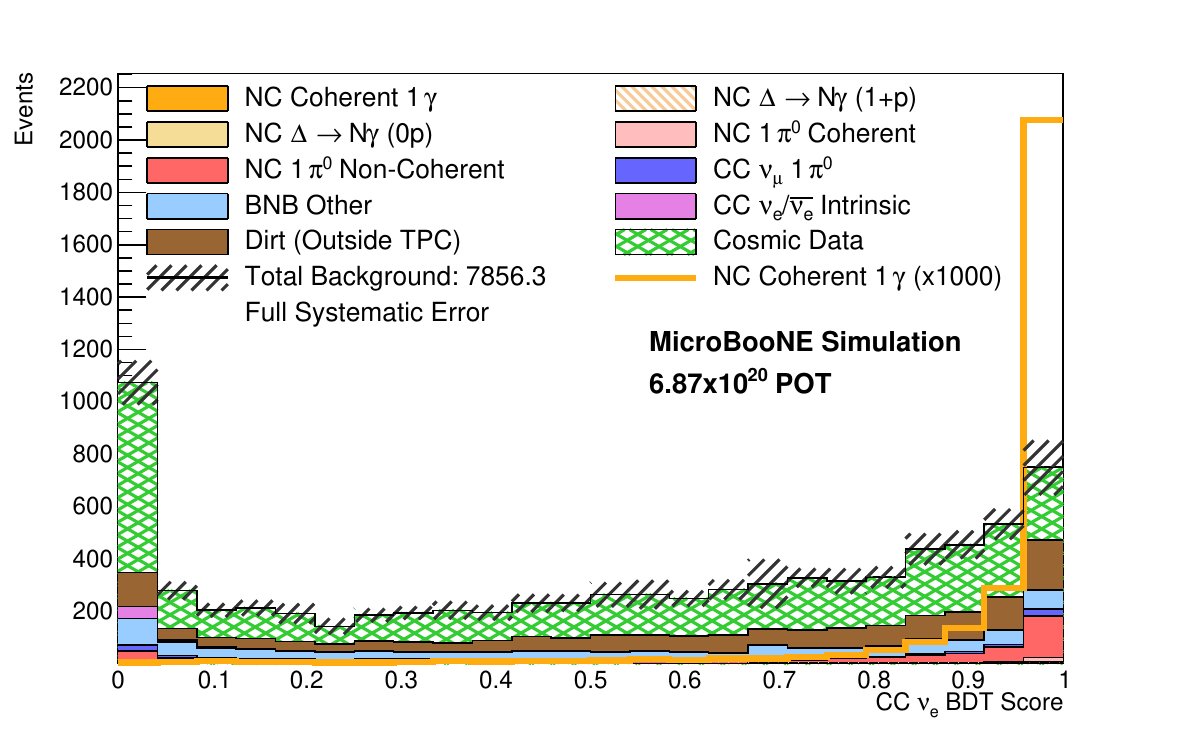}
        \caption{CC $\nu_{e}$ BDT}
  \end{subfigure} 
  \begin{subfigure}{0.49\textwidth}
        \includegraphics[width =\textwidth]{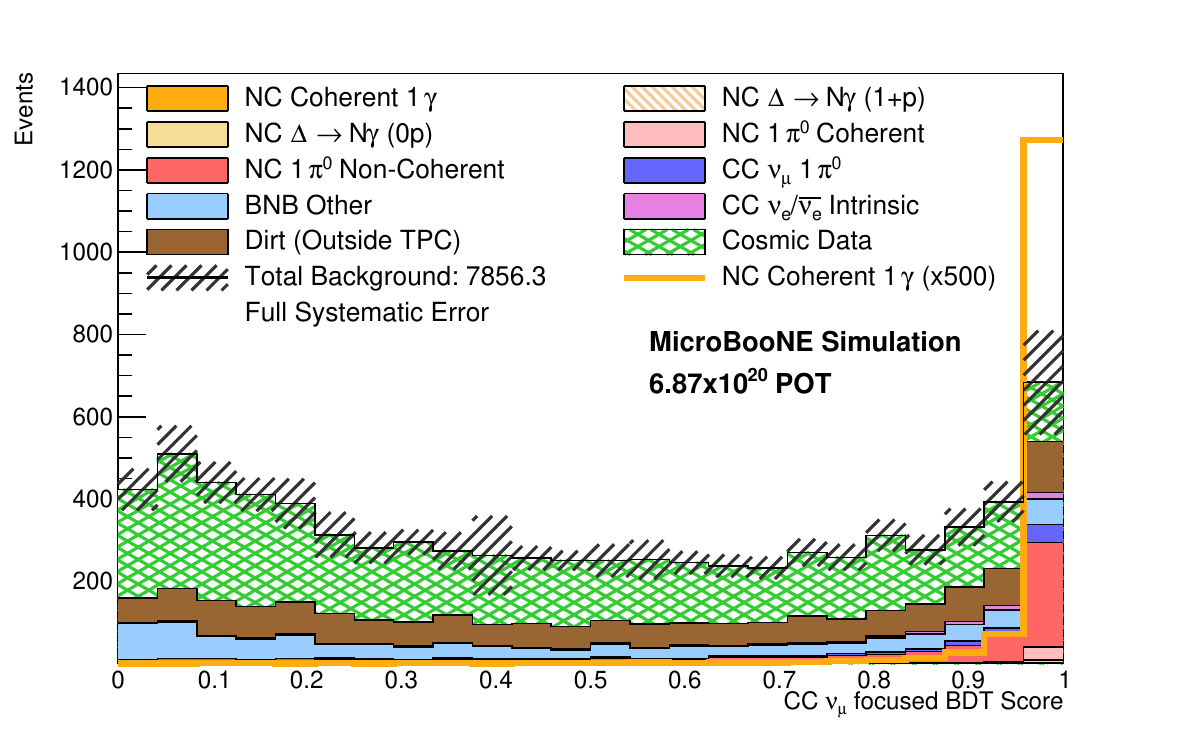}
        \caption{CC $\nu_{\mu}$ focused BDT}
  \end{subfigure} 
  \begin{subfigure}{0.49\textwidth}
        \includegraphics[width = \textwidth]{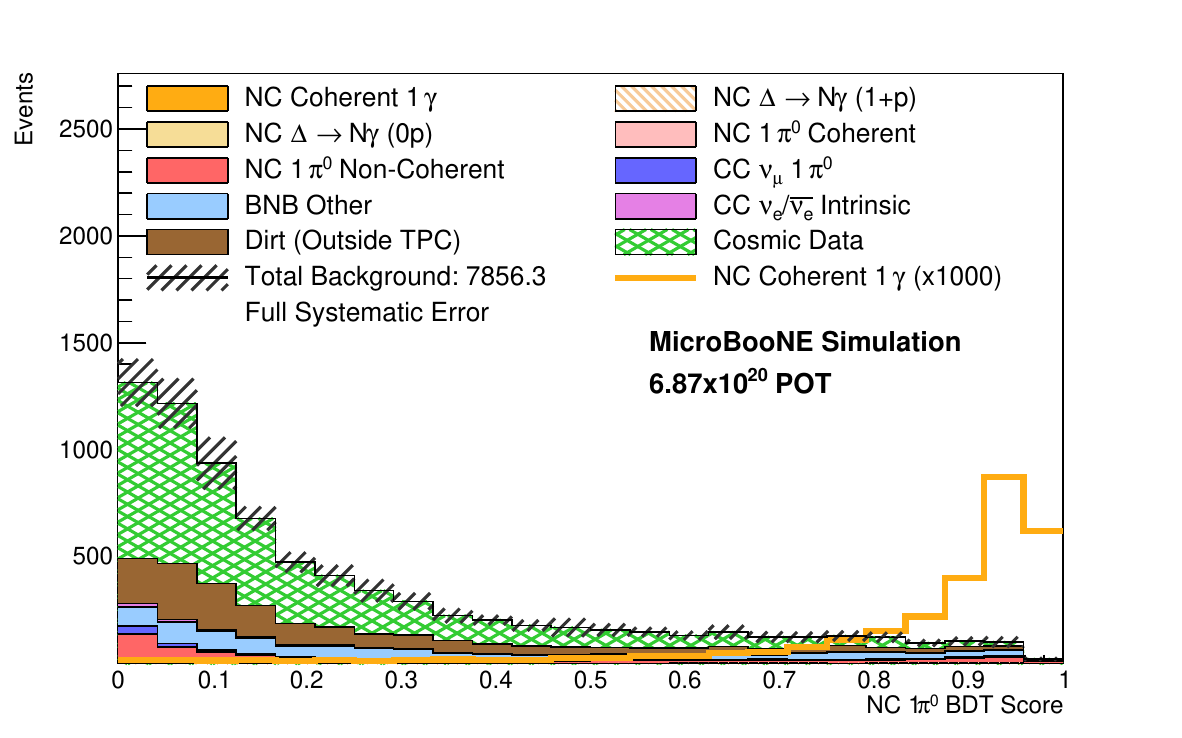}
        \caption{NC $1\pi^0$ BDT}
  \end{subfigure} 
  \caption[The BDT classifier scores for the four event-level BDTs employed in the NC coherent $1\gamma$ analysis, after pre-selection requirements are applied.]{The BDT classifier scores for (a) the Cosmic BDT; (b) CC $\nu_{e}$ BDT; (c) CC $\nu_{\mu}$ focused BDT and (d) NC $1\pi^0$ BDT. Higher scores indicate more NC coherent $1\gamma$ signal-like events. The stacked histograms show the predicted background distribution, while the yellow hollow histogram represents the distribution for the NC coherent $1\gamma$ signal, both after preselection. The signal histogram is significantly scaled for visual purposes by 1000 times in (a), (b), (d) and by 500 times in (c). }
  \label{fig:bdt_responses}
\end{figure*}

The outputs of the event-level BDTs are scores in the range of [0,1]; a higher event-level BDT score indicates the event is more NC coherent $1\gamma$ signal like. Figure~\ref{fig:bdt_responses} shows the predicted BDT score distributions after the preselection requirement. The NC coherent $1\gamma$ signal is scaled up in order to highlight the separation between the signal and the targeted background in the BDT responses. 

Cut positions for these four event-level BDTs are optimized towards maximizing the statistical significance of the NC coherent $1\gamma$ signal, i.e. $\frac{N_{\text{sig}}}{\sqrt{N_{\text{bkg}}}}$~\footnote{Primary event-level BDT optimization with neutrino flux and cross-section uncertainty taken into account is explored and found to prone to cuts with very low MC statistics remaining. Since low MC statistics make it impossible for reliable systematic uncertainty evaluation, this optimization approach is not adopted.}. The possible cut values are explored simultaneously on all four BDTs and the signal significance is evaluated after every choice of cut values. Significant degeneracy is found in the four-dimensional space of possible cuts. A set of cuts chosen to maximize signal selection efficiency within the degenerate region is shown in Tab.~\ref{tab:BDT_cuts}. 
A selection of events pass requirements on these four BDTs if their BDT scores are higher than the corresponding cut values and comprise the ``semi-final'' selection. 

\begin{table}[h!]
    \centering
    \begin{tabular}{|c|c|}
    \hline 
      Event-level BDTs   & Optimized BDT Score Cut\\
      \hline 
       Cosmic  & 0.990 \\
        CC $\nu_{e}$ & 0.885 \\
       CC $\nu_{\mu}$  & 0.992 \\
       NC $1\pi^0$  & 0.891 \\
       \hline 
    \end{tabular}
    \caption[Final cut positions on the four primary event-level BDTs in the NC coherent $1\gamma$ analysis.]{Final cut positions on the four primary event-level BDTs, optimized based on the signal statistical significance ($\frac{N_{\text{sig}}}{\sqrt{N_{\text{bkg}}}}$).}
    \label{tab:BDT_cuts}
\end{table}

\section{Cluster-level Boosted Decision Tree Based Selection}\label{ch5:clusterBDT}
Cluster-level BDTs are designed to identify activities of interest that are missed by Pandora pattern recognition. There are two cluster-level BDTs employed in this analysis. First, a second-shower veto (SSV) BDT targets missing second shower from $\pi^0$ decay in $\pi^0$ backgrounds. Second, a proton-stub veto (PSV) BDT aims to identify proton activity near the shower vertex, in order to remove non-coherent backgrounds. Both BDTs build on hits that are not reconstructed in 3D by Pandora: individual scattered hits are clustered together to form candidate clusters of interest which get classified by the BDTs. We will discuss the clustering mechanism first before describing the specifics of each BDT. 
\subsection{Hit Clustering}\label{ch5:hit_clustering}
Each reconstructed hit consists of an ADC waveform and associated wire and timing information. Nearby neighboring hits are grouped together on a plane-by-plane basis; this is referred to as ``hit clustering''. Hits with summed ADC lower than 25 are considered ``noise'' hits and ignored during clustering formation. The separation of two hits on the same plane is defined by the following metric: 
\begin{equation}
    D = \sqrt{((w_1-w_2)*0.3\text{cm})^2 + (\frac{t_1-t_2}{25\text{cm}^{-1}})^2}
\end{equation}
where $w_1, w_2$ are the wire number of two hits, $t_1, t_2$ are the associated time tick (1 tick = 0.5~$\mu$s) information and where the normalization factors are chosen to make the wire spacing and time interval difference comparable in physical extent. 

The clustering is performed using the density-based spatial clustering of applications with noise (DBSCAN) algorithm~\cite{dbscan} with two parameters: 
\begin{itemize}
    \item $N_{\text{minPts}}$: the minimum number of hits required to form a cluster.
    \item $\epsilon$: the maximum distance between two hits. This distance requirement is used in both initial cluster formation, which requires at least $N_{\text{minPts}}$ hits all within a distance of $\epsilon$ from each other, and cluster expansion, where an additional hit is absorbed as long as it is within a distance of $\epsilon$ from at least one hit in an existing cluster. 
\end{itemize}
The result of DBSCAN is a group of clusters on three planes that are robust to spatial outlier hits and arbitrarily shaped complying with the restriction from $\epsilon$.

\subsection{Second shower veto BDT}
The SSV BDT targets the second shower from the $\pi^0$ decay in the detector, which has an expected topology of a sparse cascade of hits near the primary reconstructed shower. Thus, the candidate clusters for the second shower are formed with $N_{\text{minPts}} = 8$ and $\epsilon = 4$ cm. Undesirable clusters can result from electronic noise along a single wire during continuous time intervals and correlated noises across several wires at the same time. To avoid clustering on such noise, the candidate clusters are required to span at least four (4) time ticks and three different wires. Further, the spatial variance of hits within a cluster along any single axis cannot exceed 99.9\% of the total spatial variance of the cluster, which is evaluated through principal component analysis (PCA). 
This further reduces the mis-clustering of noise and yields one or more second shower candidate clusters per plane per event.

\begin{figure}[h!]
    \centering
    \includegraphics[width=0.7\linewidth]{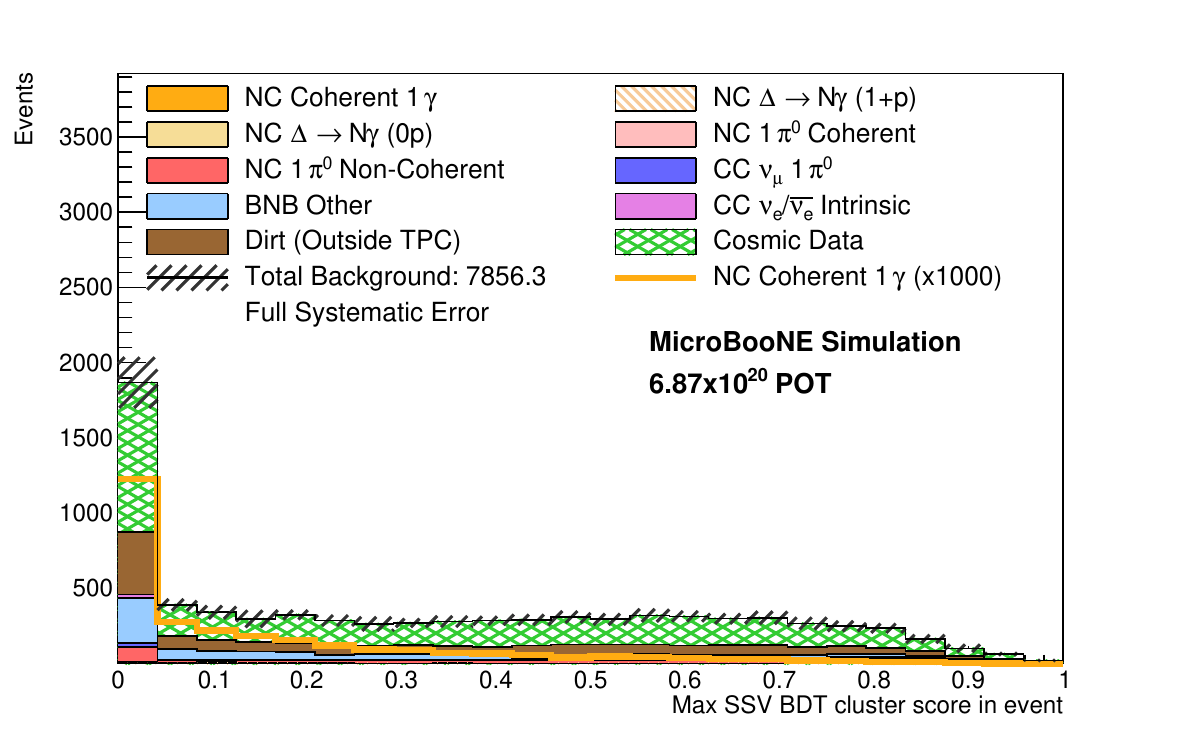}
    \caption[Distribution of the maximum SSV BDT score of all clusters formed in events passing the preselection requirements]{Distribution of the maximum SSV BDT score of all clusters formed in events passing the preselection requirements. Nominal predictions for backgrounds are stacked in the colored histograms, while the signal prediction is scaled by a factor of 1000 and plotted separately in the yellow hollow histogram. Higher score means a higher probability of a second shower present in the event. Note that, if no second shower candidate cluster is formed in an event, the maximum SSV BDT score for this event is set to zero. }
    \label{fig:maxSSV}
\end{figure}
The reconstructed calorimetric and spatial variables associated with second shower candidate clusters are the inputs to the SSV BDT. More specifically, second shower candidate clusters that are truth-matched to a different photon from the primary reconstructed shower in the NC $1\pi^0$ sample serve as the training signal, while second shower candidate clusters formed in simulated NC coherent $1\gamma$ events are used as training background. 
The output of the BDT is a score assigned to each cluster from 0 to 1. The higher the score, the higher the probability that the cluster originates from a photon from $\pi^0$ decay. The result of SSV BDT is not directly used in event selection; instead, high-level variables such as the maximum SSV BDT score of all clusters on each plane are input variables to the NC $1\pi^0$ BDT. Figure~\ref{fig:maxSSV} shows the distribution of the maximum SSV BDT scores of all clusters formed in events.

\subsection{Proton stub veto BDT}\label{subsec:psv_bdt}
Besides the SSV score, another handle is leveraged to aid in the rejection of background: tiny proton traces. MicroBooNE's detection threshold for protons is 40 MeV in KE; however, hand scanning of NC $\Delta$ radiative decay events shows visible true proton-like activity on one or more planes near the reconstructed shower in some $1\gamma0p$ events. 
These proton-like activities usually sit very close to the backward projection of the shower direction. The PSV BDT is designed to identify such indications of missed protons, through which non-coherent backgrounds can be removed, specifically misidentified NC backgrounds with protons exiting the nucleus. 

The building block of the PSV BDT is proton candidate clusters, similarly to the SSV BDT. Since most protons that are missed by the reconstruction are low energy, and proton tracks are expected to be very straight with dense energy deposition, proton candidate clusters are required to be dense by setting $\epsilon = 1$ cm during the DBSCAN clustering. Driven by handscan observations, $N_{\text{minPts}}$ is set to 1 to allow the capture of tiny clusters originating from proton activities. This allows small noise hits to leak in, leading to $\mathcal{O}(100)$ proton candidate clusters per event. Thus the PSV BDT must be efficient in correctly identifying true proton clusters from substantial amount of background clusters arising from other activity. Two handles help solve this problem: the geometric relation between true proton candidate clusters and the reconstructed shower, and the distinct calorimetric profile of proton track. 

\begin{figure}[h!]
    \centering
    \includegraphics[width=0.7\linewidth, trim = 3cm 4cm 5cm 1cm, clip]{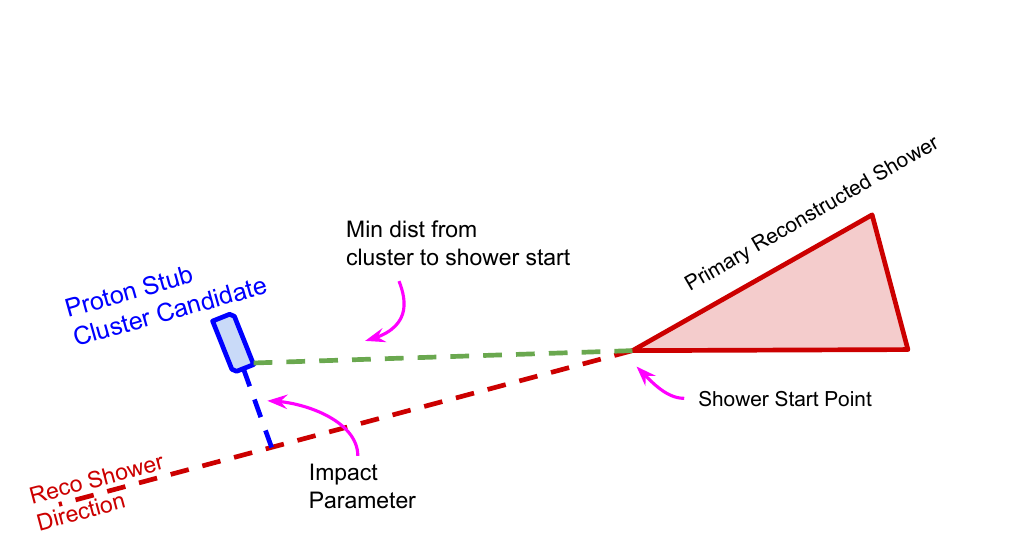}
    \caption[Cartoon showing the relative position of a proton candidate cluster to the primary reconstructed shower on a plane.]{Cartoon showing the relative position of a proton candidate cluster to the primary reconstructed shower on a plane. Two important variables for the PSV BDT are also highlighted: the minimum distance between the proton cluster and the reconstructed shower start, and the impact parameter of the candidate cluster to the shower direction. If the proton stub candidate is indeed from a proton that comes from the same vertex as the reconstructed shower, the impact parameter is expected to be close to zero, assuming the reconstructed shower direction is accurate.}
    \label{fig:psv_cartoon}
\end{figure}

The cartoon in Fig.~\ref{fig:psv_cartoon} shows a potential geometric relation between a proton candidate cluster and the reconstructed shower on a plane. In backgrounds that have true protons exiting the nucleus and reconstructed shower from a true photon, the proton cluster is expected to intersect the line of backward projection of the reconstructed shower. This means the minimum perpendicular distance from the candidate cluster to the direction of the primary reconstructed shower, i.e. the impact parameter of the cluster, should be zero for the proton candidate clusters. 
Additionally, the PSV BDT also takes advantage of the straightness and the Bragg peak signature expected for a proton track to identify true proton clusters.

The simulated NC $\Delta$ radiative decay sample is used to train the PSV BDT: proton candidate clusters in the NC $\Delta$ sample that are truly from protons are used as training signals, while the remaining candidate clusters are the training background. Unlike the SSV BDT which is not used as a selection cut, 
a cut is applied directly on the derived PSV BDT output to further improve the signal-to-background ratio after the semi-final selection.

The derived PSV BDT output used is the maximum PSV BDT score of all proton candidate clusters on plane 0 and plane 2 (abbreviated as ``maximum PSV score on Planes 02''). The reason for not including clusters formed on plane 1 is due to the fact that an excess amount of proton candidate clusters is found in data in the open dataset~\cite{coherent_sp_TN,coherent_psv_excess}. The excess is localized in specific regions on plane 1 and found to be due to mismatch of run periods used in the simulation overlay sample and the open data. This is successfully resolved by requiring the overlay sample in simulation and the collected data to come from overlapping run periods. However, to be conservative, PSV BDT score information on plane 1 is dropped. 

\begin{figure}[h!]
    \centering
    \includegraphics[width=0.7\linewidth]{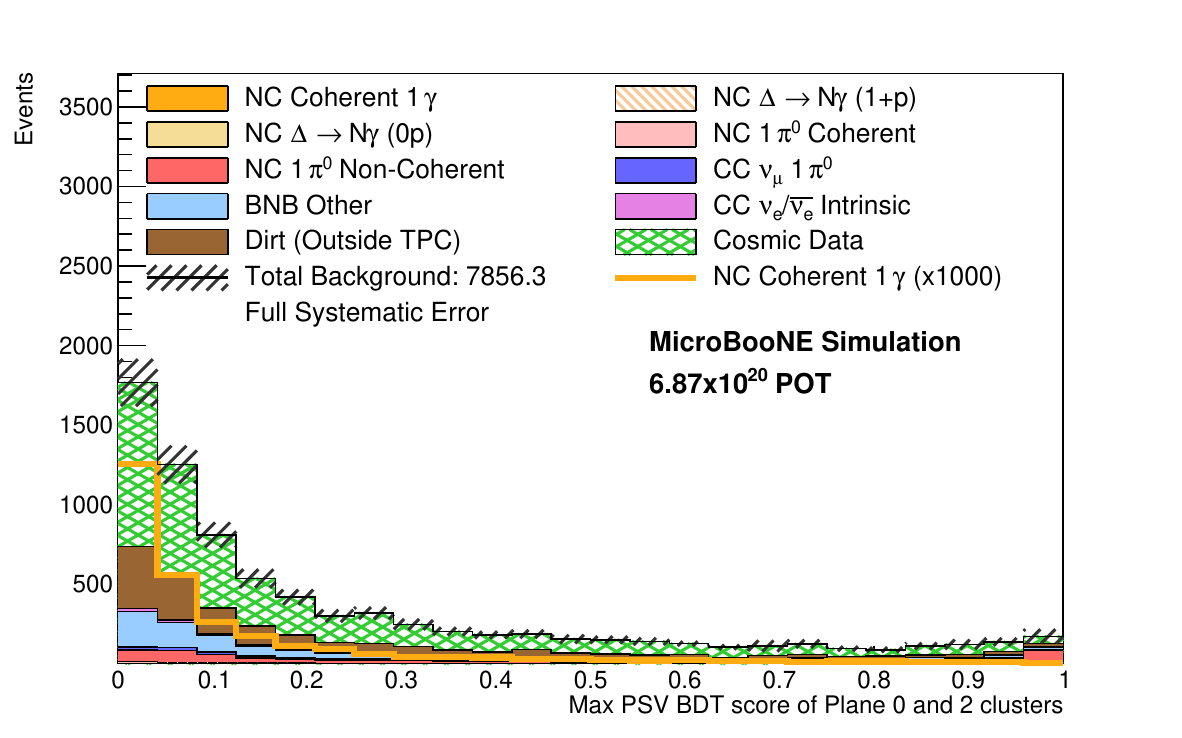}
    \caption[The maximum PSV score among all clusters on plane 0 and plane 2 for events passing the preselection requirements.]{The maximum PSV score among all clusters on plane 0 and plane 2 for events passing the preselection requirements. Higher score indicates increasing confidence there is proton exiting the nucleus that’s missed by Pandora. Nominal predictions for backgrounds are stacked in colored histograms, while the signal prediction is scaled by a factor of 1000 and plotted separately in the yellow hollow histogram. }
    \label{fig:maxPSV}
\end{figure}

\begin{figure}[h!]
    \centering
    \includegraphics[width=0.7\linewidth]{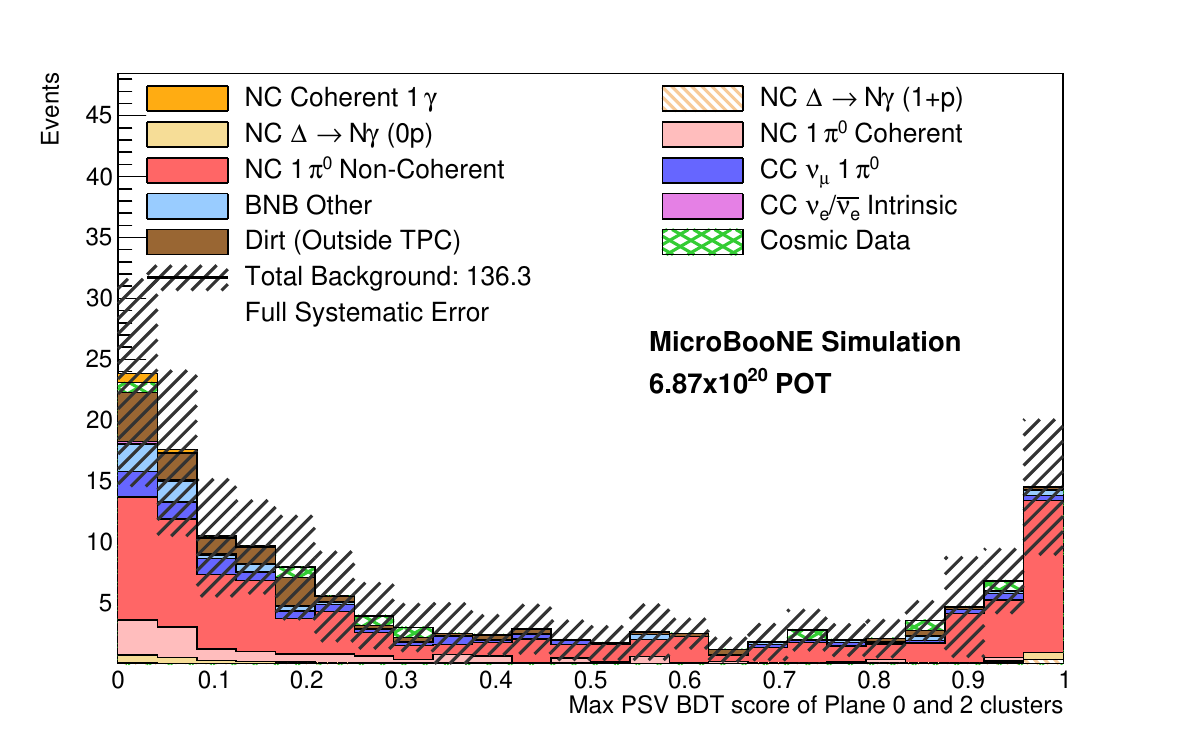}
    \caption[The maximum PSV score among all clusters on plane 0 and plane 2 for events passing the three event-level BDTs (cosmic, CC $\nu_{e}$ and CC $\nu_{\mu}$ focused BDTs).]{The maximum PSV score among all clusters on plane 0 and plane 2 for events passing the three event-level BDTs (cosmic, CC $\nu_{e}$ and CC $\nu_{\mu}$ focused BDTs). Nominal prediction for the signal is on top of the background stacked histograms shown as orange filled histogram, which can be seen on the first bin on the left.}
    \label{fig:maxPSV_photonRich}
\end{figure}

Figure~\ref{fig:maxPSV} shows the distribution of the maximum PSV score on Planes 02 for the predicted background after preselection. As expected, the NC coherent $1\gamma$ signal and most of the background pile up on the left, while the right corner is most populated by the NC non-coherent $1\pi^0$. The distribution of the same variable with the cosmic, CC $\nu_{e}$ and CC $\nu_{\mu}$-focused BDT requirements applied is shown in Fig.~\ref{fig:maxPSV_photonRich}. The resulting sample is a photon-rich sample dominated by the NC non-coherent $1\pi^0$s. The peak near 1 is mostly populated by NC non-coherent $1\pi^0$ background, highlighting the strong separation power of the PSV BDT. The NC coherent $1\gamma$ signal and majority of the NC coherent $1\pi^0$ background cluster at low BDT score regions due to their coherent nature, as expected.

The requirement on the maximum PSV score on Planes 02 is applied to events in the semi-final selection, and the cut value is optimized to maximize the signal significance with the neutrino flux and neutrino interaction uncertainties (discussed in Sec.~\ref{sec:sys_error}) included and constrained from the $2\gamma$ samples (discussed in Sec.~\ref{sec:pi0_constraint}). The optimized cut position for the maximum PSV score on Planes 02 is found to be 0.2. Events that have a maximum PSV score on Planes 02 less than the cut value (0.2) comprise the final signal selection. 

\begin{figure}[h!]
    \centering
    \includegraphics[width=0.7\linewidth]{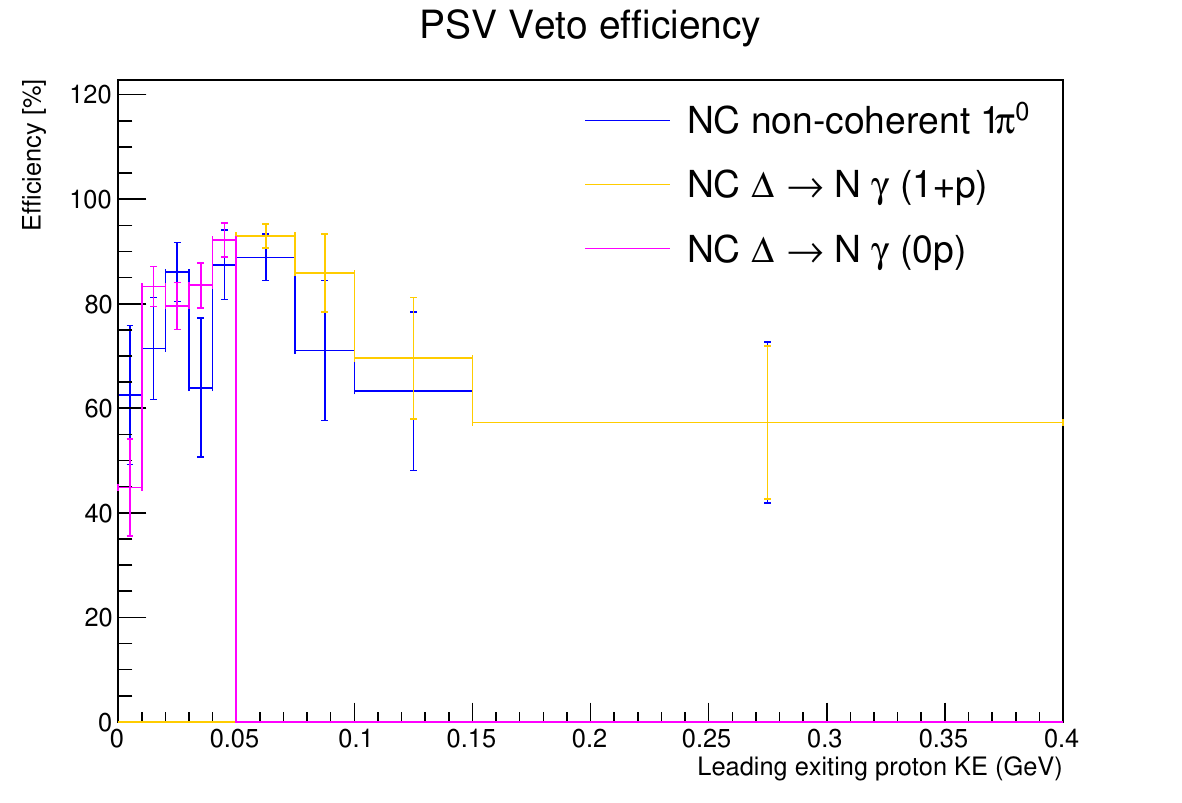}
    \caption[The rejection efficiency of the PSV BDT as a function of the KE of the leading proton exiting the nucleus.]{The rejection efficiency of the PSV BDT as a function of the KE of the leading proton exiting the nucleus. Note that, NC $\Delta \rightarrow N\gamma$ (0p) contains events with exiting protons (with KE less than 0.05 GeV), hence the pink line representing the rejection efficiency on NC $\Delta \rightarrow N\gamma$ (0p) only extends to 0.05 GeV. The rejection efficiency on the complementary NC $\Delta \rightarrow N\gamma$ (1+p) sample instead starts from proton KE of 0.05 GeV (in yellow). The error bar represents the uncertainty arising from finite MC statistics.}
    \label{fig:psv_veto_eff}
\end{figure}

Figure~\ref{fig:psv_veto_eff} shows the rejection efficiency of the PSV BDT on simulated NC non-coherent $1\pi^0$, NC $\Delta \rightarrow N\gamma$ (1+p) and $0p$. The rejection efficiency is evaluated by calculating the fraction of events that pass the PSV BDT requirement in each bin, relative to selected events passing the topological, preselection, and cosmic, CC $\nu_{e}$ and CC $\nu_{\mu}$ BDT requirements. The PSV yields a high rejection efficiency on protons at low energies on all three samples, with an average $\sim70\%$ rejection for protons with KE less than 50 MeV.

\section{Final Selections}
\begin{figure}[h!]
    \centering 
    \begin{subfigure}{0.55\textwidth}
        \includegraphics[width = \textwidth]{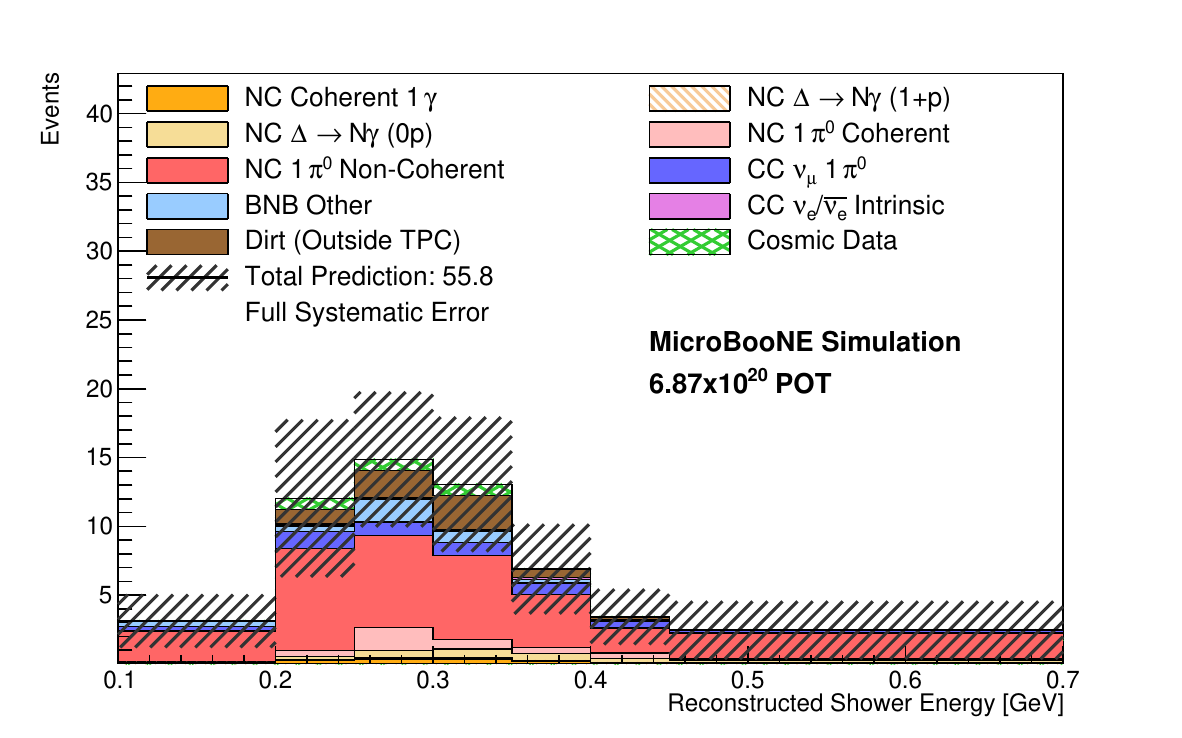}
        \caption{Reconstructed Shower Energy}
        \label{fig:cosmic_bdt}
  \end{subfigure} 
  \begin{subfigure}{0.55\textwidth}
        \includegraphics[width = \textwidth]{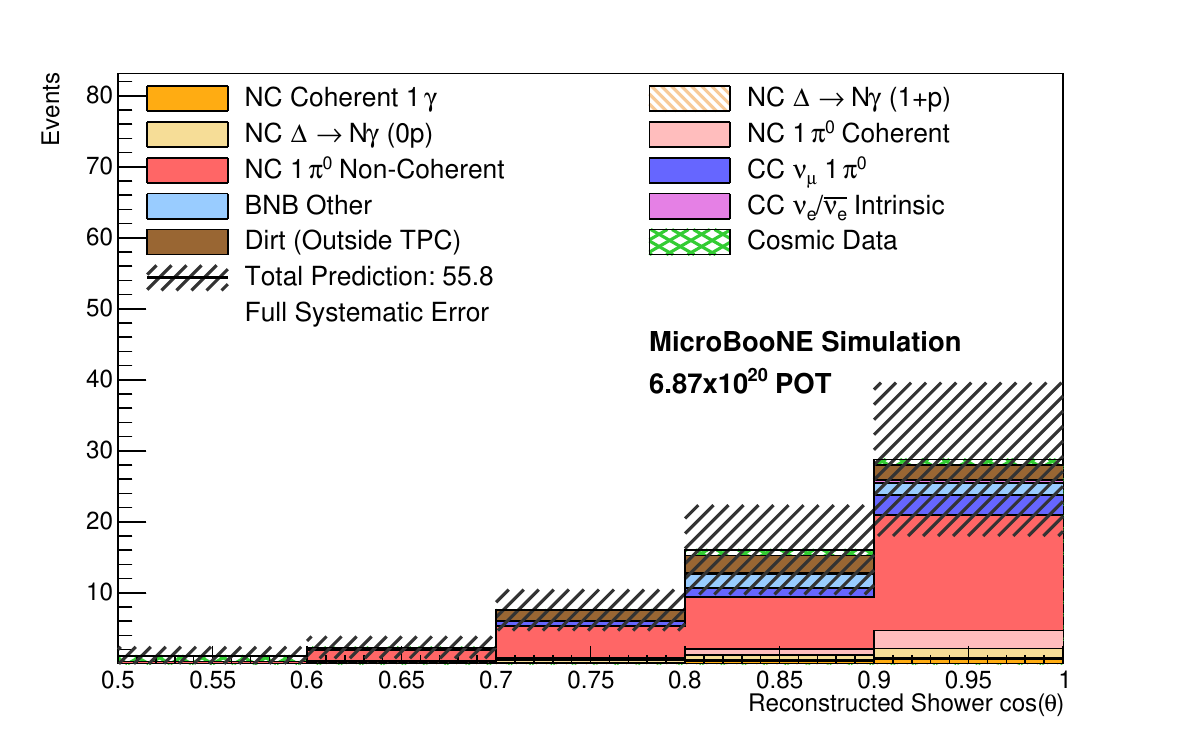}
        \caption{Reconstructed Shower cos($\theta$)}
  \end{subfigure} 
  \begin{subfigure}{0.55\textwidth}
        \includegraphics[width = \textwidth]{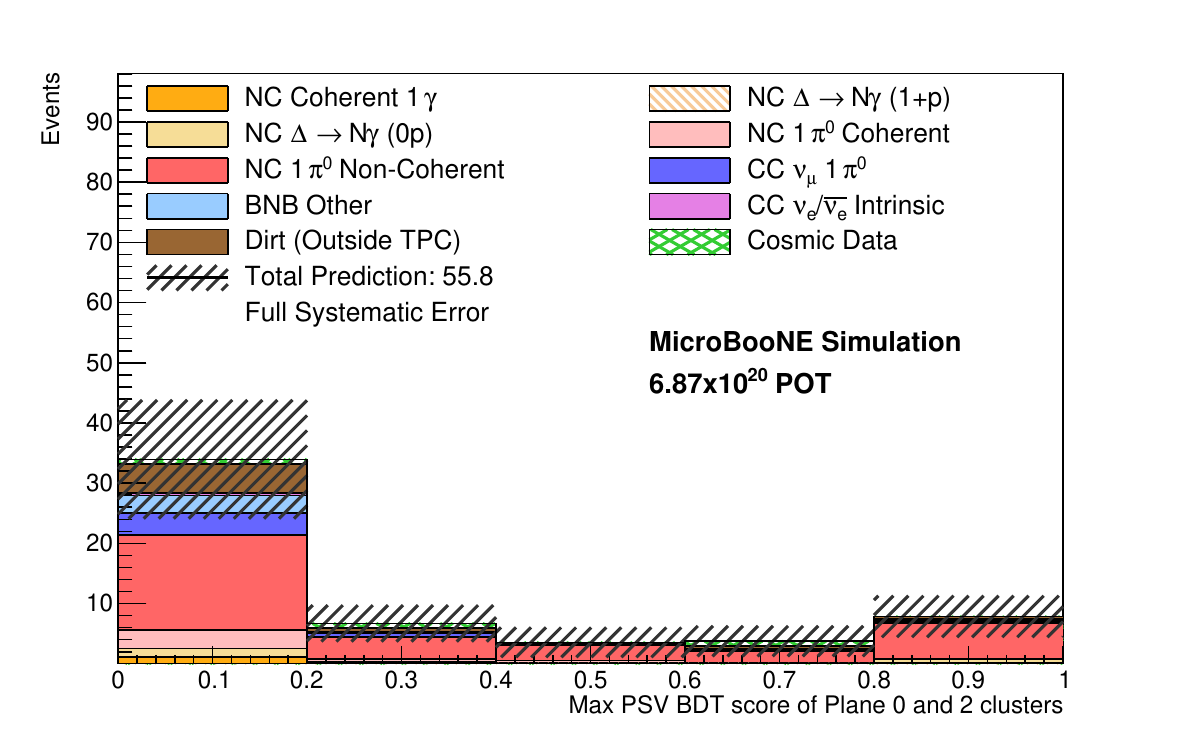}
        \caption{Maximum PSV Score on Planes 02}
  \end{subfigure} 
  \caption[Distributions of semi-final selection in three variables of interest: (a) reconstructed shower energy; (b) reconstructed shower cos($\theta$); and (c) the maximum PSV score on Planes 02.]{Distributions of semi-final selection in: (a) reconstructed shower energy; (b) reconstructed shower cos($\theta$); and (c) the maximum PSV score on Planes 02. The selected events have reconstructed shower with energy peak at $\sim300$ MeV in very forward directions, which is expected for the NC coherent $1\gamma$ signal. The distribution of the maximum PSV score on Planes 02 suggests the PSV cut at 0.2 could reject a significant fraction of NC non-coherent $1\pi^0$ background.}
  \label{fig:semi_final}
\end{figure}

Figure~\ref{fig:semi_final} shows the prediction at the semi-final stage in three variables that highlight the reconstructed shower energy and angle, as well as the PSV features of signal and backgrounds. The selection has a reconstructed shower distribution with an energy peak of around 300 MeV and a very forward angle, which is expected for the NC coherent $1\gamma$ signals. As expected, the NC non-coherent $1\pi^0$ is the most dominant background, and the PSV distribution suggests that an additional requirement on the PSV variable could further reject a large fraction of it. 

\begin{figure}[h!]
\centering
    \begin{subfigure}{0.6\textwidth}
        \includegraphics[page=1,width = \textwidth,trim=0cm 0cm 0cm 1cm, clip]{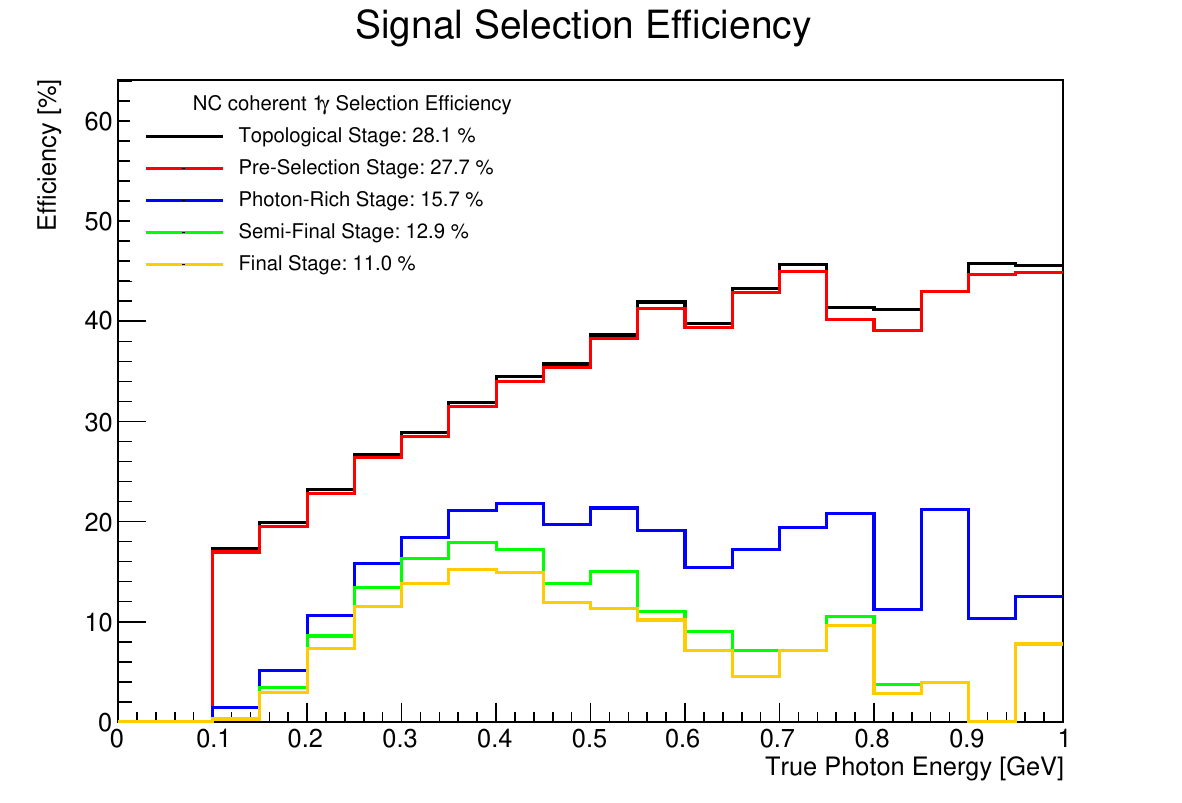}
        \caption{Selection Efficiency v.s. True Photon Energy}
        \label{}
  \end{subfigure} 
  \begin{subfigure}{0.6\textwidth}
        \includegraphics[page=3, width = \textwidth,trim=0cm 0cm 0cm 1cm, clip]{figures/chapter5/COH_signal_selection_efficiency.pdf}
        \caption{Selection Efficiency v.s. True Photon $\cos{\theta}$}
        \label{}
  \end{subfigure} 
    \caption[Selection efficiencies at various stages of the analysis for the NC coherent 1$\gamma$ signal.]{Efficiencies at various stages of the selection for the NC coherent 1$\gamma$ signal as a function of (a) true photon energy and (b) true photon angle with respect to the neutrino beam, highlighting the focus on the photon phase space targeted by this analysis. The ``photon-rich stage'' refers to the stage where the three event-level BDTs (cosmic, CC $\nu_{e}$ and CC $\nu_{\mu}$ focused BDTs) are applied. The efficiencies are calculated using the selected NC coherent $1\gamma$ within the range plotted for each variable with respect to the total prediction of NC coherent $1\gamma$.}
    \label{fig:sig_eff_final}
\end{figure}

Table~\ref{tab:final_prediction} shows all backgrounds and signal predictions for the semi-final and final selection, normalized to $6.868\times 10^{20}$ POT, expected for Runs 1-3. At the semi-final stage, there are a total of 55.8 predicted background and 1.3 signal events, leading to a signal-to-background ratio of 1:42. The PSV requirement further improves the signal-to-background ratio to 1:30 by filtering out almost half of the NC non-coherent $1\pi^0$ background. Figure~\ref{fig:sig_eff_final} shows the selection efficiency for the simulated NC coherent $1\gamma$ signal in the active TPC at different stages, as a function of the true energy and angle of the outgoing photon. 

\begin{table}[h!]
    \centering
    \begin{tabular}{lrr}
    \hline\hline
        Stage & Semi-Final & Final \\ \hline
        NC coherent $1\gamma$ (Signal) & 1.3 & 1.1 \\
        NC $\Delta\rightarrow N\gamma$ (1+p) & 0.3 & 0.1 \\
        NC $\Delta\rightarrow N\gamma$ (0p) & 2.5 & 1.3 \\
        NC $1\pi^0$ Non-Coherent & 29.9 & 15.8 \\
        NC $1\pi^0$ Coherent & 3.8 & 3.1 \\
        CC $\nu_\mu$ $1\pi^0$ & 5.0 & 3.6\\
        CC $\nu_e$ and $\bar{\nu}_e$ & 0.6 & 0.4\\
        BNB Other & 3.5 &  2.9 \\
        Dirt (outside TPC) & 6.4 & 4.8 \\
        Cosmic Ray Data& 2.4 & 0.8 \\
        \hline
        Total Prediction (Unconstr.) & 55.8 & 34.0 \\ 
        Total Prediction (Constr.) & 45.8 & 29.0 \\ 
    \hline\hline
    \end{tabular}
    \caption[The expected event rates in the semi-final and final selection for all event types, normalized to the Runs 1-3 data POT.]{The expected event rates in the semi-final and final selection for all event types, normalized to the Runs 1-3 data POT ($6.868 \times 10^{20}$). The constrained total prediction is evaluated with the conditional constraint from the $2\gamma1p$ and $2\gamma0p$ samples.}
    \label{tab:final_prediction}
\end{table}

The selection efficiencies for all background categories at different analysis stages are included in Tab.~\ref{tab:final_eff}, calculated relative to predictions after the topological selection. The event-level BDTs successfully reject all types of backgrounds by a large fraction, especially the intrinsic CC $\nu_{e}$, BNB Other, dirt and cosmic backgrounds. Transitioning from semi-final stage to final selection stage, the cut on the PSV BDT acts mostly on the NC $\Delta \rightarrow N\gamma$ (1+p), NC $\Delta \rightarrow N \gamma$ (0p) and NC non-coherent 1$\pi^{0}$ backgrounds, with corresponding rejection efficiency of 68.8\%, 47.9\% and 47.4\% respectively. Note that NC $\Delta \rightarrow N \gamma$ (0p) consists of NC $\Delta$ radiative decay events with exiting neutrons out of the nucleus as well as NC $\Delta$ radiative decay events with exiting protons with $\text{KE} < 50$ MeV, the latter of which is specifically targeted by the PSV. Overall, compared to the topological selection, the final selection achieves 97.5\% rejection efficiency on NC non-coherent 1$\pi^0$ and 99.99\% rejection efficiency on the cosmic background while keeping signal efficiency at 39\%. 

\begin{table*}[hbt!]
\centering
\begin{tabular}{|c|c|c|c|}
\hline
Category & Pre-Selection Eff [\%] & Semi-Final Eff. [\%] & Final Eff. [\%] \\
\hline\hline
NC Coherent 1 $\gamma$ & 98.59 & 45.85 & 39.09\% \\ 
NC $\Delta \rightarrow N \gamma$ (0p) & 98.30 & 19.18 & 9.99 \\ \hline \hline
NC $\Delta \rightarrow N\gamma$ (1+p) & 97.13 & 7.99 & 2.49 \\  
NC 1 $\pi^{0}$ Coherent & 96.91 &  8.38 & 6.83 \\ 
NC 1 $\pi^{0}$ Non-Coherent & 96.39 & 4.75 & 2.50 \\ 
CC $\nu_{\mu} 1 \pi^{0}$ & 90.21 & 3.07 & 2.24\\ 
BNB Other & 82.17 &  0.30 & 0.25 \\ 
CC $\nu_{e}/\bar{\nu_{e}}$ Intrinsic & 94.59 & 0.69 & 0.47 \\ 
Dirt (Outside TPC) & 61.67 &  0.25 & 0.18 \\ 
Cosmic Data & 61.01 & 0.03 & 0.01 \\ \hline
\end{tabular}
\caption[Selection efficiencies for signal and background categories at different stages of the NC coherent $1\gamma$ analysis.]{Selection efficiencies for signal and background categories at different stages of the analysis, calculated with respect to selections of events after topological requirement. }
\label{tab:final_eff}
\end{table*}

\section{Systematic Uncertainty and Evaluation}\label{sec:sys_error}
The sources of systematic uncertainties considered in this analysis are broken down into contributions from 1) neutrino flux, 2) neutrino interaction cross-sections, 3) hadron-argon reinteractions, 4) detector response, and 5) finite MC statistics. 
\subsection{Neutrino Flux Uncertainty}
The flux uncertainty includes proton delivery, hadron production, hadronic interactions and the modeling of the horn magnetic field~\cite{BNB_flux_MiniBooNE,uB_flux_note}. The intensity of the proton beam delivered to the BNB and striking position on the target is monitored, and the associated uncertainty with proton delivery is estimated to be a 2\% normalization uncertainty. The hadronic production concerns the production of secondary particles ($\pi^{\pm}$, $K^{\pm}$, $K^0_{L}$ which decay to produce neutrinos. Hadronic interaction cross-sections impact the rate of proton-Be interaction and the rate of pion absorption in the target and the horn, which ultimately affects the rate and shape of the neutrino flux. Uncertainty on the horn magnetic field modeling comes from two resources: the intrinsic variation in the horn current, and uncertainty in the modeling of the current within the inner cylinder in the horn. 

Tools and techniques developed by MiniBooNE~\cite{BNB_flux_MiniBooNE} are adopted to evaluate flux uncertainty in MicroBooNE. The ``multisim'' technique is employed for flux uncertainty evaluation where multiuniverses are generated and, in each universe, events are reweighted independently based on neutrino parentage, neutrino type, and energy. 

\subsection{Neutrino Cross-section Uncertainty}
The neutrino interaction cross-section uncertainties arise from uncertainties in the model parameters used in \textsc{genie} prediction. This includes uncertainty in NC resonant production, branching ratio and angular uncertainty on NC $\Delta$ radiative decay, uncertainties on CC quasielastic, CC resonance as well as other NC and CC interactions, and final state interactions. More details on the treatment of cross-section parameter uncertainties can be found in Ref.~\cite{uB_genie_tune}. 

Simulated events are reweighed using \textsc{genie} reweighting tools when model parameters are varied. For cross-section uncertainty evaluation, there exist universes resulting from single parameter variation for certain model parameters as well as universes with multiple parameters varied simultaneously, which allows incorporating correlations between model parameters. Due to technical limitations\footnote{While in \textsc{genie} \texttt{\detokenize{v3.0.6}} reweighing tools allow varying two parameters relevant for coherent $\pi^0$ production, the implementation of Berger-Sehgal coherent $\pi^0$ model leads to incompatibility with those weight calculators.}, the NC and CC coherent $\pi$ production cross-sections are assigned with $\pm 100$\% normalization uncertainty.

\subsection{Hadron-argon Reinteraction Uncertainty}
Hadron-argon reinteractions refer to the scattering of charged hadrons off the argon nuclei through hadronic interactions during propagation, which could significantly bend the particle trajectory or lead to additional particle production. \textsc{geant4} is used to simulate hadrons propagating through the detector medium~\cite{geant4_hadron}, and events are reweighed independently for $\pi^{+}$, $\pi^{-}$, proton interactions via \textsc{Geant4Reweight} tool~\cite{geant4_reweight}. 

\subsection{Detector Uncertainty}
The detector systematic uncertainty stems from the difference between detector simulation and the actual detector response.  There are three main different types of uncertainties: 1) uncertainty in TPC waveform from the electronic response simulation, 2) uncertainty in the light yield simulation, and 3) other detector effects concerning the distribution of drifting electrons. 

MicroBooNE employs a novel data-driven technique to evaluate the uncertainty on TPC wire waveforms, by parameterizing the differences between simulated cosmic and observed cosmic data at the level of Gaussian hits after waveform deconvolution~\cite{det_err1}. This avoids heavy computation involved in detector simulation and signal deconvolution and allows the detector uncertainty to be evaluated in a model-agnostic way. Four variations as functions of x position (drifting direction), yz position (vertical and beam direction), and the angular variables $\theta_{XZ}$ and $\theta_{YZ}$ for the particle trajectory, respectively, are considered. 

Variations associated with the light yield consider an overall 25\% reduction in the light yield, variation in the attenuation of light yield, and changes in light yield due to Rayleigh scattering length variations. Other detector effects, including the space charge effect due to the buildup of argon ions~\cite{uB_Efield_measurement,SCE}, and variations in the ion recombination model~\cite{electron_recomb_model} are accounted for through additional variations. 

Independent central value (CV) MC simulations, apart from the samples used in training the BDTs and making the final predictions, are generated together with the corresponding detector variations discussed above. The detector CV and variation samples are run through the reconstruction and analysis selection chain before being compared for detector uncertainty evaluation. 

\subsection{Uncertainty Evaluation}
The systematic uncertainties are encapsulated in covariance matrices, which describe the covariance across bins in a predicted spectrum or across different predicted spectra. One covariance matrix is constructed for each systematic error source and is built via:
\begin{equation}
    M_{ij} = \frac{1}{N}\sum_{k = 1}^{N}(n_i^k - n_i^{CV})(n_j^k - n_{j}^{CV})
\end{equation}
where $M_{ij}$ is the resulting covariance matrix describing the covariance between bin $i$ and bin $j$ in the predicted spectrum, $N$ is the number of universes generated for a given systematic error source, $n_i^k$ is the prediction at bin $i$ in the $k$th universe and $n_i^{CV}$ is the prediction at the same bin by the CV simulation. For reweighable systematics, the $n^k$ spectrum is calculated by reweighing simulated events in the selection, and for the detector systematics, the $n^k$ spectrum is the spectrum of the variation sample after applying analysis-specific requirements discussed in previous sections. 

\begin{figure}[h!]
    \centering
    \includegraphics[width=0.6\linewidth]{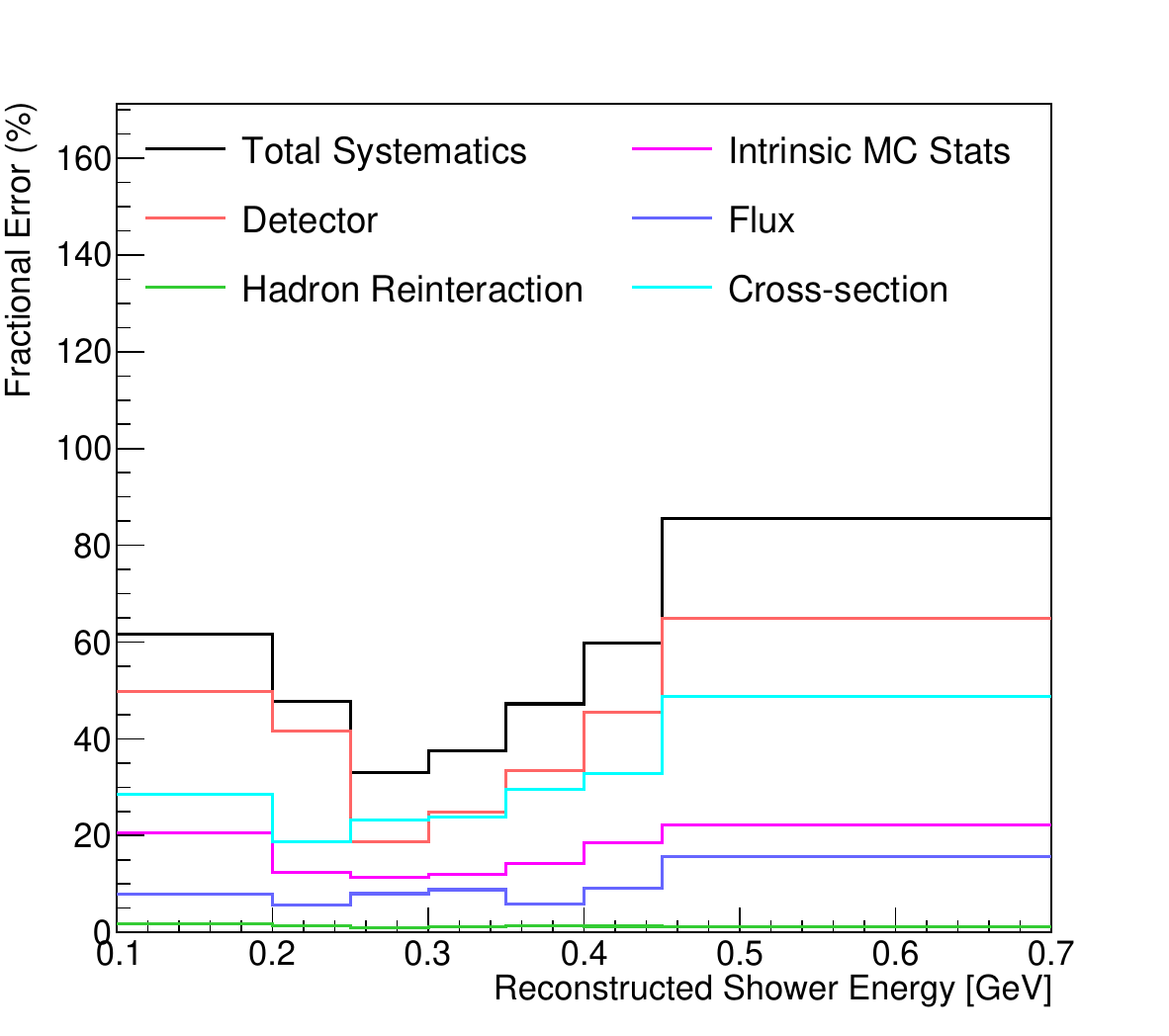}
    \caption[Overall systematic uncertainty for NC coherent $1\gamma$ analysis at the semi-final stage.]{Overall systematic uncertainty for selection at the semi-final stage, as a function of the reconstructed shower energy. Contribution from each source is highlighted in different colors. }
    \label{fig:semi_final_error_budget}
\end{figure}

\begin{figure}[h!]
    \centering
    \includegraphics[width=0.6\linewidth]{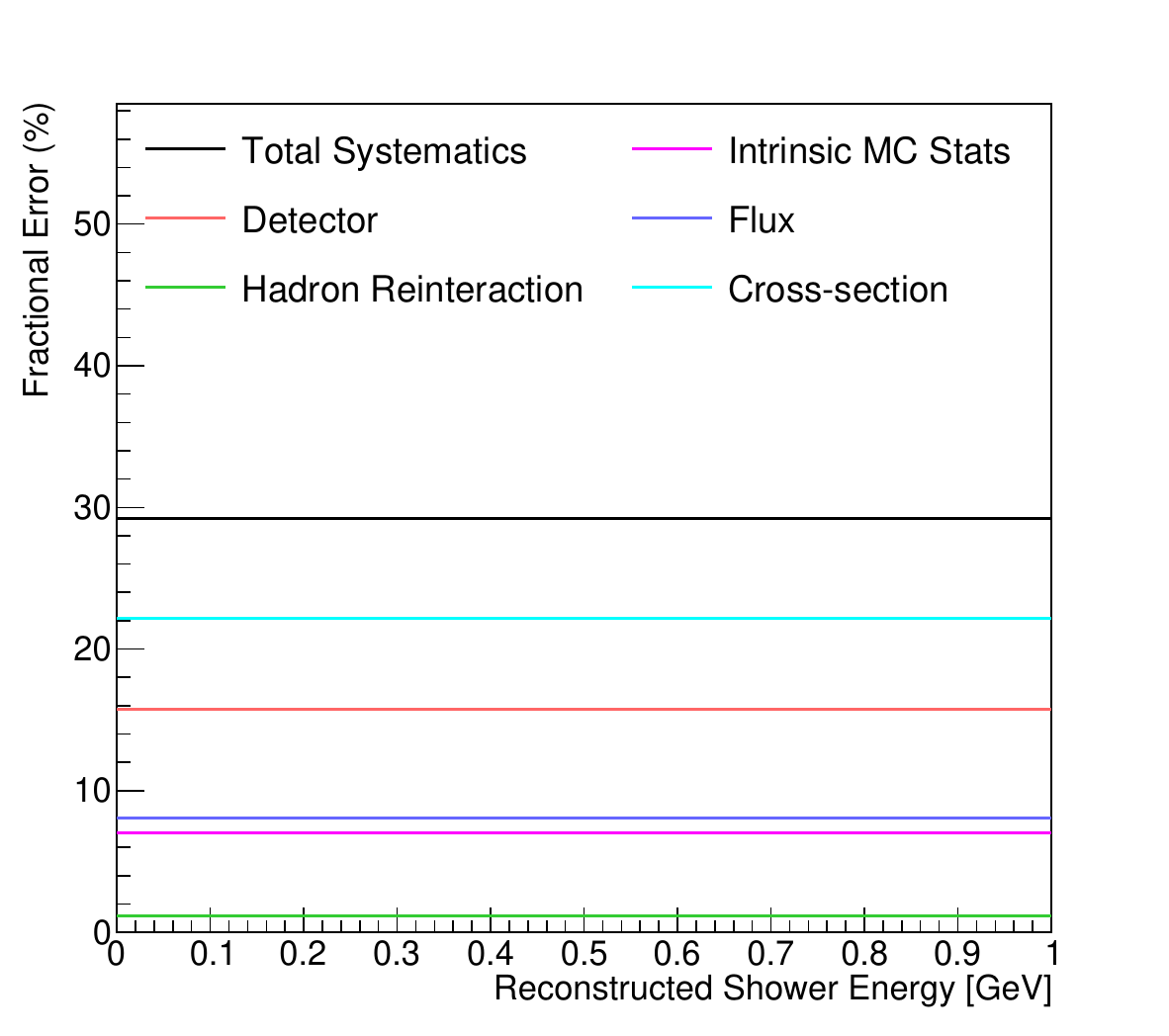}
    \caption[Overall systematic uncertainty for NC coherent $1\gamma$ analysis at final selection.]{Overall systematic uncertainty for the final selection, with contribution from each source highlighted in different colors. Driven by the low statistics available at the final stage, a single normalization bin is used. }
    \label{fig:final_error_budget}
\end{figure}

The breakdown of uncertainty contributions into the four main categories for the semi-final selection is shown in Fig.~\ref{fig:semi_final_error_budget} as a function of the reconstructed shower energy. The dominant uncertainty contribution at the semi-final stage is the uncertainty in detector modeling, followed by the uncertainty in cross-section modeling. Figure~\ref{fig:final_error_budget} shows the size of systematic errors for the final selection; cross-section and detector uncertainty remain the two largest contributions.

\section{NC 1$\pi^0$ constraint}\label{sec:pi0_constraint}
The goal of this analysis is to search for the NC coherent $1\gamma$ process from the observed data. This is quantified by a normalization scaling factor $x$ of the nominal SM predicted rate for this process. Extraction of parameter $x$ and its range is achieved by fitting the MC prediction to the observed data while allowing the scaling factor $x$ to vary. In order to yield higher sensitivity to parameter $x$, external high-statistic measurements are used in the fit to constrain the predicted background and systematic uncertainty in the signal region, in a similar manner to the NC $\Delta$ analysis in Ch.~\ref{ch3:ncdelta_LEE} and other MicroBooNE's LEE analyses~\cite{uboone_WC_eLEE, uboone_pelee, uboone_DLEE, uboone_combined_eLEE}.

The NC $1\pi^0$ selections, originally developed in the NC $\Delta\rightarrow N\gamma$ analysis for background validation and constraint, are used to constrain the $\pi^0$ background in this analysis as well, with recent updates on the analysis selection requirements and $\approx$20\% more data available for the Runs 1-3~\cite{MOU_epem}. 
\begin{figure}[h!]
    \centering 
    \begin{subfigure}{0.49\textwidth}
        \includegraphics[width = \textwidth,trim=0cm 1.5cm 0cm 0cm, clip]{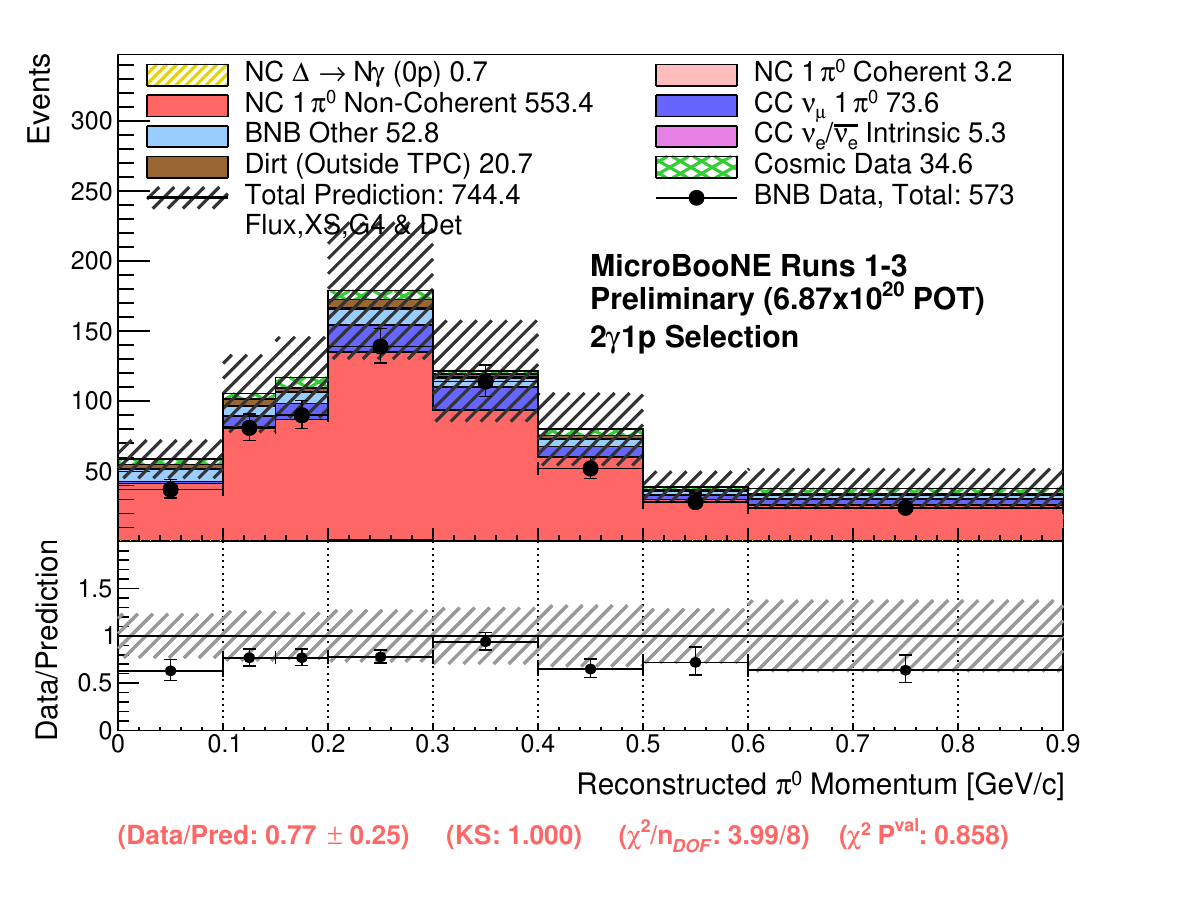}
        \caption{$2\gamma1p$}
  \end{subfigure} 
  \begin{subfigure}{0.49\textwidth}
        \includegraphics[width = \textwidth,trim=0cm 1.5cm 0cm 0cm, clip]{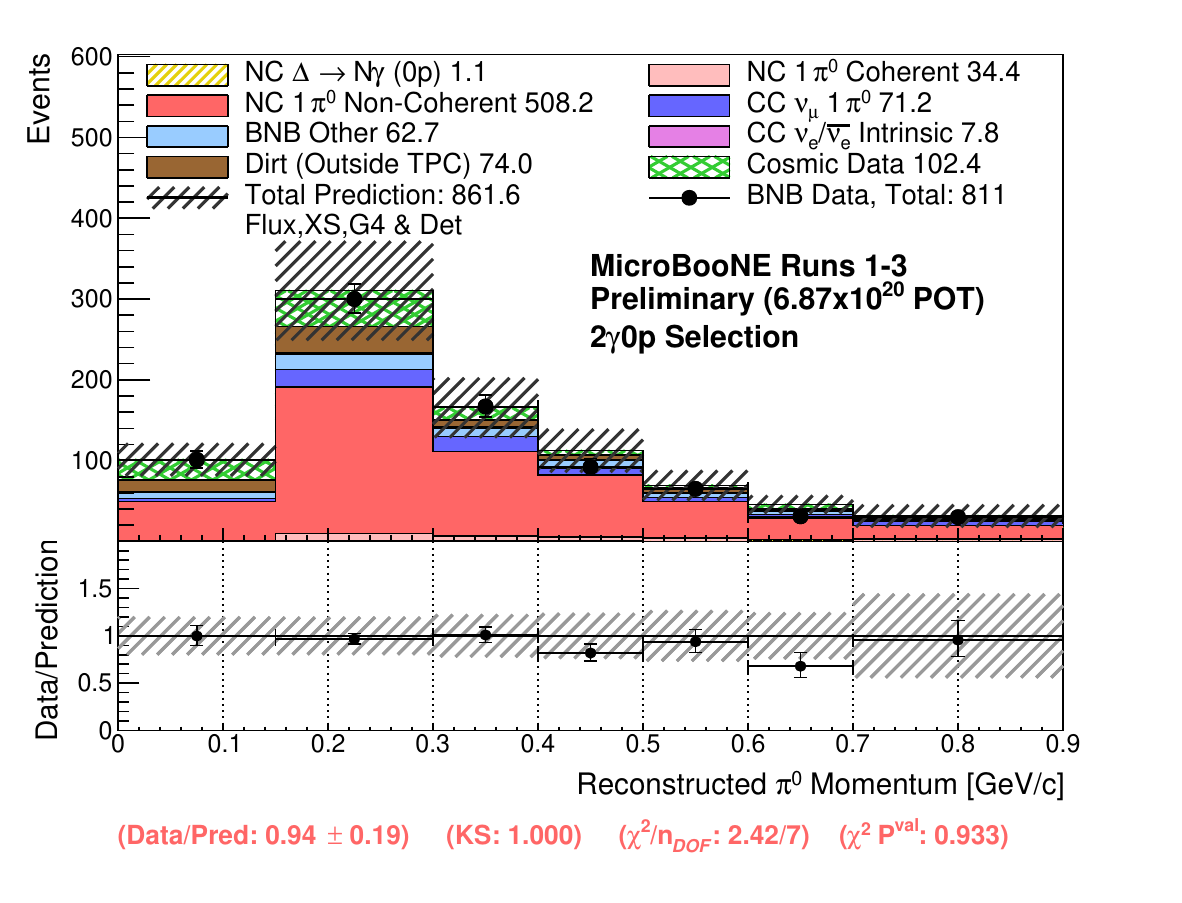}
        \caption{$2\gamma0p$}
  \end{subfigure} 
  \caption[Distributions of the $2\gamma1p$ and $2\gamma0p$ constraining samples in reconstructed $\pi^0$ momentum.]{Distributions of the $2\gamma1p$ and $2\gamma0p$ constraining samples in reconstructed $\pi^0$ momentum, shown in binning used in the simultaneous fit. A mild deficit in data is observed in the $2\gamma1p$ sample, but in general, good data-MC agreement is seen within systematic uncertainty.}
  \label{fig:2g}
\end{figure}

\begin{figure}[h!]
    \centering
    \includegraphics[width=0.6\linewidth, trim=0cm 0cm 0cm 1cm,clip]{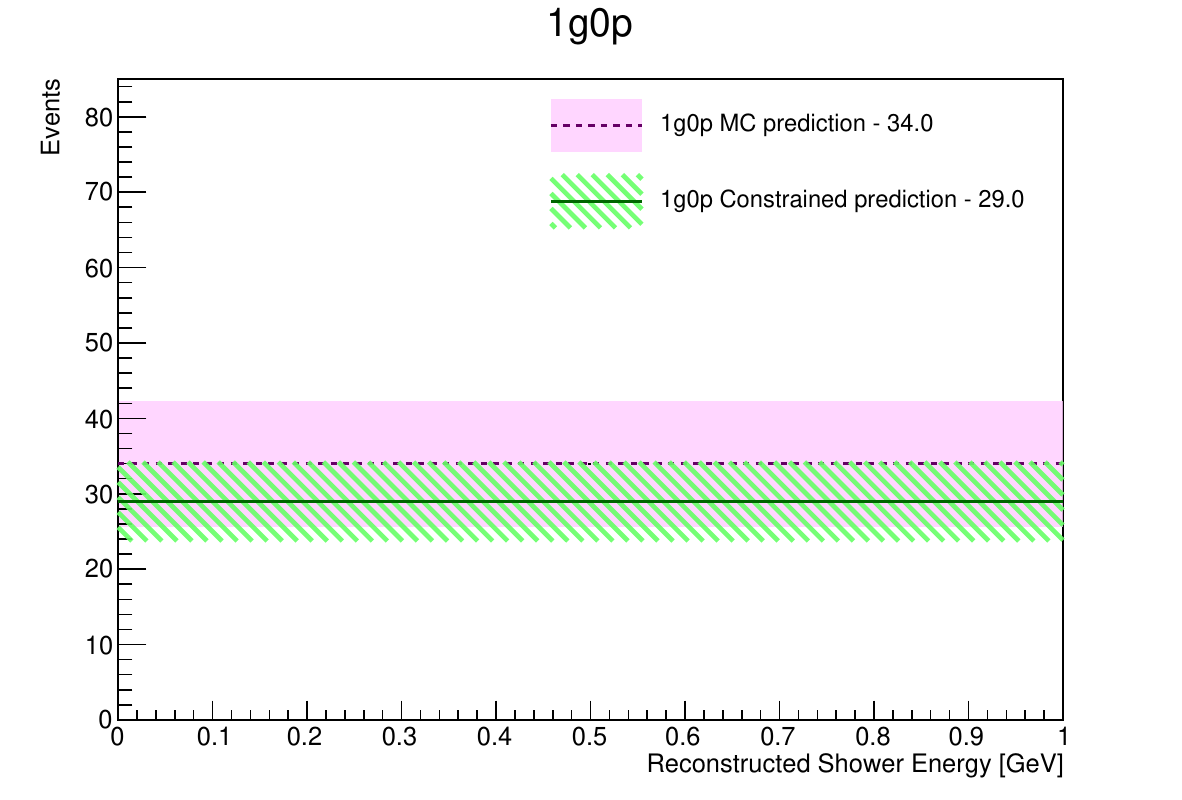}
    \caption[Final distribution of reconstructed shower energy in the signal sample with full systematic uncertainty, before and after the $2\gamma$ constraint.]{Final distribution of reconstructed shower energy in the signal sample with full systematic uncertainty, before and after the conditional constraint from $2\gamma$ selections are applied. The red dashed line and the pink band represent the MC CV prediction and full systematic error; the green solid line and green band represent the prediction and systematic uncertainty after applying the constraint from high-statistic $2\gamma1p$ and $2\gamma0p$ samples.}
    \label{fig:2g_constraint}
\end{figure}

Three samples (signal selection, $2\gamma1p$ and $2\gamma0p$ selection) are fitted simultaneously in order to maximize the sensitivity to the NC coherent $1\gamma$ process. 
The distributions of the $2\gamma$ samples used in the fit are shown in Fig.~\ref{fig:2g}. The effect of the constraint from the $2\gamma$ samples is illustrated in Fig.~\ref{fig:2g_constraint} following the conditional constraint approach~\cite{conditional_constraint} described in Sec.~\ref{ch3:glee_fit}. The 2$\gamma$ constraint reduces the MC prediction from 34.0 to 29.0 for the signal region with a 25\% reduction in the systematic uncertainty.

\section{Results with Runs 1-3 Data}

\subsection{Sideband}\label{sec:sideband}
The analysis is designed as a blind analysis, where 10\% of the total data during the first run periods is open and utilized to develop the analysis and validate the agreement between data and MC in variables of interest and input variables to all BDTs involved. The backgrounds involved in this analysis are similar to those of the NC $\Delta\rightarrow N\gamma$ analysis, where the NC $\pi^0$ and BNB Other backgrounds with single shower topology are validated~\cite{gLEE_internal_note}. The performance of the shower energy reconstruction and validation of NC $\pi^0$ modeling are demonstrated in the $2\gamma1p$ and $2\gamma0p$ NC $\pi^0$ measurements (as in Secs.~\ref{ch2.1:energy_corr} and~\ref{ch3:glee_2p_rate}). 

To validate the performance of the PSV BDT, a sideband that is proton-rich is isolated by reversing the cut on the NC $\pi^0$ BDT while relaxing cuts on other event-level BDTs. A total of 850.0 events are predicted, of which 54.0\% events are predicted to have protons exiting the nucleus (regardless of proton KE). 
Consistency between data and MC is inspected through the goodness-of-fit test that incorporates systematic uncertainties and Combined-Neyman-Pearson statistical uncertainty~\cite{CNP}. 

\begin{figure}[hbt!]
    \centering 
    \begin{subfigure}{0.55\textwidth}
        \includegraphics[width = \textwidth,trim=0cm 1.5cm 0cm 0cm, clip]{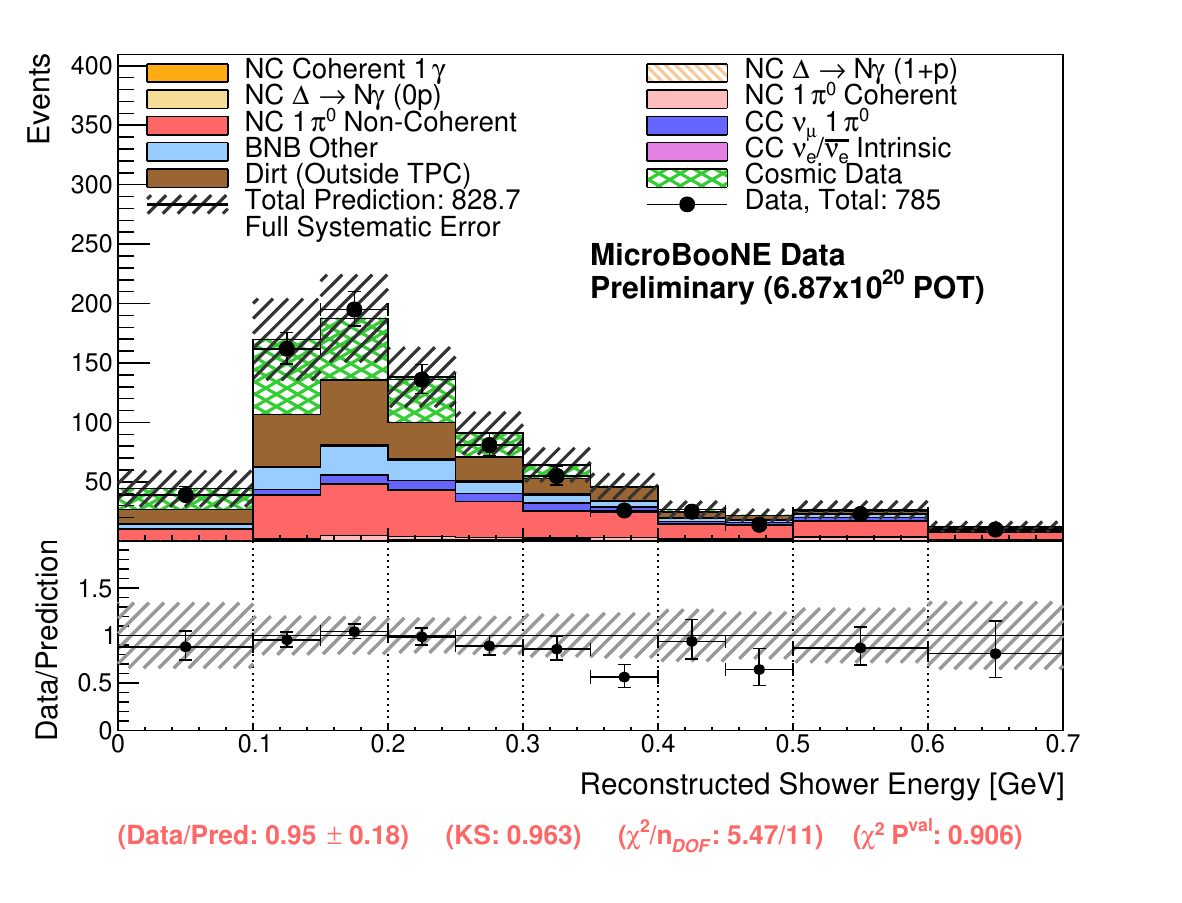}
        \caption{Reconstructed Shower Energy}
        \label{fig:cosmic_bdt}
  \end{subfigure} 
  \begin{subfigure}{0.55\textwidth}
        \includegraphics[width = \textwidth,trim=0cm 1.5cm 0cm 0cm, clip]{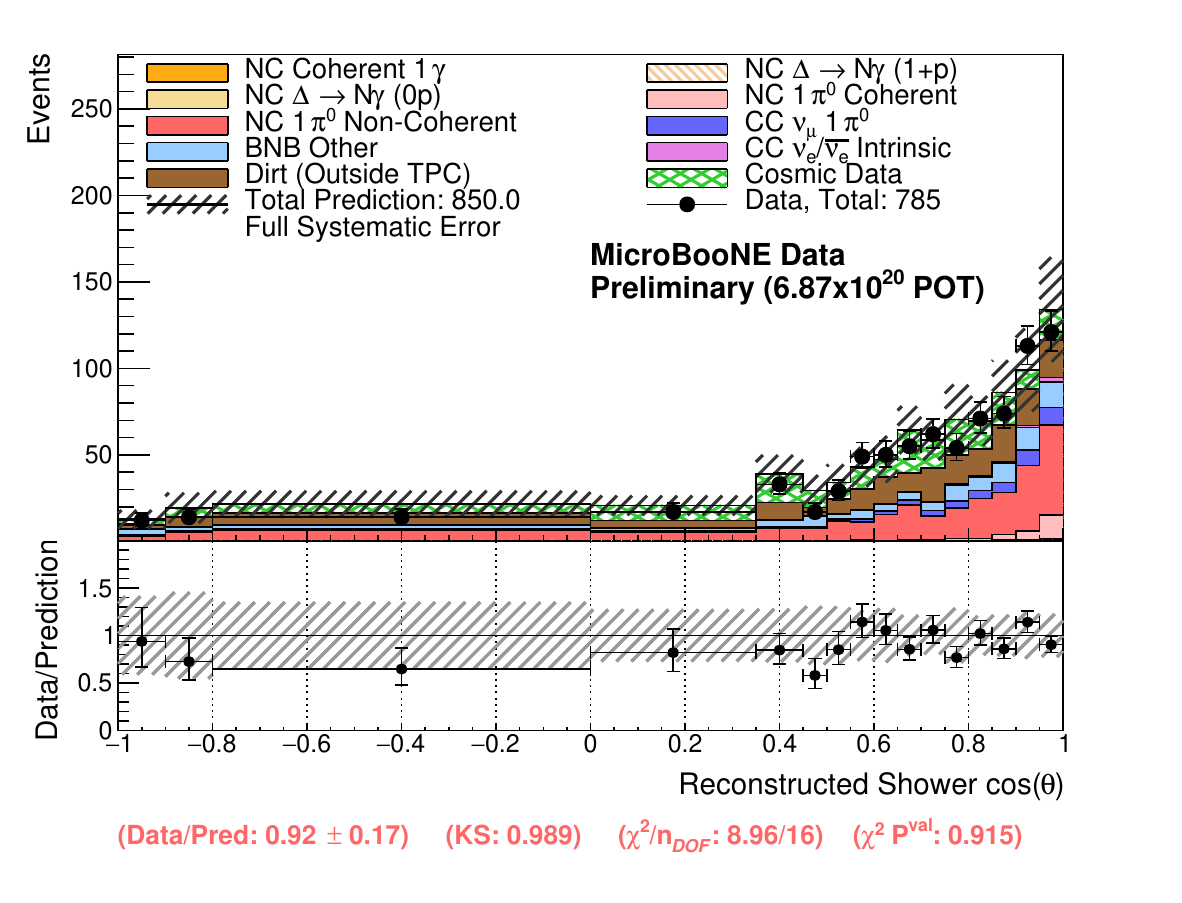}
        \caption{Reconstructed Shower cos($\theta$)}
  \end{subfigure} 
  \begin{subfigure}{0.55\textwidth}
        \includegraphics[width = \textwidth,trim=0cm 1.5cm 0cm 0cm, clip]{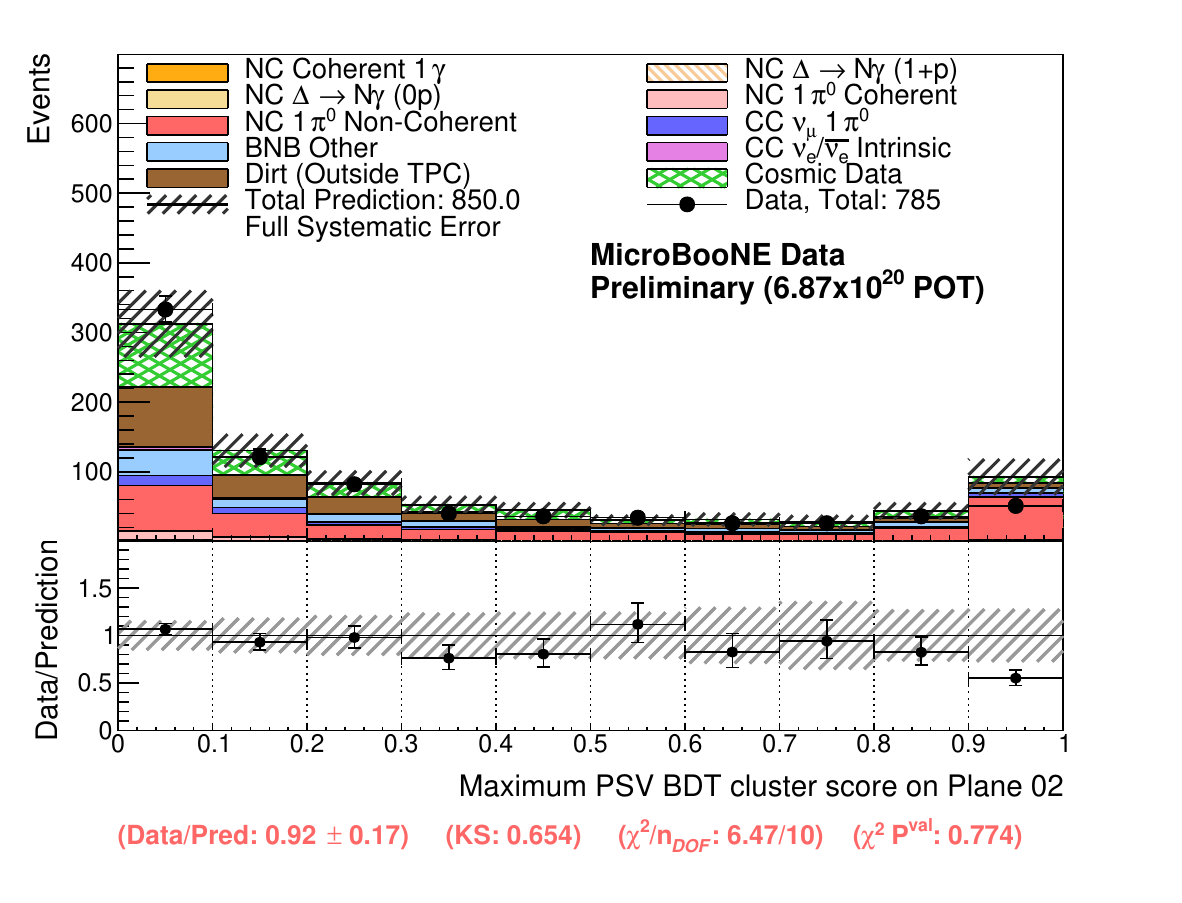}
        \caption{Maximum PSV score on Plane 02}
  \end{subfigure} 
  \caption[Distributions of the sideband sample in: (a) reconstructed shower energy; (b) reconstructed shower cos($\theta$); and (c) the maximum PSV score on Plane 02.]{Distributions of the sideband sample in: (a) reconstructed shower energy; (b) reconstructed shower cos($\theta$); and (c) the maximum PSV score on Plane 02. A mild data deficit is observed in the highest PSV score bin, but data is found to be consistent with prediction within systematic uncertainty over all three variables.}
  \label{fig:sideband}
\end{figure}
Figure~\ref{fig:sideband} shows the distributions of the reconstructed shower energy, angle, and the maximum PSV score on Planes 02 for the sideband sample. A mild deficit in data is observed in the region of high PSV score; the subsample of events with high PSV score (above 0.8) is isolated and is found to be flat in kinematic variables~\cite{coherent_sp_TN}. Overall, the data-MC agreement for the PSV variable is good, with corresponding $p^{\text{value}} = 0.78$ when full systematic uncertainty is taken into account. Consistency in event-level and cluster-level variables including all BDT input variables is inspected, and good agreement is observed within systematic uncertainty.

\subsection{Semi-final and Final selection Results}\label{sec:final_results}
\begin{figure}[hbt!]
    \centering 
    \begin{subfigure}{0.55\textwidth}
        \includegraphics[width = \textwidth,trim=0cm 1.5cm 0cm 0cm, clip]{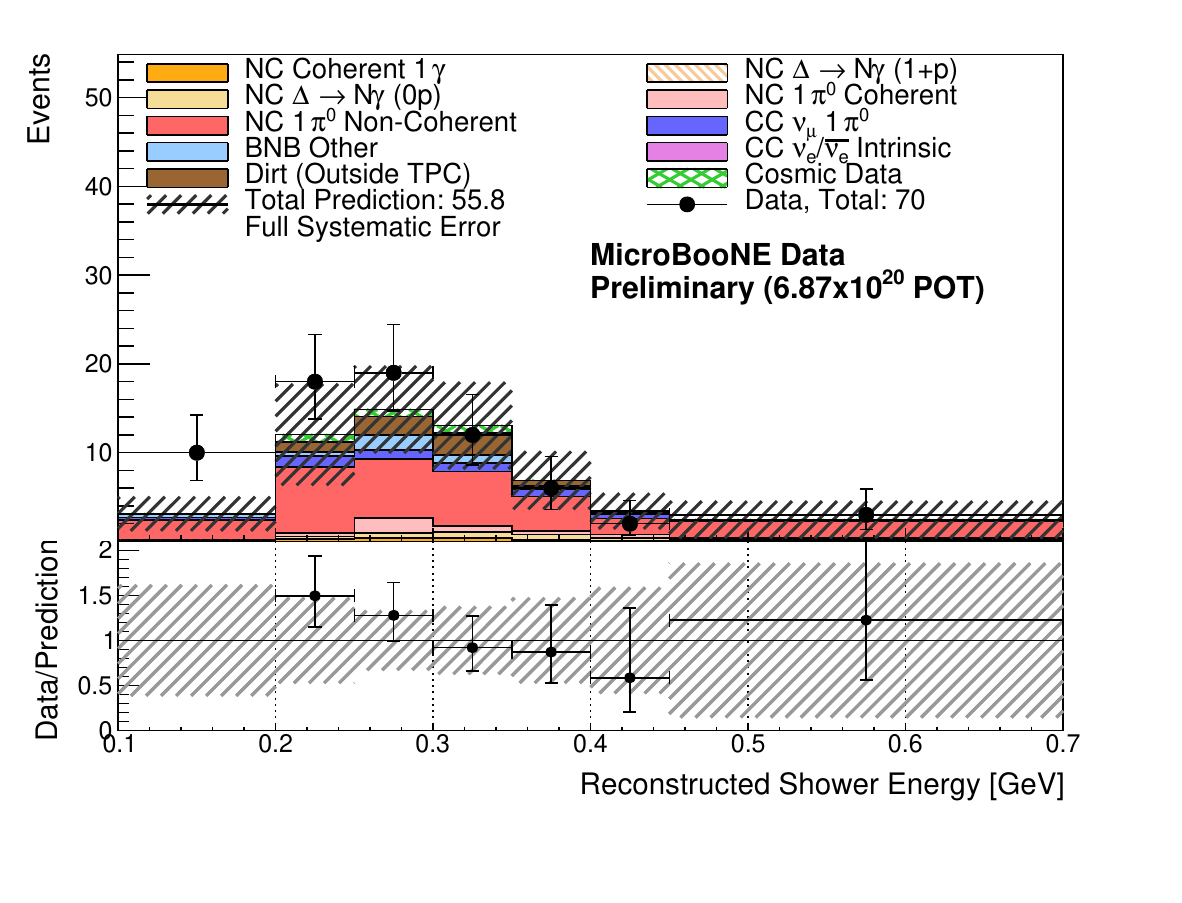}
        \caption{Reconstructed Shower Energy}
        \label{fig:cosmic_bdt}
  \end{subfigure} 
  \begin{subfigure}{0.55\textwidth}
        \includegraphics[width = \textwidth,trim=0cm 1.5cm 0cm 0cm, clip]{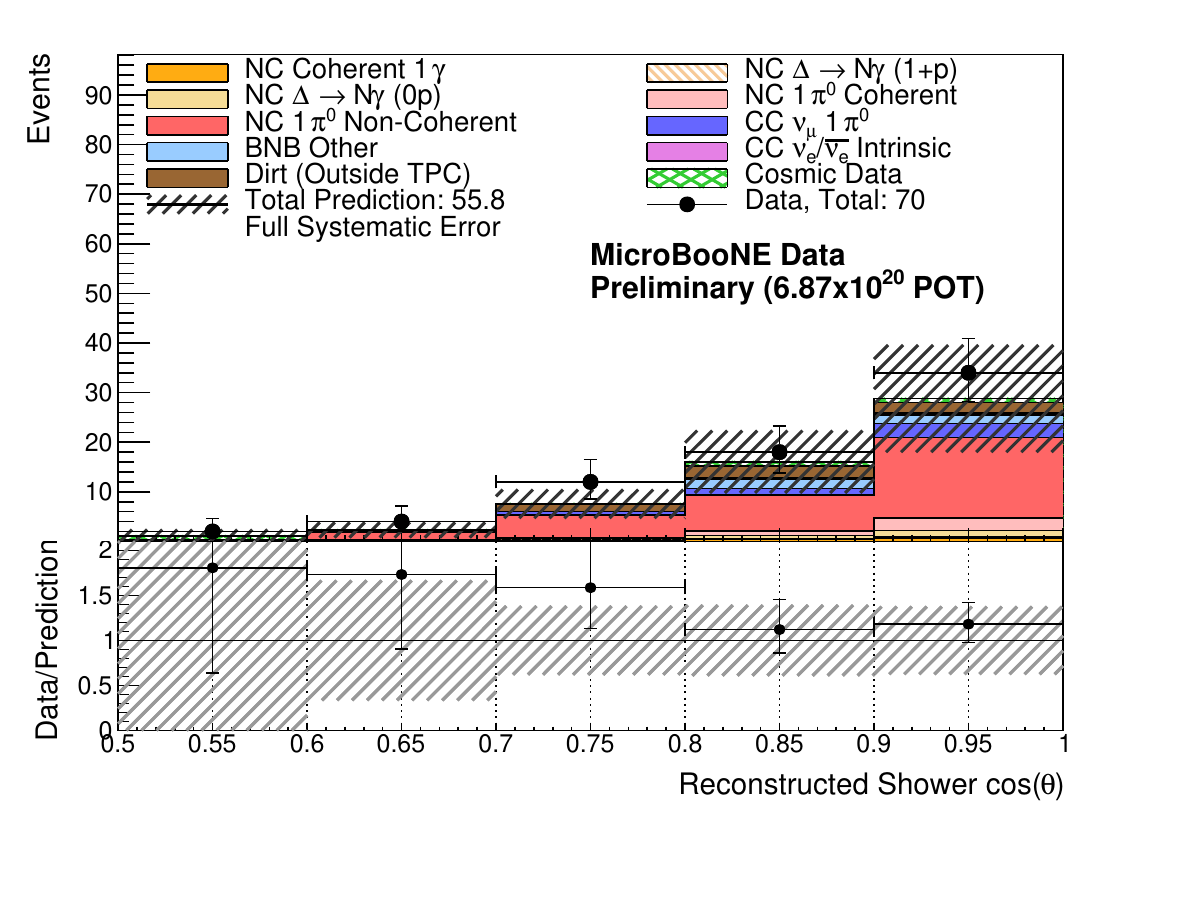}
        \caption{Reconstructed Shower cos($\theta$)}
  \end{subfigure} 
  \begin{subfigure}{0.55\textwidth}
        \includegraphics[width = \textwidth,trim=0cm 1.5cm 0cm 0cm, clip]{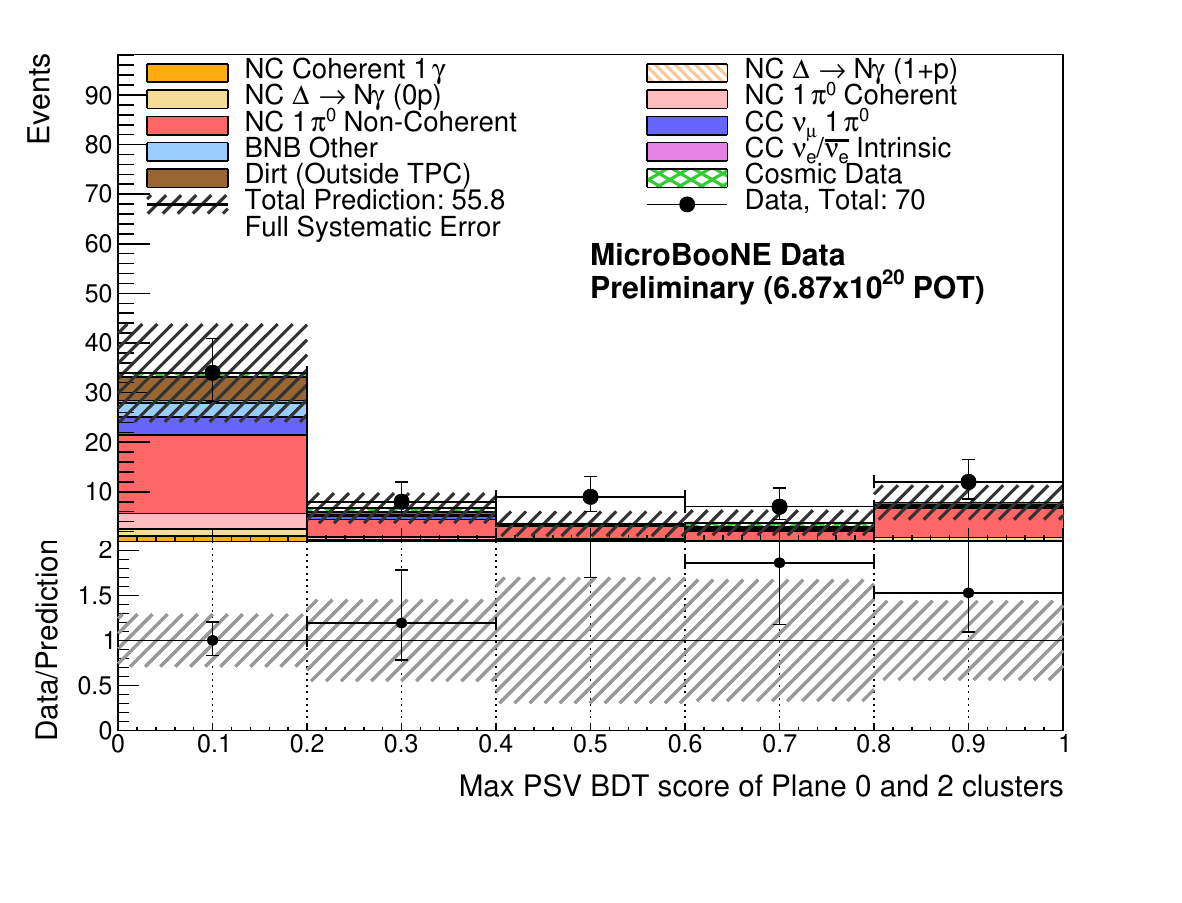}
        \caption{Maximum PSV score on Plane 02}
  \end{subfigure} 
  \caption{Distributions of selected semi-final sample in: (a) reconstructed shower energy; (b) reconstructed shower cos($\theta$); and (c) the maximum PSV score on Plane 02.}
  \label{fig:semi_final_data}
\end{figure}

For the semi-final selection, 70 events are observed in data compared to an MC prediction of 55.8, leading to a data-MC ratio of $1.25\pm0.32 (\text{sys})$. Distributions of the reconstructed shower energy, angle (cos$\theta$) and the maximum PSV score on Planes 02 are shown in Fig.~\ref{fig:semi_final_data}. A data excess is observed in the reconstructed shower energy region [100, 300] MeV, especially for showers with energy less than 200 MeV. The significance of the excess in the first energy bin is calculated to be 2.64$\sigma$ (2.78$\sigma$) before (after) the $2\gamma$ constraint. Hand scanning of data events in this bin suggests that all selected events are good showers, with 4 out of 10 containing visible second shower, suggesting a $\pi^0$ origin. Overall, the agreement between data-MC is good according to the goodness-of-fit tests. 

\begin{figure}[hbt!]
    \centering
    \includegraphics[width=0.6\linewidth, trim=0cm 0cm 0cm 1cm,clip]{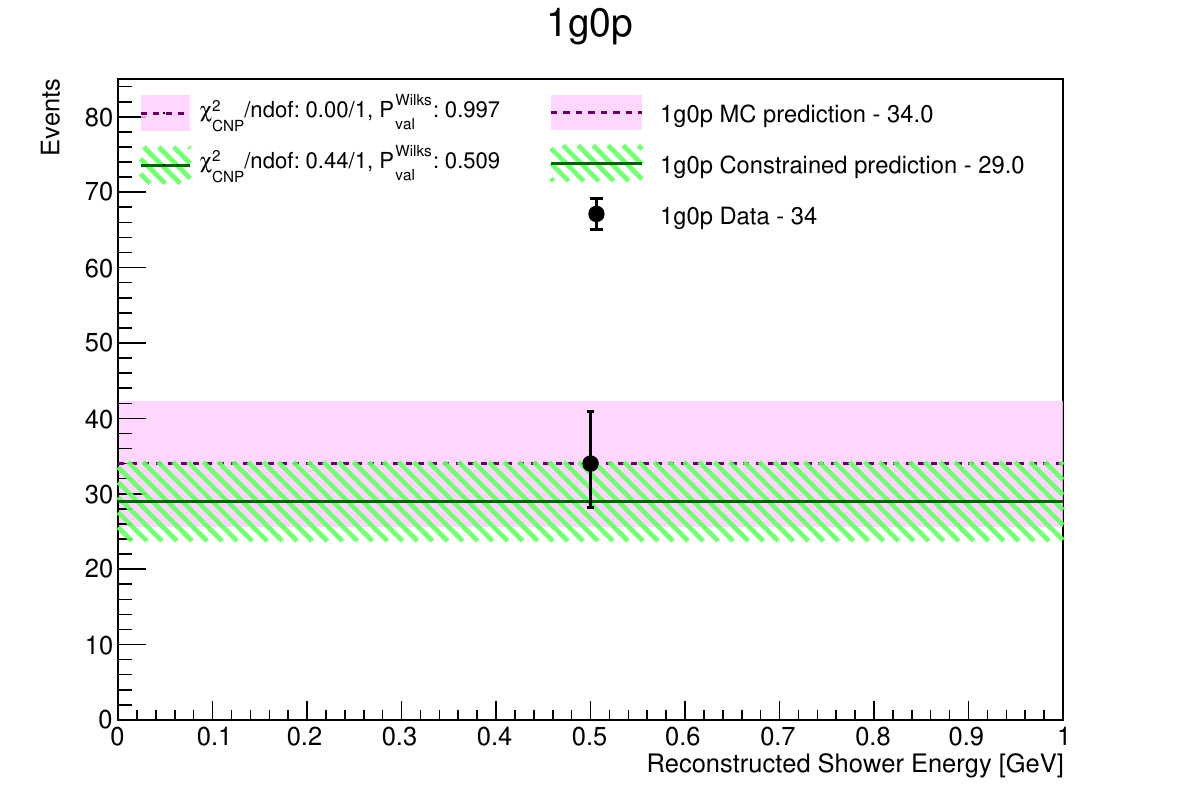}
    \caption[Comparison of observed data and MC CV predictions for the final selection.]{Comparison of observed data and MC CV predictions for the final selection. The red (green) line and surrounding band represent the MC prediction and associated systematic uncertainty before (after) the $2\gamma$ constraint is applied.}
    \label{fig:final_data}
\end{figure}

For the final selection, 34 data events are observed, compared to a constrained prediction of $29.0 \pm 5.2 \text{(syst)} \pm 5.4 \text{(stats)}$ and an unconstrained prediction of $34.0 \pm 8.4 \text{(syst)} \pm 5.8 \text{(stats)}$. The comparison between data and MC together with results from goodness-of-fit tests are shown in Fig~\ref{fig:final_data}. Data-MC agreement slightly worsens after the constraint from $2\gamma$ channels is applied, from $p^{\text{value}} = 1.00$ to $p^{\text{value}} = 0.51$.

\begin{figure}[hbt!]
    \centering
    \includegraphics[width=0.6\linewidth,trim=10cm 0cm 0cm 0cm, clip]{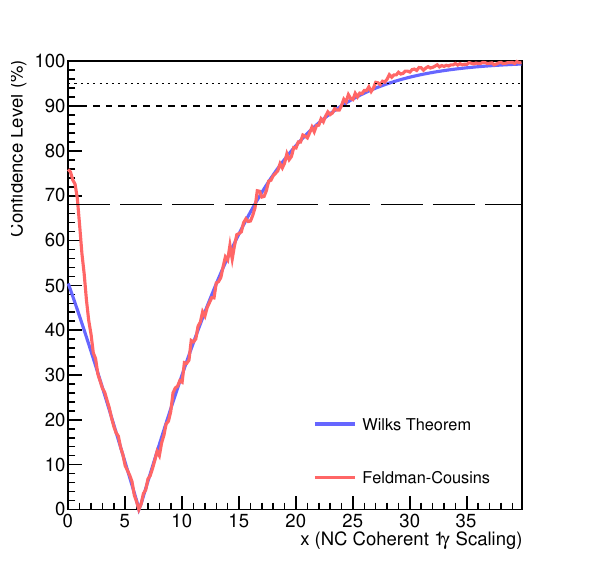}
    \caption[The confidence level on NC coherent $1\gamma$ normalization scaling factor $x$ derived from observed data in the final selection.]{The confidence level on NC coherent $1\gamma$ normalization scaling factor $x$ derived from observed data in the final selection. The C.L. evaluated with the Feldman-Counsins method is shown in pink with C.L. assuming the validity of Wilks' theorem overlaid in purple.}
    \label{fig:FC_corrected_fit}
\end{figure}

The observed data in the final selection is fit to extract a normalization scaling factor for NC coherent $1\gamma$ process $x$, with full systematic uncertainty except the GENIE cross-section uncertainty for the signal process. The best-fit for $x$ is found to be $x = 6.20$ with $\chi^2/ndf = 9.7/15$. Figure~\ref{fig:FC_corrected_fit} shows the confidence level (C.L.) for different values of $x$ as is evaluated using data and the Feldman-Counsins (FC) frequentist method~\cite{feldman_cousins}, with the C.L. under the assumption of Wilks' theorem overlaid. Curves from FC and Wilks theorem agree with each other in the majority of the parameter space and start to differ in the parameter region near the boundary, which is expected. Given the observed data, the bound at $90\%$ C.L. is placed at $x=24.0$ with $95\%$ C.L bound at $x=28.0$ for the NC coherent $1\gamma$. Note that the noisy confidence level curve for the FC method is due to the insufficient amount of pseudo-experiments used. 

\clearpage
\section{Summary}
This chapter has presented the world's first search for the neutrino-induced NC coherent single photon process with the first three years of data from the MicroBooNE detector. Differences in the shower kinematic properties between the shower originating from coherent photons and that from other backgrounds are leveraged to yield effective background removal, leading to 99.97\% and $>95$\% rejection efficiency on the cosmic background and NC non-coherent 1$\pi^0$ relative to topological selection. Furthermore, tools utilizing low-level information are developed to identify and reject low-energy proton traces near the vertex, which further reduces the dominant NC non-coherent 1$\pi^0$ background by 48\%. This search yields a signal-to-background ratio of $\sim 1:30$ in the final selection, and good agreement between data and MC is observed with MicroBooNE Runs 1-3 data. This leads to the world's first experimental limit on the cross-section of neutrino-induced NC coherent single photon production on argon nucleus below 1 GeV, of $1.49 \times 10^{-41} \text{cm}^2$ at 90\% CL, corresponding to 24.0 times the prediction in Ref.~\cite{coh_gamma_model}.

The proton veto tool developed in this analysis has shown great potential in rejecting low-energy protons and can be easily adapted to other coherent interaction searches. Furthermore, it indicates the way forward for enhancing the reconstruction of low-energy particles. While this search has limited sensitivity, the selections developed can be applied to future LArTPC experiments for a search of this process with high statistics. An example of this is the upcoming SBND experiment which is expected to collect $\mathcal{O}(20)$ times more data~\cite{SBND} than MicroBooNE, bringing the limit on this process to a few times of the SM predicted rate.

%% file: Conclusion.tex
\begin{center}
\pagebreak
\vspace*{5\baselineskip}
\textbf{\large Conclusion}
\end{center}

This thesis presents MicroBooNE's searches for anomalously large production of neutrino-induced NC single photon events, specifically NC $\Delta$ production followed by $\Delta \rightarrow N\gamma$ decay, and NC coherent single photon production. Both searches make use of the first three years of MicroBooNE data, and yield the world-leading constraints on the two processes. 

The NC $\Delta \rightarrow N\gamma$ decay is one of the dominant photon backgrounds at low energy in MiniBooNE's $\nu_{e}$ measurement. 
A model-dependent search for NC $\Delta \rightarrow N\gamma$ decay is performed. Two selections are developed and topological and kinematic properties of final state particles from the NC $\Delta \rightarrow N\gamma$ decay are leveraged to achieve excellent signal-background separation. Additional, independent high-statistics NC $\pi^0$ measurements are developed, and utilized to validate and constrain the dominant $\pi^0$ background as well as systematic uncertainties involved in this analysis. With Runs 1–3 data, good consistency is observed between data and simulation, which leads to a bound on the normalization $x_{\Delta}$ of the nominally predicted SM NC $\Delta$ radiative decay rate of $x_{\Delta} < 2.3$ times at the 90\% CL. This is more than a 50-fold improvement over the bound on single-photon production in the sub-GeV neutrino energy range from the T2K experiment~\cite{T2K_NC1gamma}. The measurement disfavors the candidate photon interpretation of the MiniBooNE LEE as a factor of $3.18$ times the nominal NC $\Delta$ radiative decay rate at the 94.8\% CL, while favoring the nominal prediction. 

The NC coherent single photon process contributes subdominantly to the total NC single photon production. It is an extremely rare SM-predicted process with a predicted rate less than $\frac{1}{10}$ of the NC $\Delta \rightarrow N\gamma$ decay rate. Due to the limited handles available for signal-background differentiation, from the single-shower-only topology of the signal, the search for NC coherent $1\gamma$ has limited sensitivity, with a final signal-to-background ratio of $1:30$. Good agreement between data and simulation is observed with Runs 1–3 data, and a world-leading limit on the cross-section of this process is placed at $1.49 \times 10^{–41} \text{cm}^2$ at 90\% CL, corresponding to 24.0 times the SM predicted cross-section.

For possible future improvements, since both analyses are statistics-limited, improved sensitivities are expected with MicroBooNE's full five years of data. Future LArTPC experiments that expect much higher statistics should offer better constraints of these processes, such as the SBND experiment~\cite{SBND}. The two leading systematic uncertainties in both analyses are detector systematics and neutrino background cross-section uncertainty. Thus, precise neutrino interaction measurements, a better understanding and a more complete and consistent implementation of model parameters should significantly improve the sensitivity. The current approach for the detector systematic uncertainty evaluation in MicroBooNE faces the issue of limited statistics in the simulated detector variation samples, especially for rare event searches like the ones presented in this thesis, thus leading to a conservative overestimation of the detector uncertainty. Therefore, increasing the statistics in the detector variations samples or improvements in the detector systematic evaluation approach may yield better modeling of the detector uncertainty and improve the sensitivity. 

Specifically focusing on the NC coherent $1\gamma$ search, it is seen that the majority ($\sim70$\%) of the signal is removed at the topological selection, almost half of which is due to reconstruction inefficiency. Thus improving the shower reconstruction efficiency (particularly for forward-going showers with energy of $\sim 300$ MeV) is critical in enhancing the search sensitivity in the future. The developed proton veto tool demonstrated high efficiency in identifying proton activity. Although the proton veto tool only identifies potential proton activity in the detector at this stage, it has the potential to be adopted in the reconstruction chain and assist in low-energy proton reconstruction in the future.

%% file: appendix.tex
\begin{appendices}

\addtocontents{toc}{\protect\renewcommand{\protect\cftchappresnum}{\appendixname\space}}
\addtocontents{toc}{\protect\renewcommand{\protect\cftchapnumwidth}{6em}}

\chapter{Possible Explanations for the MiniBooNE LEE}\label{app:miniboone_lee_interpretations}
This appendix discusses briefly some possible resolutions for the MiniBooNE LEE. For a more comprehensive overview please refer to Ref. ~\cite{whitepaper_sterile}.

\section{Underestimated Background}
As shown in Fig.~\ref{fig:MiniBooNE_data_mc_energy}, main contributions in the low energy region where the excess is pronounced can be divided into two categories: interactions of intrinsic $\nu_{e}$ in the beam and interactions that generate photon as final-state particle which is mis-reconstructed as $\nu_{e}$ CCQE events. It's possible that the excess observed is due to underestimated background in either or both categories. 

\begin{figure}[h!]
\centering
\includegraphics[page=7, trim = 1cm 19.5cm 11cm 1.6cm, clip=true, width=0.6\textwidth]{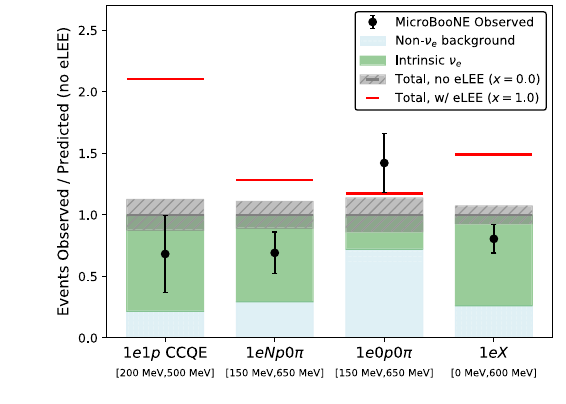}
\caption[Ratio of observed data to predicted $\nu_{e}$ candidate events in MicroBooNE analyses testing the electron hypothesis of MiniBooNE LEE.]{Ratio of observed to predicted $\nu_{e}$ candidate events in each analysis’s signal-enhanced neutrino energy range. The observed data-MC ratio and statistical error are shown in black. The grey solid line and dashed region shows the prediction and associated systematic uncertainty while relative contributions from non-$\nu_{e}$ backgrounds are shown in light blue and intrinsic $\nu_{e}$’s in green. 
The expected ratio assuming enhanced $\nu_{e}$ rate enough to explain MiniBooNE LEE is shown in red. Taken from Fig. 4 of Ref. \cite{uboone_combined_eLEE}}
\label{fig:MicroBooNE_eLEE}
\end{figure}

Regarding the intrinsic $\nu_{e}$ contribution, MicroBooNE has explicitly tested the hypothesis that there is anomalous $\nu_{e}$ rate in the neutrino beam, using three distinctive reconstruction techniques and targeting three different event topologies \cite{uboone_combined_eLEE, uboone_DLEE, uboone_pelee, uboone_WC_eLEE}. Figure \ref{fig:MicroBooNE_eLEE} is reproduced summarizing the ratio between observed data and MC prediction for different analyses with first three years of data. No excess of $\nu_{e}$ events is seen across all analyses with exception of $1e0p0\pi$ topology which has the lowest purity of $\nu_{e}$ interaction. Thus a consistent picture with the nominal $\nu_{e}$ rate expectation is seen, and the hypothesis that anomalous $\nu_{e}$ CC interaction is fully responsible for the LEE is rejected by MicroBooNE at $>97$\% confidence level for the three analyses with high $\nu_{e}$ interaction purity. 

Regarding the mis-reconstructed photon contribution, the dominant contribution is from NC $\pi^0$s, which has been constrained by MiniBooNE's \textit{in-situ} high-statistic NC $\pi^0$ measurements \cite{PhysRevD.81.013005}. The second highest contribution is from $\Delta$ resonance production and following $\Delta \rightarrow \text{N}\gamma$ radiative decay. $\Delta$ contribution is also constrained through its correlation with NC $\pi^0$ events in MiniBooNE. Dirt events are these originating outside the detector with final-state particles scattered in, and their contribution can be measured using MiniBooNE data \cite{MiniBooNE_dirt}. While NC $\Delta$ radiative decay is indirectly constrained, it has never measured directly and MiniBooNE's latest result on $\nu_{e}$ CCQE measurement shows that a flat scaling of 3.18 on the NC $\Delta$ radiative decay fits the shape and normalization of the LEE well \cite{MiniBooNE_combine_osc}. MicroBooNE has examined the hypothesis that the MiniBooNE LEE is caused by an enhancement of 3.18 on the NC $\Delta$ production and radiative decay making use of first three years of data \cite{uboone_delta_LEE}, and the observed data disfavors this hypothesis at 94.8\% confidence level while in favor of the nominal prediction. This is discussed in more details in Sec. \ref{ch3:ncdelta_LEE}.

\section{Sterile Neutrino Oscillation}
With MiniBooNE's mission being to investigate the $\Delta m^2 \sim \mathcal{O}(1) \text{eV}^2$ prefered by LSND result, it's natural to check if neutrino oscillation can provide sufficient description to MiniBooNE LEE and how MiniBooNE result can help constrain the parameter space of different oscillation models. 

As mentioned above MiniBooNE is a short-baseline experiment with $L/E \sim \mathcal{O}(1)$ m/MeV and is sensitive to oscillation with $\Delta m^2 \sim \mathcal{O}(1) \text{eV}^2$, the simplest model to accomodate this $\Delta m^2$ range is to introduce additional sterile neutrino heavier than the three SM neutrinos, which we refer to (3+1) model. In this model, the three SM neutrinos and sterile neutrino mix with four mass eigenstates, and the PMNS matrix is extended from $3\times 3$ to $4\times 4$. By assuming that the mass of $\nu_4$ is much larger than $\nu_1, \nu_2$ and $\nu_3$, for short-baseline experiment we can safely ignore the contribution from $\Delta m^2_{12}$ $\Delta m^2_{13}$ and $\Delta m^2_{23}$ during oscillation probability calculation, and the neutrino oscillation probability can be simplified to:
\begin{align}
    P(\nu_{\alpha} \rightarrow \nu_{\alpha}) &\simeq 1 - 4|U_{\alpha 4}|^2(1-|U_{\alpha 4}|^2)\sin^2(\frac{\Delta m^2_{41} L}{4E}) \equiv 1 - \sin^2(2\theta_{\alpha\alpha})\sin^2(\frac{\Delta m^2_{41} L}{4E})\\
    P(\nu_{\alpha} \rightarrow \nu_{\beta}) &\simeq 4|U_{\alpha 4}|^2|U_{\beta 4}|^2\sin^2(\frac{\Delta m^2_{41} L}{4E}) \equiv \sin^2(2\theta_{\alpha\beta})\sin^2(\frac{\Delta m^2_{41} L}{4E})
\end{align}
where $\alpha$, $\beta$ denote the weak flavor of neutrinos; $U_{\alpha 4}$ ($U_{\beta 4}$) denotes the mixing between $\nu_{\alpha}$ ($\nu_{\beta}$) and the fourth mass eigenstate $\nu_4$; $\Delta m^2_{41}$ the mass-square difference between $\nu_4$ and other three mass eigenstates, and $\sin^2(2\theta_{\alpha\beta})$ are defined as mixing angles. 

\begin{figure}[h!]
\centering
\includegraphics[trim = 1cm 16cm 11cm 1.5cm, clip=true, width=0.6\textwidth]{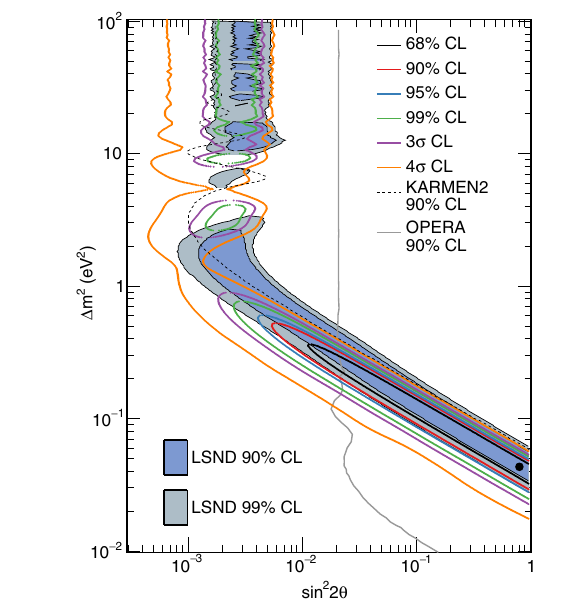}
\caption[MiniBooNE allowed regions for $\nu_{\mu} \rightarrow \nu_{e}$ oscillation within a
two-neutrino oscillation model using combined neutrino mode data sets.]{MiniBooNE allowed regions for $\nu_{\mu} \rightarrow \nu_{e}$ oscillation within a
two-neutrino oscillation model using combined neutrino mode data sets for events with $200 < E_{\nu}^{\text{QE}} < 3000$ MeV. The black point
shows the MiniBooNE best fit point at $(\sin^2{2\theta}, \Delta m^2) = (0.807;0.043 \text{ev}^2)$. Taken from Fig. 20 of Ref. \cite{MiniBooNE_combine_osc}}
\label{fig:miniboone_osc}
\end{figure}

\begin{figure}[h!]
\centering
\includegraphics[page=31, trim = 5cm 17.8cm 5cm 1.5cm, clip=true, width=0.49\textwidth]{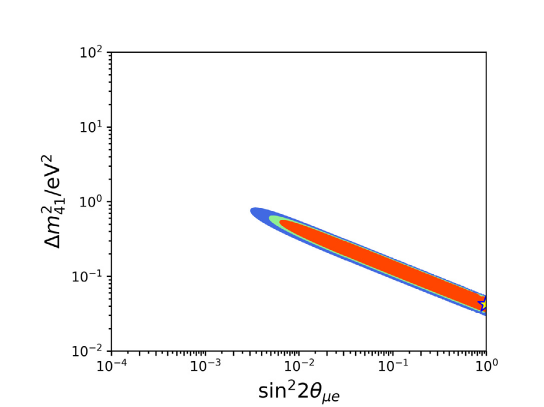}
\includegraphics[page=31, trim = 5cm 11.2cm 5cm 8.5cm, clip=true, width=0.49\textwidth]{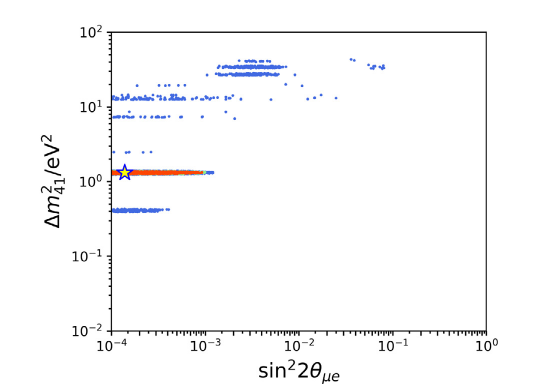}
\caption[Best-fit oscillation parameters and confidence intervals using frequentist method for (3+1) global fits to neutrino oscillation datasets.]{Best-fit oscillation parameters and confidence intervals using frequentist method for (3+1) global fits to only appearance (left) data sets and disappearance (right) datasets. Taken from Fig. 18 of Ref. \cite{global_fits_1}.}
\label{fig:3plus1_tension}
\end{figure}

MiniBooNE has performed neutrino oscillation fits to its data assuming only two neutrino hypothesis. The best-fit oscillation parameter and sensitivity curve is shown in Fig. \ref{fig:miniboone_osc}. As showns in Fig. \ref{fig:MiniBooNE_data_excess}, oscillation with (3+1) model is not sufficient to cover the large excess in the low energy region. Moreover, independent efforts have been put into global fits to (3+1) model with data from various short baseline oscillation experiments \cite{global_fits_1, global_fits_2, global_fits_3}. Consistent tension is found between the appearance data and disappearance data, as highlighted in Fig. \ref{fig:3plus1_tension}.

\begin{figure}[h!]
\centering
\includegraphics[page=4, trim = 1cm 19.5cm 11cm 1.5cm, clip=true, width=0.6\textwidth]{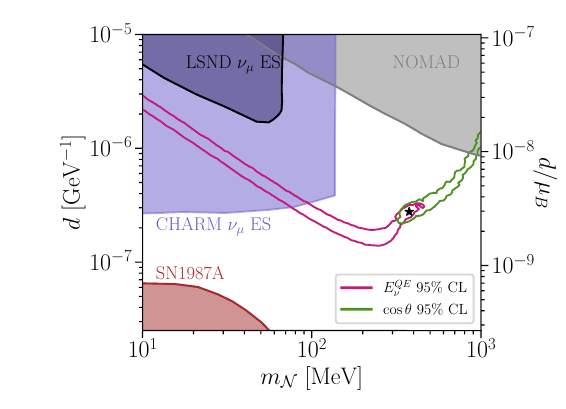}
\caption[Preferred regions of model parameters to explain the MiniBooNE excess for a proposed (3+N) model.]{Preferred regions to explain the MiniBooNE excess in $E_{\nu}^{\text{QE}}$ (pink) and $\cos{\theta}$ (green) as a function of model parameters: dipole coupling and heaviest neutrino mass. Taken from Fig. 2 of Ref. \cite{3plus1plusDecay}.}
\label{fig:3plus1_decay_sensitivity}
\end{figure}

\begin{figure}[h!]
\centering
\includegraphics[page=4, trim = 11cm 23cm 6cm 1.5cm, clip=true, width=0.49\textwidth]{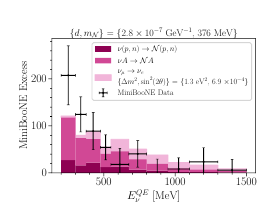}
\includegraphics[page=4, trim = 16cm 23cm 1cm 1.5cm, clip=true, width=0.49\textwidth]{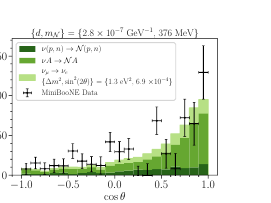}
\caption[Comparison between MiniBooNE observed excess and prediction of a representative point of a (3+1+decay) model.]{Comparison between MiniBooNE observed excess and prediction of a representative point of the (3+1+decay) model in $E_{\nu}^{\text{QE}}$ (left)) and $\cos{\theta}$ (right). Taken from Fig. 3 of Ref. \cite{3plus1plusDecay}.}
\label{fig:3plus1_decay_spectra}
\end{figure}

This motivates explorations of more general model like (3+N) model where there are N sterile neutrinos beyond three active ones. However the same issue would remain, as observed $\nu_{e}$ appearance would always imply sizable $\nu_{\mu}$ disappearance which is not found experimentally. More complex models have been also proposed to ``mitigate'' this tension. For example, reference \cite{3plus1plusDecay} considers a model with three active neutrinos and three sterile neutrinos, where the lowest mass eigenstate with mass $\sim \mathcal{O}(1)$eV would make observable neutrino oscillation signals at MiniBooNE, while the highest mass eigenstates with mass $\sim \mathcal{O}(100)$MeV will decay into active neutrino and photon, which will be mis-reconstructed and contribute to prediction at low-energy region. A region of parameter space is found to produce consistent spectra with both $E_{\nu}^{\text{QE}}$ and $\cos{\theta}$ distributions MiniBooNE observed shown in Fig.~\ref{fig:3plus1_decay_sensitivity} and prediction at selected point is compared to the MiniBooNE observed excess in Fig. \ref{fig:3plus1_decay_spectra} which shows good agreement.

\section{Heavy Sterile Neutrino Decay}
\begin{figure}[h!]
\centering
\includegraphics[page=6, trim = 1cm 19.8cm 1cm 1.5cm, clip=true, width=\textwidth]{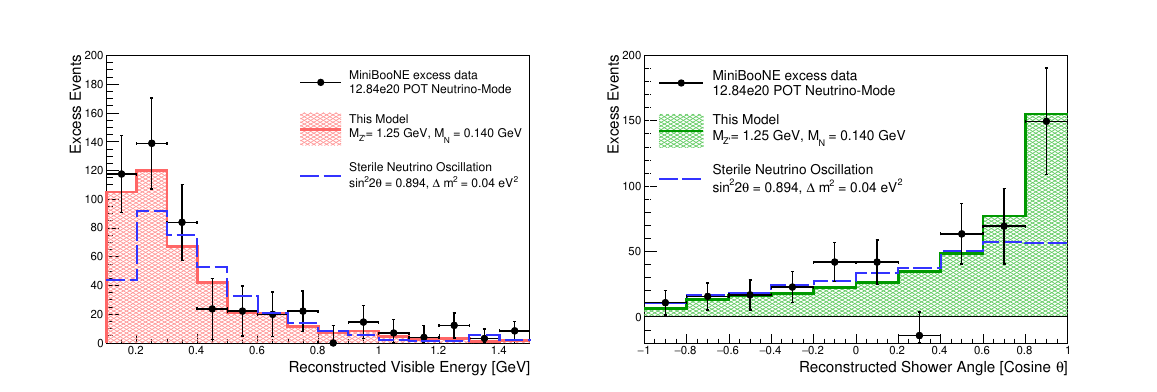}
\caption[Comparison between the MiniBooNE excess, and prediction from the model in Ref. \cite{dark_Z_epem1}.]{Comparison between the MiniBooNE excess, and prediction from the model in Ref. \cite{dark_Z_epem1} in both reconstructed visible energy (left) and reconstructed shower $\cos{\theta}$ (right), for a 0.14GeV sterile neutrino and 1.25 GeV $Z^{'}$. Taken from Fig. 2 of Ref. \cite{dark_Z_epem1}.}
\label{fig:dark_Zprime}
\end{figure}

Difficulty to explain MiniBooNE LEE with only neutrino oscillations also leads to proposal of more exotic models, such as heavy neutrino production and decay through mixing, transition magnetic moment or through dark photon discussed here. In these models, heavy neutrino are usually generated near or in the detector to be consistent with the timing distribution of MiniBooNE LEE \cite{MiniBooNE_combine_osc}. Inside the detector, heavy neutrino will decay promptly into active neutrino, of which $\nu_{e}$ will undergo CCQE interaction and contribute to the LEE \cite{ heavy_nu_decay_to_active}, or into active (or lighter sterile) neutrino and photon ($\gamma$) where $\gamma$ will be mis-reconstructed \cite{3plus1plusDecay, heavy_nu_decay_to_photon1, heavy_nu_decay_to_photon2}, or decay into $e^+e^-$ pairs \cite{dark_Z_epem1, dark_Z_epem2} which if sufficiently co-linear or asymmetric only one will be reconstructed and enter MiniBooNE's $\nu_{e}$ CCQE selection. Heavy neutrino production and decay into $e^+e^-$ pair mediated by dark photon with mass $\sim $MeV-GeV sufficiently produces more events in the low energy region while not violating the flat angle distribution seen by MiniBooNE, as highlighted in Fig. \ref{fig:dark_Zprime}. Analysis targeting a group of dark photon models is currently under active development in MicroBooNE.

\chapter{Weight Correction for Simulated NC coherent $1\gamma$ Signal}\label{app:signal_weight}
The POT normalization procedure described in Eq.~\ref{eq_POT} assumes the NC coherent $1\gamma$ simulation has contributions from all neutrino flavors, to be consistent with the flux uncertainty evaluation.  However, only $\nu_{\mu}$-induced NC coherent $1\gamma$ events were simulated with \textsc{genie}, and the shape of cross-section averaged flux $S(E_{\nu})*F(E_{\nu})$ slightly varies when contributions from all four neutrino flavors are considered instead of only $\nu_{\mu}$ contribution. As a result, each simulated signal event should carry a weight to take care of the shape difference between the two distributions. 

\begin{figure}[h]
\centering 
\begin{subfigure}[b]{0.49\textwidth}
         \centering
         \includegraphics[width=\textwidth]{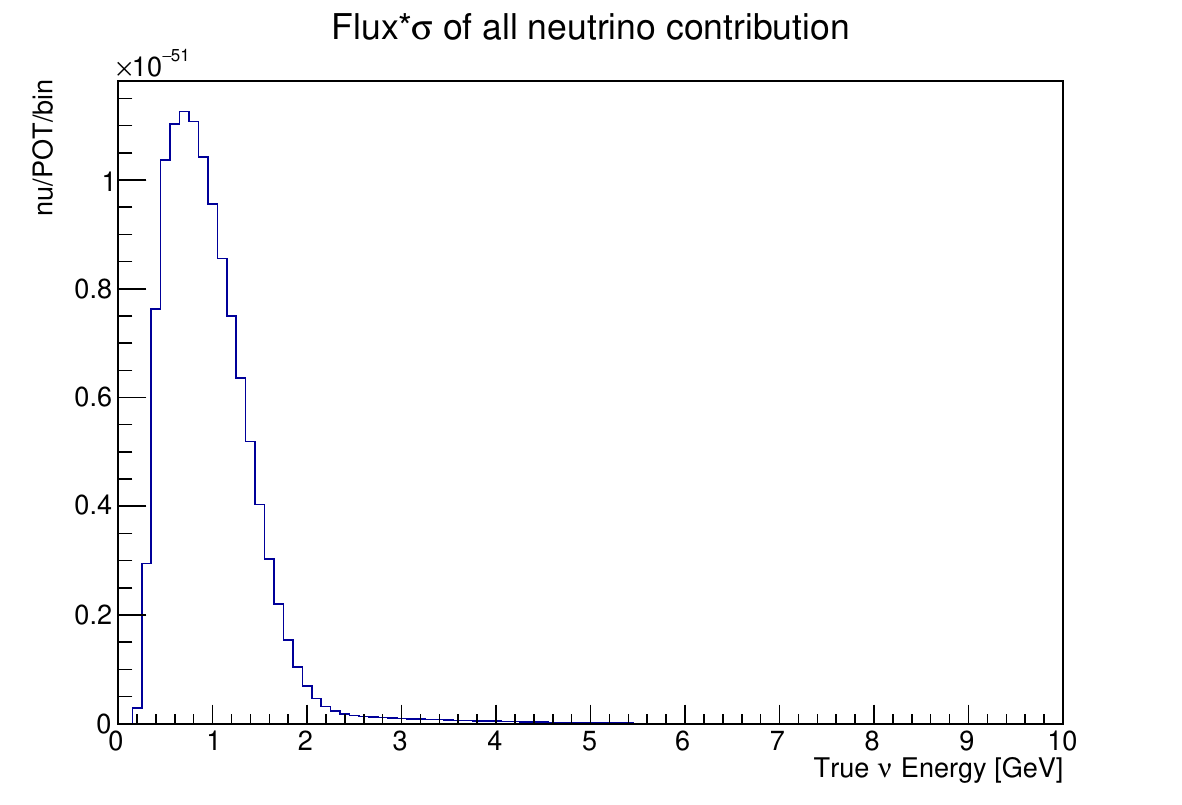}
         \caption{}
         \label{fig:averaged_flux_sum}
\end{subfigure}
\hfill
\begin{subfigure}[b]{0.49\textwidth}
         \centering
         \includegraphics[width=\textwidth]{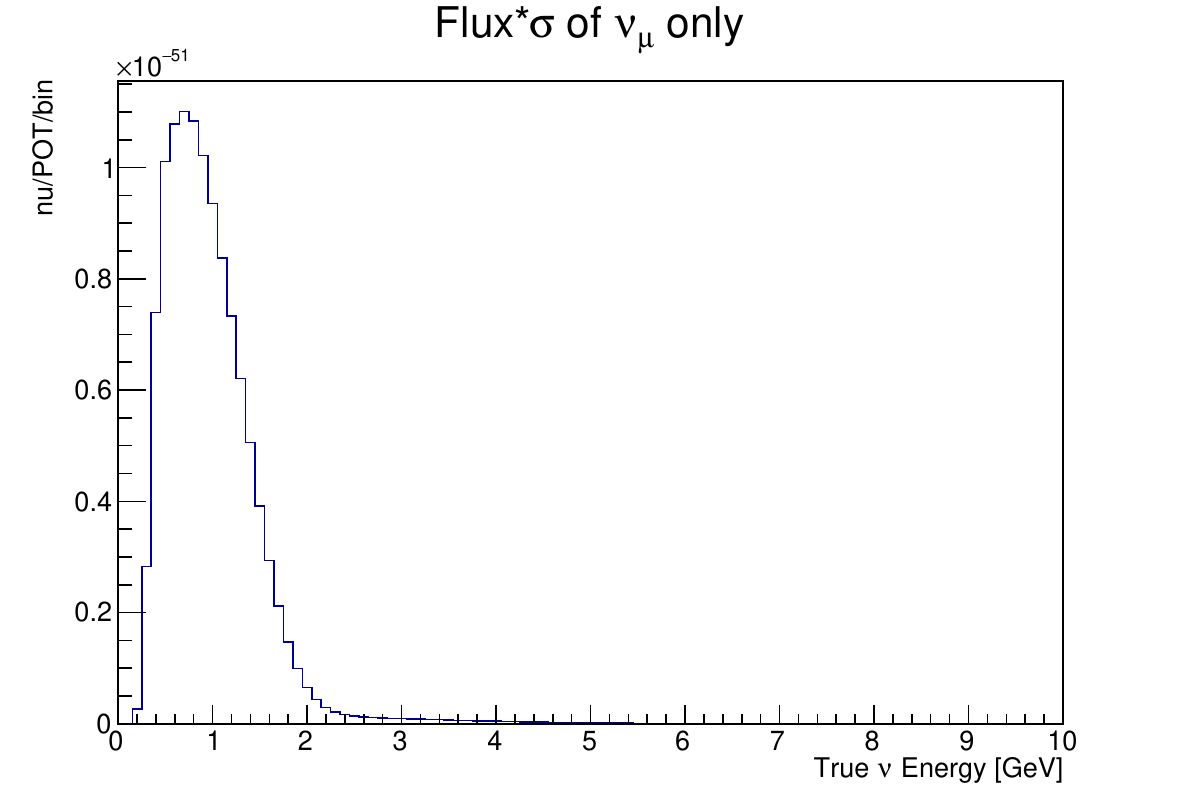}
         \caption{}
         \label{fig:averaged_flux_numu}
\end{subfigure}
\caption[Cross-section averaged neutrino flux for NC coherent $1\gamma$ process.]{Cross-section averaged neutrino flux. (a): summed cross-section averaged flux distribution for $\nu_{\mu}$, $\overline{\nu_{\mu}}$, $\nu_{e}$ and $\overline{\nu_{e}}$ in the beam (b) cross-section averaged flux distribution for $\nu_{\mu}$ only. }
\label{fig:signal_xs_averaged_flux}
\end{figure}

\begin{figure}[h]
\centering 
         \includegraphics[width=0.49\textwidth]{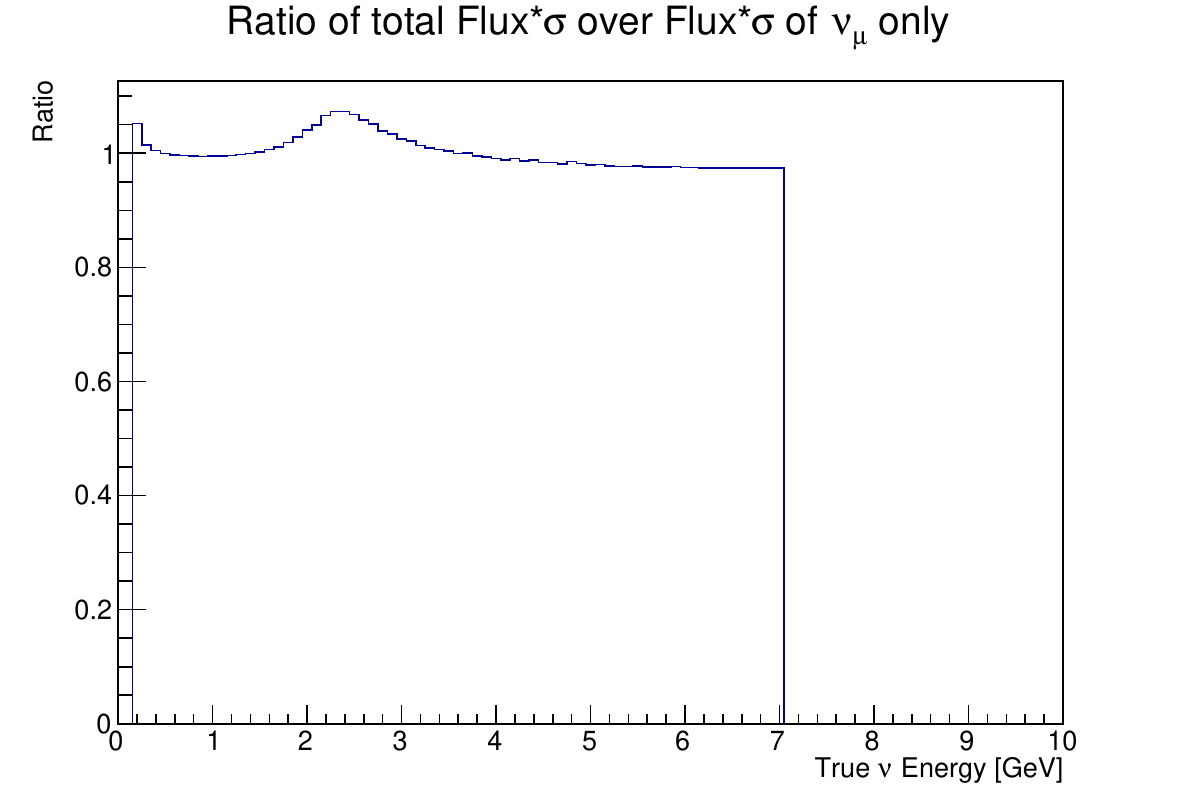}
\caption{The ratio between Fig.~\ref{fig:averaged_flux_sum} and Fig.~\ref{fig:averaged_flux_numu} after area-normalizing two distributions. }
\label{fig:averaged_flux_ratio}
\end{figure}

Figure~\ref{fig:signal_xs_averaged_flux} shows the distribution of cross-section averaged neutrino flux when the contributions from all neutrino flavors are taken into account, and only $\nu_{\mu}$ contribution respectively. These two distributions are area-normalization and the bin-to-bin ratio of Fig.~\ref{fig:averaged_flux_sum} over Fig.~\ref{fig:averaged_flux_numu} is calculated and applied as weight to every signal event according to their true neutrino energy. The resulting bin-wise weight is shown in Fig.~\ref{fig:averaged_flux_ratio}.

\end{appendices}